\documentclass{vldb}
\pdfoutput=1

\usepackage{graphicx}
\usepackage{balance}
\usepackage{times}
\newif\ifsigalternate

\usepackage{amsmath}
\usepackage{amssymb}
\usepackage{array}
\usepackage{balance}
\usepackage{booktabs}
\usepackage{boxedminipage}
\usepackage{blindtext}
\usepackage{bm}
\usepackage{caption}
\usepackage{color}
\usepackage{epstopdf}
\usepackage{fancybox}
\usepackage{fixltx2e}
\usepackage{float}
\usepackage{graphicx}
\usepackage{hyperref}
\usepackage[utf8x]{inputenc}
\usepackage{listings}
\usepackage{multirow}
\usepackage[noeepic]{qtree}
\usepackage{ragged2e}
\usepackage{soul}
\usepackage{setspace}
\usepackage{subcaption}
\usepackage{url}
\usepackage{verbatim}

\usepackage{tikz}
\usetikzlibrary{arrows}
\usetikzlibrary{backgrounds}
\usetikzlibrary{calc}

	\usepackage{subcaption}
	\usepackage{listings, lstautogobble}
\usepackage{algorithm}
\usepackage{algorithmicx}
\usepackage[noend]{algpseudocode}

\makeatletter
\newcommand*\bigcdot{\mathpalette\bigcdot@{.5}}
\newcommand*\bigcdot@[2]{\mathbin{\vcenter{\hbox{\scalebox{#2}{$\m@th#1\bullet$}}}}}
\makeatother
    \newcommand{\minus}{\scalebox{0.75}[1.0]{\( - \)}}

\newcolumntype{P}[1]{>{\centering\arraybackslash}p{#1}}

\def\InnerSep{1.8pt}

\usepackage{listings, lstautogobble}
\usepackage{xcolor}
\newcommand{\squishlist}{
  \begin{list}{$\bullet$}
   {
     \setlength{\itemsep}{0pt}
     \setlength{\parsep}{2pt}
     \setlength{\topsep}{1.0pt}
     \setlength{\partopsep}{0pt}
     \setlength{\leftmargin}{1.8em}
     \setlength{\labelwidth}{1em}
     \setlength{\labelsep}{0.5em}
   }
}

\newcommand{\squishend}{
   \end{list}
}

\renewcommand{\footnotesize}{}

\newtheorem{definition}{Definition}[section]
\newtheorem{example}[definition]{Example}

\usepackage{footnote}
\makesavenoteenv{tabular}
\graphicspath{ {images/} }

\vldbTitle{Optimizing Subgraph Queries by Combining Binary and Worst-Case Optimal Joins}
\vldbAuthors{Amine Mhedhbi, Semih Salihoglu}
\vldbDOI{https://doi.org/10.14778/xxxxxxx.xxxxxxx}
\vldbVolume{12}
\vldbNumber{xxx}
\vldbYear{2019}

\usepackage{alphalph}

\begin{document}

\newif\iflong
\longtrue

\title{Optimizing Subgraph Queries by Combining\\Binary and Worst-Case Optimal Joins}

\numberofauthors{2}

\author{
	\alignauthor{
		Amine Mhedhbi\\
		\affaddr{University of Waterloo}\\
		\email{amine.mhedhbi@uwaterloo.ca}
	}
	\alignauthor{
		Semih Salihoglu\\
		\affaddr{University of Waterloo}\\
		\email{semih.salihoglu@uwaterloo.ca}
	}
}

\maketitle

\begin{abstract}
\label{sec:abstract}

We study the problem of optimizing subgraph queries using the new worst-case optimal join plans. Worst-case optimal plans evaluate queries by matching one query vertex at a time using multiway intersections.  The core problem in optimizing worst-case optimal plans is to pick an ordering of the query vertices to match. We design a cost-based optimizer that (i) picks efficient query vertex orderings for worst-case optimal plans; and (ii) generates {\em hybrid plans} that mix traditional binary joins with worst-case optimal style multiway intersections. Our cost metric combines the cost of binary joins with a new cost metric called {\em intersec\-tion-cost}. The plan space of our optimizer contains plans that are not in the plan spaces based on tree decompositions from prior work.  In addition to our optimizer, we describe an {\em adaptive technique} that changes the orderings of the worst-case optimal sub-plans during query execution.
We demonstrate the effectiveness of the plans our optimizer picks and adaptive technique through extensive experiments. Our optimizer is integrated into the Graphflow DBMS.

\end{abstract}
\vspace{-8pt}
\section{Introduction}
\label{sec:introduction}

Subgraph queries, which find instances of a query subgraph \linebreak $Q(V_Q, E_Q)$ in an input graph $G(V, E)$, are a fundamental class of queries supported by graph databases. Subgraph queries appear in many applications where graph patterns reveal valuable information. For example, Twitter searches for diamonds in their follower network for recommendations~\cite{gupta:magicrecs}, clique-like structures in social networks indicate communities~\cite{newman:community}, and cyclic patterns in transaction networks indicate fraudulent activity~\cite{bodaghi:fraud, neo4j:fraud}. 

\begin{figure*}[t!]
	\centering
	\captionsetup{justification=centering}
    \begin{subfigure}[t]{0.315\textwidth}
    \centering
    \begin{tikzpicture}[
  level 1/.style={sibling distance=1.8cm,level distance=1.9cm},
  level 2/.style={sibling distance=1.8cm,level distance=1.6cm},
  level 3/.style={sibling distance=2.05cm,level distance=1.2cm},
  level 4/.style={sibling distance=1.75cm,level distance=1cm}]
  \node[draw,dashed,inner sep=\InnerSep] {
    \begin{tikzpicture}[scale=0.4,solid, transform shape,->,>=stealth', shorten >=1pt, auto,node distance=2cm, thick, main node/.style={circle,draw,font=\sffamily\Huge\bfseries}]
    \node[main node] (1) {$a_1$};
    \node[main node] (2) [above right of=1] {$a_2$};
    \node[main node] (3) [below right of=1] {$a_3$};
    \node[main node] (4) [below right of=2] {$a_4$};
    \path[every node/.style={font=\sffamily\small}]
    (1) edge (2) edge (3)
    (2) edge (4)
    (3) edge (4)
    (2) edge (3) ;
    \end{tikzpicture}
  }
    child {node[draw,dashed,inner sep=\InnerSep] {
    \begin{tikzpicture}[scale=0.4,solid, transform shape,->,>=stealth', shorten >=1pt, auto,node distance=2cm, thick, main node/.style={circle,draw,font=\sffamily\Huge\bfseries}]
    \node[main node] (1) {$a_1$};
    \node[main node] (2) [above right of=1] {$a_2$};
    \node[main node] (3) [below right of=1] {$a_3$};
    \node[main node] (4) [below right of=2] {$a_4$};
    \path[every node/.style={font=\sffamily\small}]
    (1) edge (2) edge (3)
    (3) edge (4)
    (2) edge (3) ;
\end{tikzpicture}
  }
    child {node[draw,dashed,inner sep=\InnerSep] {
    \begin{tikzpicture}[scale=0.4,solid, transform shape,->,>=stealth', shorten >=1pt, auto,node distance=2cm, thick, main node/.style={circle,draw,font=\sffamily\Huge\bfseries}]
        \node[main node] (2) {$a_2$};
        \node[main node] (1) [above right of=2] {$a_1$};
        \node[main node] (3) [below right of=1] {$a_3$};
        \path[every node/.style={font=\sffamily\small}]
        (1) edge (2) edge (3)
        (2) edge (3) ;
        \end{tikzpicture}
    }
      child {node[draw,dashed,inner sep=\InnerSep] {
    \begin{tikzpicture}[scale=0.4,solid, transform shape,->,>=stealth', shorten >=1pt, auto,node distance=2cm, thick, main node/.style={circle,draw,font=\sffamily\Huge\bfseries}]
        \node[main node] (1) {$a_1$};
        \node[main node] (2) [right of=1] {$a_2$};
        \node[main node] (3) [left of=1] {$a_3$};
        \path[every node/.style={font=\sffamily\small}]
        (1) edge (2)
        (1) edge (3) ;
        \end{tikzpicture}
      }
      child {node[draw,dashed,inner sep=\InnerSep] {
    \begin{tikzpicture}[scale=0.4,solid, transform shape,->,>=stealth', shorten >=1pt, auto,node distance=2cm, thick, main node/.style={circle,draw,font=\sffamily\Huge\bfseries}]
        \node[main node] (1) {$a_1$};
        \node[main node] (2) [right of=1] {$a_2$};
        \path[every node/.style={font=\sffamily\small}]
        (1) edge (2) ;
        \end{tikzpicture}
      }}
      child {node[draw,dashed,inner sep=\InnerSep] {
    \begin{tikzpicture}[scale=0.4,solid, transform shape,->,>=stealth', shorten >=1pt, auto,node distance=2cm, thick, main node/.style={circle,draw,font=\sffamily\Huge\bfseries}]
        \node[main node] (1) {$a_1$};
        \node[main node] (3) [right of=1] {$a_3$};
        \path[every node/.style={font=\sffamily\small}]
        (1) edge (3) ;
        \end{tikzpicture}
      }}
      }
      child {node[draw,dashed,inner sep=\InnerSep] {
    \begin{tikzpicture}[scale=0.4,solid, transform shape,->,>=stealth', shorten >=1pt, auto,node distance=2cm, thick, main node/.style={circle,draw,font=\sffamily\Huge\bfseries}]
        \node[main node] (2) {$a_2$};
        \node[main node] (3) [right of=2] {$a_3$};
        \path[every node/.style={font=\sffamily\small}]
        (2) edge (3) ;
        \end{tikzpicture}
      }}
    }
    child {node[draw,dashed,inner sep=\InnerSep] {
    \begin{tikzpicture}[scale=0.4,solid, transform shape,->,>=stealth', shorten >=1pt, auto,node distance=2cm, thick, main node/.style={circle,draw,font=\sffamily\Huge\bfseries}]
        \node[main node] (3) {$a_3$};
        \node[main node] (4) [right of=3] {$a_4$};
        \path[every node/.style={font=\sffamily\small}]
        (3) edge (4);
        \end{tikzpicture}
    }}
    }
    child {node[draw,dashed,inner sep=\InnerSep] {
    \begin{tikzpicture}[scale=0.4,solid, transform shape,->,>=stealth', shorten >=1pt, auto,node distance=2cm, thick, main node/.style={circle,draw,font=\sffamily\Huge\bfseries}]
        \node[main node] (2) {$a_2$};
        \node[main node] (4) [ right of=1] {$a_4$};
        \path[every node/.style={font=\sffamily\small}]
        (2) edge (4) ;
        \end{tikzpicture}
    }};
\end{tikzpicture}
	\caption{BJ plan.}
	\label{fig:ldbj-plan-example}
	\end{subfigure}
    \begin{subfigure}[t]{0.15\textwidth}
    \centering
    \begin{tikzpicture}[
    level 1/.style={sibling distance=2.4cm,level distance=2cm},
    level 2/.style={sibling distance=1cm,level distance=1.4cm},
    level 3/.style={sibling distance=1cm}]]
  \node[draw,dashed,inner sep=\InnerSep] {
    \begin{tikzpicture}[scale=0.4,solid, transform shape,->,>=stealth', shorten >=1pt, auto,node distance=2cm, thick, main node/.style={circle,draw,font=\sffamily\Huge\bfseries}]
    \node[main node] (1) {$a_1$};
    \node[main node] (2) [above right of=1] {$a_2$};
    \node[main node] (3) [below right of=1] {$a_3$};
    \node[main node] (4) [below right of=2] {$a_4$};
    \path[every node/.style={font=\sffamily\small}]
    (1) edge (2) edge (3)
    (2) edge (4)
    (3) edge (4)
    (2) edge (3) ;
    \end{tikzpicture}
  }
    child {node[draw,dashed,inner sep=\InnerSep] {
    \begin{tikzpicture}[scale=0.4,solid, transform shape,->,>=stealth', shorten >=1pt, auto,node distance=2cm, thick, main node/.style={circle,draw,font=\sffamily\Huge\bfseries}]
        \node[main node] (2) {$a_2$};
        \node[main node] (1) [above right of=2] {$a_1$};
        \node[main node] (3) [below right of=1] {$a_3$};
        \path[every node/.style={font=\sffamily\small}]
        (1) edge (2) edge (3)
        (2) edge (3) ;
        \end{tikzpicture}
      }
      child {node[draw,dashed,inner sep=\InnerSep] {
    \begin{tikzpicture}[scale=0.4,solid, transform shape,->,>=stealth', shorten >=1pt, auto,node distance=2cm, thick, main node/.style={circle,draw,font=\sffamily\Huge\bfseries}]
        \node[main node] (2) {$a_2$};
        \node[main node] (3) [right of=2] {$a_3$};
        \path[every node/.style={font=\sffamily\small}]
        (2) edge (3) ;
        \end{tikzpicture}
      }}
    }
    ;
\end{tikzpicture}
	\caption{WCO plan.}
	\label{fig:wco-plan-example}
	\end{subfigure}\hspace{-4mm}
    \begin{subfigure}[t]{0.2\textwidth}
    \centering
    \begin{tikzpicture}[
    level 1/.style={sibling distance=1.95cm,level distance=2cm},
    level 2/.style={sibling distance=1.5cm,level distance=1.4cm},
    level 3/.style={sibling distance=1.5cm,level distance=1.4cm}]
    \node[draw,dashed,inner sep=\InnerSep] {
    	\begin{tikzpicture}[scale=0.4,solid, transform shape,->,>=stealth', shorten >=1pt, auto,node distance=2cm, thick, main node/.style={circle,draw,font=\sffamily\Huge\bfseries}]
        \node[main node] (1) {$a_1$};
        \node[main node] (2) [above right of=1] {$a_2$};
        \node[main node] (3) [below right of=1] {$a_3$};
        \node[main node] (4) [below right of=2] {$a_4$};
        \path[every node/.style={font=\sffamily\small}]
        (1) edge (2) edge (3)
        (2) edge (4)
        (3) edge (4)
        (2) edge (3) ;
        \end{tikzpicture}
    }
    child {node[draw,dashed,inner sep=\InnerSep] {
    \begin{tikzpicture}[scale=0.4,solid, transform shape,->,>=stealth', shorten >=1pt, auto,node distance=2cm, thick, main node/.style={circle,draw,font=\sffamily\Huge\bfseries}]
        \node[main node] (2) {$a_2$};
        \node[main node] (1) [above right of=2] {$a_1$};
        \node[main node] (3) [below right of=1] {$a_3$};
        \path[every node/.style={font=\sffamily\small}]
        (1) edge (2) edge (3)
        (2) edge (3) ;
        \end{tikzpicture}
    }
      child {node[draw,dashed,inner sep=\InnerSep] {
    \begin{tikzpicture}[scale=0.4,solid, transform shape,->,>=stealth', shorten >=1pt, auto,node distance=2cm, thick, main node/.style={circle,draw,font=\sffamily\Huge\bfseries}]
        \node[main node] (2) {$a_2$};
        \node[main node] (3) [right of=2] {$a_3$};
        \path[every node/.style={font=\sffamily\small}]
        (2) edge (3) ;
        \end{tikzpicture}
      }}
    }
    child {node[draw,dashed,inner sep=\InnerSep] {
    \begin{tikzpicture}[scale=0.4,solid, transform shape,->,>=stealth', shorten >=1pt, auto,node distance=2cm, thick, main node/.style={circle,draw,font=\sffamily\Huge\bfseries}]
        \node[main node] (2) {$a_2$};
        \node[main node] (4) [above right of=2] {$a_4$};
        \node[main node] (3) [below right of=4] {$a_3$};
        \path[every node/.style={font=\sffamily\small}]
        (2) edge (4)
        (3) edge (4)
        (2) edge (3) ;
        \end{tikzpicture}
    }
      child {node[draw,dashed,inner sep=\InnerSep] {
    \begin{tikzpicture}[scale=0.4,solid, transform shape,->,>=stealth', shorten >=1pt, auto,node distance=2cm, thick, main node/.style={circle,draw,font=\sffamily\Huge\bfseries}]
        \node[main node] (2) {$a_2$};
        \node[main node] (3) [right of=2] {$a_3$};
        \path[every node/.style={font=\sffamily\small}]
        (2) edge (3) ;
        \end{tikzpicture}
      }}
    }
    ;
\end{tikzpicture}
	\caption{Hybrid plan.}
	\label{fig:hybrid-plan-example}
	\end{subfigure}
	\begin{subfigure}[t]{0.3\textwidth}
    \centering
    \begin{tikzpicture}[
		level 1/.style={sibling distance=1.95cm,level distance=1.8cm},
		level 2/.style={sibling distance=3.12cm,level distance=1.4cm},
		level 3/.style={sibling distance=1.58cm,level distance=1.4cm}]
		\node[draw,dashed,inner sep=\InnerSep] {
		\begin{tikzpicture}[scale=0.4,solid, transform shape,->,>=stealth', shorten >=1pt, auto,node distance=2cm, thick, main node/.style={circle,draw,font=\sffamily\Huge\bfseries}]
			\node[main node] (1) {$a_1$};
			\node[main node] (2) [right of=1] {$a_2$};
			\node[main node] (3) [right of=2] {$a_3$};
			\node[main node] (4) [below of=3] {$a_4$};
			\node[main node] (5) [left of=4]  {$a_5$};
			\node[main node] (6) [left of=5]  {$a_6$};
			\path[every node/.style={font=\sffamily\small}]
			(1) edge (2)
			(2) edge (3)
			(3) edge (4)
			(4) edge (5)
			(5) edge (6)
			(6) edge (1) ;
		\end{tikzpicture}
	}
	child {node[draw,dashed,inner sep=\InnerSep] {
		\begin{tikzpicture}[scale=0.4,solid, transform shape,->,>=stealth', shorten >=1pt, auto,node distance=2cm, thick, main node/.style={circle,draw,font=\sffamily\Huge\bfseries}]
			\node[main node] (1) {$a_1$};
			\node[main node] (2) [right of=1] {$a_2$};
			\node[main node] (3) [right of=2] {$a_3$};
			\node[main node] (4) [below of=3] {$a_4$};
			\node[main node] (5) [left of=4]  {$a_5$};
			\path[every node/.style={font=\sffamily\small}]
			(1) edge (2)
			(2) edge (3)
			(3) edge (4)
			(4) edge (5) ;
		\end{tikzpicture}
	}
	child {node[draw,dashed,inner sep=\InnerSep] {
		\begin{tikzpicture}[scale=0.4,solid, transform shape,->,>=stealth', shorten >=1pt, auto,node distance=2cm, thick, main node/.style={circle,draw,font=\sffamily\Huge\bfseries}]
			\node[main node] (1) {$a_1$};
			\node[main node] (2) [right of=1] {$a_2$};
			\node[main node] (3) [right of=2] {$a_3$};
			\path[every node/.style={font=\sffamily\small}]
			(1) edge (2)
			(2) edge (3) ;
		\end{tikzpicture}
	}
	child {node[draw,dashed,inner sep=\InnerSep] {
		\begin{tikzpicture}[scale=0.4,solid, transform shape,->,>=stealth', shorten >=1pt, auto,node distance=2cm, thick, main node/.style={circle,draw,font=\sffamily\Huge\bfseries}]
			\node[main node] (1) {$a_1$};
			\node[main node] (2) [right of=1] {$a_2$};
			\path[every node/.style={font=\sffamily\small}]
			(1) edge (2);
		\end{tikzpicture}
	}}
	child {node[draw,dashed,inner sep=\InnerSep] {
		\begin{tikzpicture}[scale=0.4,solid, transform shape,->,>=stealth', shorten >=1pt, auto,node distance=2cm, thick, main node/.style={circle,draw,font=\sffamily\Huge\bfseries}]
			\node[main node] (2) {$a_2$};
			\node[main node] (3) [right of=2] {$a_3$};
			\path[every node/.style={font=\sffamily\small}]
			(2) edge (3) ;
		\end{tikzpicture}
	}}
	}
	child {node[draw,dashed,inner sep=\InnerSep] {
			\begin{tikzpicture}[scale=0.4,solid, transform shape,->,>=stealth', shorten >=1pt, auto,node distance=2cm, thick, main node/.style={circle,draw,font=\sffamily\Huge\bfseries}]
			\node[main node] (3) {$a_3$};
			\node[main node] (4) [right of=3] {$a_4$};
			\node[main node] (5) [right of=4] {$a_5$};
			\path[every node/.style={font=\sffamily\small}]
			(3) edge (4)
			(4) edge (5) ;
		\end{tikzpicture}
	}
	child {node[draw,dashed,inner sep=\InnerSep] {
		\begin{tikzpicture}[scale=0.4,solid, transform shape,->,>=stealth', shorten >=1pt, auto,node distance=2cm, thick, main node/.style={circle,draw,font=\sffamily\Huge\bfseries}]
			\node[main node] (3) {$a_3$};
			\node[main node] (4) [right of=3] {$a_4$};
			\path[every node/.style={font=\sffamily\small}]
			(3) edge (4);
		\end{tikzpicture}
	}}
	child {node[draw,dashed,inner sep=\InnerSep] {
		\begin{tikzpicture}[scale=0.4,solid, transform shape,->,>=stealth', shorten >=1pt, auto,node distance=2cm, thick, main node/.style={circle,draw,font=\sffamily\Huge\bfseries}]
			\node[main node] (4) {$a_4$};
			\node[main node] (5) [right of=4] {$a_5$};
			\path[every node/.style={font=\sffamily\small}]
			(4) edge (5) ;
		\end{tikzpicture}
	}}
	}
	}
	;
	\end{tikzpicture}
	\caption{Non-GHD Hybrid plan.}
	\label{fig:non-ghd-hybrid-plan-example}
	\end{subfigure}
%
	\vspace{-5pt}
	\caption{Example plans. The subgraph on the top box of each plan is the actual query.}
	\label{fig:ldbj-wco-hybrid-example}
\end{figure*}
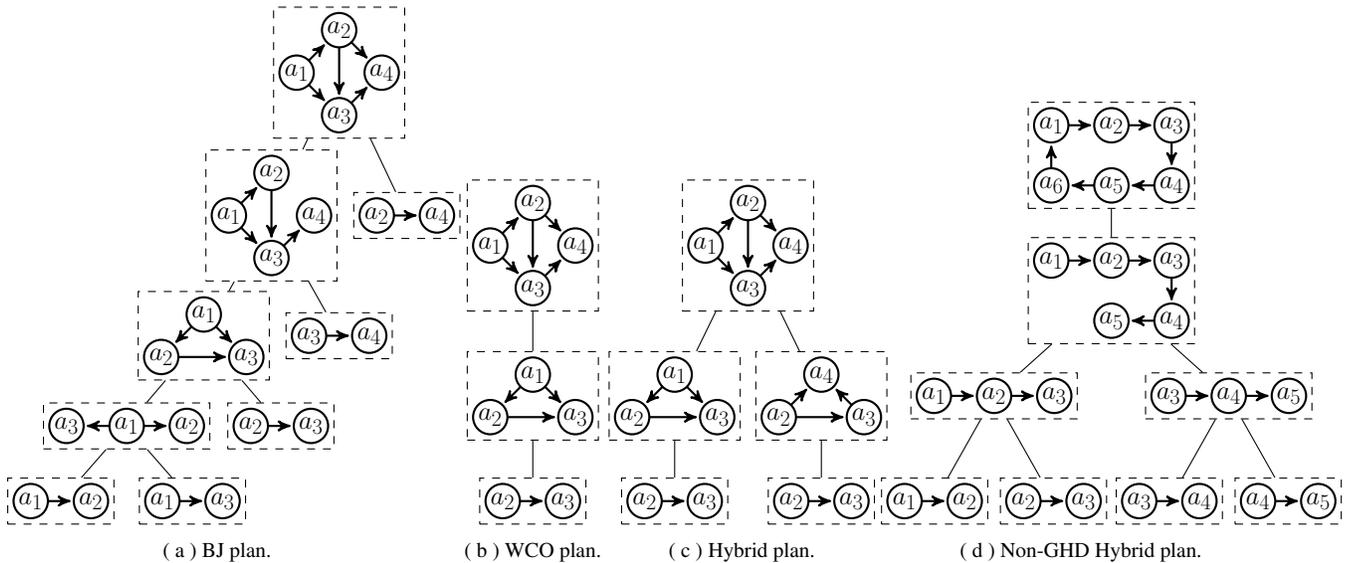

As observed in prior work~\cite{aberger:eh, ammar:bigjoin}, a subgraph query $Q$ is equivalent to a multiway self-join query that contains one $E$($a_i$,$a_j$) (for {\bf E}dge) relation for each 
$a_i$$\rightarrow$$a_j$ $\in$ $E_Q$.
The top box in Figure~\ref{fig:ldbj-plan-example} shows an example query, which we refer to as {\em diamond-X}. This query can be represented as:

\centerline{$ Q_{DX} = E_1 \bowtie E_2 \bowtie E_3 \bowtie E_4 \bowtie E_5 $}
\noindent where $E_1(a_1,a_2)$, $E_2(a_1,a_3)$, $E_3(a_2,a_3)$, $E_4(a_2,a_4)$, and  \linebreak $E_5(a_3,a_4)$ are copies of $E$($a_i$,$a_j$). We study evaluating a general class of subgraph queries where $V_Q$ and $E_Q$ can have labels. For labeled queries, the edge table corresponding to the query edge $a_i$$\rightarrow$$a_j$ contains only the edges in $G$ that are consistent with the labels on $a_i$, $a_j$, and $a_i$$\rightarrow$$a_j$. Subgraph queries are evaluated with two main approaches:
\squishlist
\item {\em Query-edge(s)-at-a-time} approach executes a sequence of binary joins to evaluate $Q$. Each binary join effectively matches a larger subset of the query edges of $Q$ in $G$ until $Q$ is matched.
\item {\em Query-vertex-at-a-time} approach picks a {\em query vertex ordering} $\sigma$ of $V_Q$ and matches $Q$ one query vertex at a time according to $\sigma$, using a multiway join operator that performs multiway intersections. This is the computation performed by the recent worst-case optimal join algorithms~\cite{ngo:survey, ngo:nprr, veldhuizen:lftj}. In graph terms, this computation intersects one or more adjacency lists of vertices to extend partial matches by one query vertex. 
\squishend

\noindent  We refer to plans with only binary joins as {\em BJ} plans, with only intersections as {\em WCO} (for {\bf w}orst-{\bf c}ase {\bf o}ptimal) plans, and both operations as {\em hybrid} plans.
Figures~\ref{fig:ldbj-plan-example}, \ref{fig:wco-plan-example}, and ~\ref{fig:hybrid-plan-example} show an example of each plan for the diamond-X query.

Recent theoretical results~\cite{atserias:agm, ngo:nprr}  showed that BJ plans can be suboptimal on cyclic queries and have asymptotically worse runtimes than the worst-case (i.e., maximum) output sizes of these queries. This worst-case output size is now known as a query's {\em AGM bound}.
These results also showed that WCO plans correct for this sub-optimality. However, this theory has two shortcomings. First, the theory gives no advice as to {\em how to pick a good query vertex ordering} for WCO plans. Second, the theory ignores plans with binary joins, which have been shown  to be efficient on many queries by decades-long research in databases as well as several recent work in the context of subgraph queries~\cite{aberger:eh, lai:seed}.

We study how to generate efficient plans for subgraph queries using a mix of worst-case optimal-style multiway intersections and binary joins. We describe a cost-based optimizer we developed for the Graphflow DBMS~\cite{kankanamge:graphflow} that generates BJ plans, WCO plans, as well as hybrid plans. Our cost metric for WCO plans capture the various runtime effects of query vertex orderings we have identified. Our plans are significantly more efficient than the plans generated by prior solutions using WCO plans that are either based on heuristics or have limited plan spaces. The optimizers of both native graph databases, such as Neo4j~\cite{neo4j}, as well as those that are developed on top of RDBMSs, such as SAP's graph database~\cite{sap:graph-story}, are often cost-based. As such, our work gives insights into how to integrate the new worst-case optimal join algorithms into the cost-based optimizers of existing systems.

\vspace{-2pt}
\subsection{Existing Approaches}
\label{subsec:existing-approaches}
Perhaps the most common approach adopted by graph databases (e.g. Neo4j), RDBMSs, and RDF systems~\cite{neumann:rdf3x, zeng:trinity}, is to evaluate subgraph queries with BJ plans. As observed in prior work~\cite{ngo:survey}, BJ plans are inefficient in highly-cyclic queries, such as cliques.
Several prior solutions, such as BiGJoin~\cite{ammar:bigjoin}, our prior work on Graphflow~\cite{kankanamge:graphflow}, and the LogicBlox system have studied evaluating queries with only WCO plans, which, as we demonstrate in this paper, are not efficient for acyclic and sparsely cyclic queries. In addition, these solutions either use simple heuristics to select query vertex orderings or arbitrarily select them.

The EmptyHeaded system~\cite{aberger:eh}, which is the closest to our work, is the only system we are aware of that mixes worst-case optimal joins with binary joins. EmptyHeaded plans are {\em generalized hypertree decompositions} (GHDs) of the input query $Q$. A GHD is effectively a join tree $T$ of $Q$, where each node of $T$  contains a sub-query of $Q$. EmptyHeaded evaluates each sub-query using a WCO plan, i.e., using only multiway intersections, and then uses a sequence of binary joins to join the results of these sub-queries. As a cost metric, EmptyHeaded uses the {\em generalized hypertree widths} of GHDs and picks a minimum-width GHD. This approach has three shortcomings: (i) if the GHD contains a single sub-query, EmptyHeaded arbitrarily picks the query vertex ordering for that query, otherwise it picks the orderings for the sub-queries using a simple heuristic; (ii) the width cost metric depends only the input query $Q$, so when running $Q$ on different graphs, EmptyHeaded always picks the same plan; and (iii) the GHD plan space does not allow plans that can perform multiway intersections after binary joins.
As we demonstrate, there are efficient plans for some queries that {\em seamlessly mix binary joins and intersections} and do not correspond to any GHD-based plan of EmptyHeaded.

\subsection{Our Contributions}
\label{subsec:contributions}
Table~\ref{table:comparison-against-prior-work} summarizes how our approach compares against prior solutions.
Our first main contribution is a dynamic programming optimizer that generates plans with both binary joins and an \textsc{Extend/Intersect} operator that extends partial matches with one query vertex. Let $Q$ contain $m$ query vertices. Our optimizer enumerates plans for evaluating each $k$-vertex sub-query $Q_k$ of $Q$, for $k$=$2,...,m$,  with two alternatives: (i) a binary join of two smaller sub-queries $Q_{c1}$ and $Q_{c2}$; or (ii) by extending a sub-query $Q_{k\text{-}1}$ by one query vertex with an intersection. This  generates all possible WCO plans for the query as well as a large space of hybrid plans which are not in EmptyHeaded's plan space. Figure~\ref{fig:non-ghd-hybrid-plan-example} shows an example hybrid plan for the 6-cycle query that is not in EmptyHeaded's plan space.

For ranking WCO plans, our optimizer uses a new cost metric called {\em intersection cost} (i-cost). I-cost represents the amount of intersection work that a plan $P$ will do using information about the sizes of the adjacency lists that will be intersected throughout $P$. For ranking hybrid plans, we  combine i-cost with the cost of binary joins. Our cost metrics account for the properties of the input graph, such as the distributions of the forward and backward adjacency lists sizes and the number of matches of different subgraphs that will be computed as part of a plan. Unlike EmptyHeaded, this allows our optimizer to pick different plans for the same query on different input graphs. 
Our optimizer uses a {\em subgraph catalogue} to estimate i-cost, the cost of binary joins, and the number of partial matches a plan will generate. The catalogue contains information about: (i) the adjacency list size distributions of input graphs; and (ii) selectivity of different intersections on small subgraphs. 
\begin{savenotes}
\begin{table}[t!]
\captionsetup{justification=centering}
	\centering
	\begin{tabular}{>{\raggedright\arraybackslash}m{1.3cm} >{\raggedright\arraybackslash}m{2cm} >{\raggedright\arraybackslash}m{4.2cm}}\toprule
		    & \textbf{Q. Vertex Ordering}             & \textbf{Binary Joins} \\ 
		            \midrule
        BiGJoin & Arbitrarily        &   No \\ 
        \midrule
                LogicBlox & Heuristics or Cost-based\footnote{LogicBlox is not open-source. Two publications describe how the system picks query vertex orderings;  a heuristics-based~\cite{NguyenABKNRR15} and  a cost-based~\cite{aref:logicblox} technique (using sampling).} & No \\
        \midrule
		 EH    & Arbitrarily   &    Cost-based: depends on $Q$\\
	\midrule
        Graphflow & Cost-based \& Adaptive        &   Cost-based: depends on $Q$ and $G$ \\ 		      
		\bottomrule
	\end{tabular}
	\vspace{-5pt}
\caption{Comparisons against solutions using worst-case optimal joins. EH stands for EmptyHeaded.}
\vspace{-12pt}
\label{table:comparison-against-prior-work}
\end{table}
\end{savenotes}

Our second main contribution is an {\em adaptive technique} for picking the query vertex orderings of WCO parts of plans during query execution. 
Consider a WCO part of a plan that extend matches of sub-query $Q_i$  into a larger sub-query $Q_k$. Suppose there are $r$ possible query vertex orderings, $\sigma_1, ..., \sigma_r$, to perform these extensions. Our optimizer tries to pick the ordering $\sigma^*$ with the lowest cumulative i-cost when extending all partial matches of $Q_i$ in $G$. However, for any specific match $t$ of $Q_i$, there may be another $\sigma_j$ that is more efficient than $\sigma^*$. Our adaptive executor re-evaluates the cost of each $\sigma_j$ for $t$ based on the actual sizes of the adjacency lists of the vertices in $t$, and picks a new ordering for $t$.

We incorporate our optimizer into Graphflow~\cite{kankanamge:graphflow} and evaluate it across a large class of subgraph queries and input graphs. We show that our optimizer is able to pick close to optimal plans across a large suite of queries and our plans, including some plans that are not in EmptyHeaded's plan space, are up to 68x more efficient than EmptyHeaded's plans. We show that adaptively picking query vertex orderings improves the runtime of some plans by up to 4.3x, in some queries improving the runtime of every plan and makes our optimizer more robust against picking bad orderings. 
\iflong
For completeness, in Appendix~\ref{app:cfl} we include comparisons against Neo4j and another subgraph matching algorithm called CFL~\cite{cfl}. Both of these baselines were not as performant as our plans in our setting. 
\else
For completeness, in Appendix~\ref{app:cfl} we include comparisons against another subgraph matching algorithm called CFL~\cite{cfl}. In the longer version of our paper~\cite{mhedhbi:sqs-tech-report}, we also compare against Neo4j. Both of these baselines were not as performant as our plans in our setting. 
\fi

Table~\ref{table:abbreviations} summarizes the abbreviations used throughout the paper.

\begin{table}[t!]
\captionsetup{justification=centering}
	\centering
	\begin{tabular}{>{\raggedright\arraybackslash}m{1cm} >{\raggedright\arraybackslash}m{2.2cm}>{\raggedright\arraybackslash}m{1cm} >{\raggedright\arraybackslash}m{2.9cm}}\toprule
		    \textbf{Abbrv.}             & \textbf{Explanation} &    \textbf{Abbrv.}             & \textbf{Explanation}\\ 
        \midrule
        BJ        &   Binary Join  &  GHD        &   Generalized Hypertree Decompositions \\ 
        \midrule
         EH        &   EmptyHeaded & QVO & Query Vertex Ordering \\  
                 \midrule
         E/I        &   Extend/Intersect & WCO & Worst-case Optimal \\        
	\bottomrule
	\end{tabular}
	\vspace{-5pt}
\caption{Abbreviations used throughout the paper.}
\vspace{-12pt}
\label{table:abbreviations}
\end{table}
\section{Preliminaries}
\label{sec:preliminaries}

We assume a subgraph query $Q(V_Q, E_Q)$ is directed, connected, and has $m$ query vertices $a_1,..., a_m$ and $n$ query edges. To indicate the directions of query edges clearly, we use the $a_i$$\rightarrow$$a_j$ and $a_i$$\leftarrow$$a_j$ notation.
We assume that all of the vertices and edges in $Q$ have labels on them, which we indicate with $l(a_i)$, and $l(a_i$$\rightarrow$$a_j)$, respectively. Similar notations are used for the directed edges in the input graph $G(V, E)$. Unlabeled queries can be thought of as labeled queries on a version of $G$ with a single edge and single vertex label.
The outgoing and incoming neighbors of each $v \in V$ are indexed in forward and backward adjacency lists. We assume the adjacency lists are partitioned first by the edge labels and then by the labels of neighbor vertices. This allows, for example, detecting a vertex $v$'s forward edges with a particular edge label in constant time.
The neighbors in a partition are ordered by their IDs, which allow fast intersections.

Generic Join \cite{ngo:survey} is a $WCO$ join algorithm that evaluates queries one attribute at a time. We describe the algorithm in graph terms; reference~\cite{ngo:survey} gives an equivalent relational description. In graph terms, the algorithm evaluates  queries one query vertex at a time with two main steps:
\squishlist
\item {\bf  Query Vertex Ordering (QVO):} Generic Join first picks an order $\sigma$ of query vertices to match. For simplicity we assume $\sigma = a_1...a_m$ and the projection of $Q$ onto any prefix of $k$ query vertices for $k=1,...,m$ is connected.
\item {\bf  Iterative Partial Match Extensions:} Let $Q_k$$=$$\Pi_{a_1,...,a_k} Q$ be a sub-query that consists of $Q$'s projection on the first $k$ query vertices $a_1...a_k$. Generic Join iteratively computes $Q_1, ..., Q_m$. Let {\em partial k-match} (k-match for short) $t$ be any set of vertices of $V$ assigned to the first $k$ query vertices in $Q$.  For $i \le k$, let $t[i]$ be the  vertex matching $a_i$ in $t$. 
To compute $Q_k$, Generic Join extends each (k--1)-match $t'$ in the result of $Q_{k\text{--}1}$ 
to a possibly empty set of k-matches 
by intersecting the forward adjacency list of $t'[i]$ for each $a_i$$\rightarrow$$a_{k}$ $\in E_Q$ and the backward adjacency list of $t[i]$ for each $a_i$$\leftarrow$$a_{k}$$\in E_Q$, where $i$$\le$$k\text{--}1$. 
Let the result of this intersection be the \emph{extension set} $S$ of $t$. 
The k-matches $t$ produces is 
the Cartesian product of $t$ with $S$.
\squishend

\section{Optimizing WCO Plans}
\label{sec:wco-plan-space}
This section demonstrates our WCO plans, the effects of different QVOs we have identified, and our i-cost metric for WCO plans. Throughout this section we present several experiments on unlabeled queries for demonstration purposes.  
The datasets we use in these experiments are described in Table~\ref{table:datasets-used} in Section~\ref{sec:evaluation}.

\subsection{WCO Plans and E/I Operator}

\begin{table}[t!]
	\centering
	\captionsetup{justification=centering}
	\begin{tabular}{lllllllll}\toprule	
		& $\sigma_1$ & $\sigma_2$ & $\sigma_3$ & $\sigma_4$ & $\sigma_5$ & $\sigma_6$ & $\sigma_7$ & $\sigma_8$ \\
		\midrule
		Cache On  & 2.4 & 2.9 & 3.2 & 3.3 & 3.3 & 3.4 & 4.4 & 6.5 \\
		\midrule
		Cache Off & 3.8 & 3.2 & 3.2 & 3.3 & 3.3 & 3.4 & 8.5 & 10.7 \\
		\bottomrule
	\end{tabular}
	\vspace{-5pt}
	\caption{Experiment on intersection cache utility for diamond-X.}
	\label{tab:plan-spectrum-caching}
	\vspace{-12pt}
\end{table}

Each query vertex ordering $\sigma$ of $Q$ is effectively a different WCO plan for $Q$. 
Figure~\ref{fig:wco-plan-example} shows an example $\sigma$, which we represent as a chain of $m$--$1$ nodes, where the $(k\text{--}1)$'th node from the bottom contains a sub-query $Q_{k}$ which is the projection of $Q$ onto the first $k$ query vertices of $\sigma$. 
We use two operators to evaluate WCO plans:

\noindent {\bf \textsc{Scan}}: Leaf nodes of plans, which match a single query edge, are evaluated with a \textsc{Scan} operator. The operator scans the forward adjacency lists in $G$ that match the labels on the query edge, and its source and destination query vertices, and outputs each matched edge $u$$\rightarrow$$v \in E$ as a 2-match.

\noindent {\bf \textsc{Extend/Intersect (E/I)}}: Internal nodes labeled $Q_{k}$$(V_{k}$,\linebreak$E_{k})$ that have a child labeled $Q_{k\text{--}1}(V_{k\text{--}1},E_{k\text{--}1})$ are evaluated with an \textsc{E/I} operator. 
The \textsc{E/I} operator takes as input (k--1)-matches and extends each tuple $t$ to one or more k-matches. The operator is configured with one or more \emph{adjacency list descriptors} (descriptors for short) and a label $l_k$ for the destination vertex. Each descriptor is an (\texttt{i}, \texttt{dir}, $l_e$) triple, where \texttt{i} is the index of a vertex in $t$, \texttt{dir} is \texttt{forward} or \texttt{backward}, and $l_e$ is the label on the query edge the descriptor represents. For each (k--1)-match $t$, the operator first computes the extension set $S$ of $t$ by intersecting the adjacency lists described by its descriptors, ensuring they match the specified edge and destination vertex labels, and then extends $t$ to $t \times S$. When there is a single descriptor, $S$ is  the list described by the descriptor. Otherwise we use iterative 2-way in-tandem intersections.

Multiple (k--1)-matches that are processed consecutively in an \textsc{E/I} operator may require the same extension set if they perform the same intersections. Our \textsc{E/I} operator caches and reuses the last extension set $S$ in such cases. We store the cached set in a flat array buffer. The intersection cache overall improves the performance of WCO plans. As a demonstrative example, Table~\ref{tab:plan-spectrum-caching} shows the runtime of all WCO plans for the diamond-X query with caching enabled and disabled on the Amazon graph. The orderings in the table are omitted. 4 of the 8 plans utilize the intersection cache and improve their run time, one by 1.9x.

\subsection{Effects of QVOs}
\label{subsec:qvo-effects}
The work done by a WCO plan is commensurate with the ``amount of intersections'' it performs. 
Three main factors affect intersection work and therefore the runtime of a WCO plan $\sigma$: (1) directions of the adjacency lists $\sigma$ intersects; (2) the amount of intermediate partial matches $\sigma$ generates; and (3) how much $\sigma$ utilizes the intersection cache. We discuss each effect next.

\vspace{-1pt}
\subsubsection{Directions of Intersected Adjacency Lists}
\label{subsubsec:directions-of-alds}
 Perhaps surprisingly, there are WCO plans that have very different runtimes {\em only because} they compute their intersections using different directions of the adjacency lists. The simplest example of this is the asymmetric triangle query $a_1$$\rightarrow$$a_2$, $a_2$$\rightarrow$$a_3$, $a_1$$\rightarrow$$a_3$. This query has 3 QVOs, all of which have the same \textsc{Scan} operator, which scans each $u$$\rightarrow$$v$  edge in $G$, followed by 3 different intersections (without utilizing the intersection cache):
\squishlist
\item $\sigma_1$:$a_1a_2a_3$: intersects both $u$ and $v$'s forward lists.
\item $\sigma_2$:$a_2a_3a_1$: intersects both $u$ and $v$'s backward lists.
\item $\sigma_3$:$a_1a_3a_2$: intersects $u$'s forward, $v$'s backward list.
\squishend
\noindent
Table \ref{table:direction-alds-experiment} shows a demonstrative experiment studying the performance of each plan on the BerkStan and LiveJournal graphs (the i-cost column in the table will be discussed in Section~\ref{subsec:i-cost} momentarily).
For example,  $\sigma_1$ is 12.1x faster than $\sigma_2$ on the BerkStan graph.
Which combination of adjacency list directions is more efficient depends on the structural properties of the input graph, e.g., forward and backward adjacency list distributions.

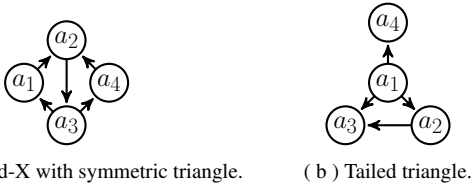
\begin{figure}[t!]
	\captionsetup{justification=centering}
	\begin{subfigure}[b]{0.3\textwidth}
		\centering
		\begin{tikzpicture}[scale=0.4, transform shape,->,>=stealth', shorten >=1pt, auto,node distance=2cm, thick, main node/.style={circle,draw,font=\sffamily\Huge\bfseries}]
		\node[main node] (1) {$a_1$};
		\node[main node] (2) [above right of=1] {$a_2$};
		\node[main node] (3) [below right of=1] {$a_3$};
		\node[main node] (4) [below right of=2] {$a_4$};
		\path[every node/.style={font=\sffamily\small}]
		(1) edge (2)
		(2) edge (3)
		(3) edge (1) edge (4)
		(4) edge (2) ;
		\end{tikzpicture}
		\caption{Diamond-X with symmetric triangle.}
		\label{fig:symmetric-diamond-x}
	\end{subfigure}
	\begin{subfigure}[b]{0.17\textwidth}
		\centering
		\begin{tikzpicture}[scale=0.4, transform shape,->,>=stealth', shorten >=1pt, auto,node distance=2cm, thick, main node/.style={circle,draw,font=\sffamily\Huge\bfseries}]
		\node[main node] (1) {$a_1$};
		\node[main node] (2) [below right of=1] {$a_2$};
		\node[main node] (3) [below left of=1] {$a_3$};
		\node[main node] (4) [above of=1] {$a_4$};
		\path[every node/.style={font=\sffamily\small}]
		(1) edge (2) edge (3) edge (4)
		(2) edge (3) ;
		\end{tikzpicture}
		\caption{Tailed triangle.}
		\label{fig:partial-matches-query}
	\end{subfigure}
	\vspace{-12pt}
	\caption{Queries used to demonstrate the effects of QVOs.}
	\vspace{-5pt}
\end{figure}

\begin{table}[t!]
	\centering
	\captionsetup{justification=centering}
	\begin{tabular}{rrrrrrrrr}\toprule
		& \multicolumn{3}{r}{BerkStan} & \multicolumn{3}{r}{Live Journal} \\
        \midrule
                QVO & time    & part. m. & i-cost & time & part. m. & i-cost \\
        \midrule
		$a_1a_2a_3$ &     2.6 & 8M     & 490M   & 64.4 & 69M      &  13.1B \\
		$a_2a_3a_1$ &    15.2 & 8M     & 55.8B  & 75.2 & 69M      &  15.9B \\
		$a_1a_3a_2$ &    31.6 & 8M     & 55.9B  & 79.1 & 69M      &  17.3B \\
		\bottomrule
	\end{tabular}
         \vspace{-5pt}
	\caption{Runtime (secs), intermediate partial matches (part. m.), and i-cost of different QVOs for the asymmetric triangle query.}
	\label{table:direction-alds-experiment}
	 \vspace{-3pt}
\end{table}

\subsubsection{Number of Intermediate Partial Matches}

Different WCO plans generate different partial matches leading to different amount of intersection work. 
Consider the {\em tailed triangle} query in Figure \ref{fig:partial-matches-query}, which can be evaluated by
two broad categories of WCO plans:
\squishlist
\item \textsc{Edge-2Path:} Some plans, such as QVO $a_1a_2a_4a_3$,  extend scanned edges $u$$\rightarrow$$v$ to 2-edge paths ($u$$\rightarrow$$v$$\leftarrow$$w$), and then close a triangle from one of 2 edges in the path. 
\item \textsc{Edge-Triangle}: Another group of plans, such as QVO \linebreak $a_1a_2a_3a_4$, extend scanned edges to triangles and then extend the triangles by one edge.
\squishend

\noindent Let $|E|$, $|2Path|$, and $|$$\bigtriangleup$$|$ denote the number of edges, 2-edge paths, and triangles. Ignoring the directions of extensions and intersections, the 
\textsc{Edge-2Path} plans do $|E|$ many extensions plus $|2Path|$ many intersections, whereas the  \textsc{Edge-Triangle} plans do $|E|$ many intersections and $|$$\bigtriangleup$$|$ many extensions. Table \ref{table:partial-matches} shows the run times of the different plans on Amazon and Epinions graphs with intersection caching disabled (again the i-cost column will be discussed momentarily). The first 3 rows are the \textsc{Edge-Triangle} plans. \textsc{Edge-Triangle} plans are significantly faster than \textsc{Edge-2Path} plans because in unlabeled queries $|2Path|$ is always at least $|$$\bigtriangleup$$|$ and often much larger. Which QVOs will generate fewer intermediate matches depend on several factors: (i) the structure of the query; (ii) for labeled queries, on the selectivity of the labels on the query; and (3) the structural properties of the input graph, e.g., graphs with low clustering coefficient generate fewer intermediate triangles than those with a high clustering coefficient.

\begin{table}[t!]
	\centering
	\captionsetup{justification=centering}
	\begin{tabular}{rrrrrrrrr}\toprule
		& \multicolumn{3}{r}{Amazon} & \multicolumn{3}{r}{Epinions} \\
		\midrule
		QVO            & time  & part. m. & i-cost & time & part. m. & i-cost  \\
		\midrule
		$a_1a_2a_3a_4$ & 0.9 & 15M    & 176M & 0.9  & 4M  & 0.9B  \\
		$a_1a_3a_2a_4$ & 1.4 & 15M    & 267M & 1.0  & 4M  & 0.9B  \\
		$a_2a_3a_1a_4$ & 2.4 & 15M    & 267M & 1.7  & 4M  & 1.0B  \\
		$a_1a_4a_2a_3$ & 4.3 & 35M    & 640M & 56.5 & 55M  & 32.5B \\
		$a_1a_4a_3a_2$ & 4.6 & 35M    & 1.4B & 72.0 & 55M & 36.5B \\
		\bottomrule
	\end{tabular}
	\vspace{-5pt}
	\caption{Runtime (secs), intermediate partial matches (part. m.), and i-cost of different QVOs for the tailed triangle query.}
	\vspace{-3pt}
	\label{table:partial-matches}
\end{table}

\begin{table}[t!]
	\centering
	\captionsetup{justification=centering}
	\begin{tabular}{rrrrrrrrrr}\toprule
		& \multicolumn{3}{r}{Amazon} & \multicolumn{3}{r}{Epinions} \\
		\midrule
		QVO & time & part. m. & i-cost & time & part. m. & i-cost \\
		\midrule
		$a_2a_3a_1a_4$ & 1.0 & 11M & 0.1B & 0.9 & 2M & 0.1B \\
		$a_1a_2a_3a_4$ & 3.0 & 11M & 0.3B & 4.0 & 2M & 1.0B \\
		\bottomrule
	\end{tabular}
	\caption{Runtime (secs), intermediate partial matches (part. m.), and i-cost of some QVOs for the symmetric diamond-X query.}
	\label{tab:intersection-cache-effect}
	\vspace{-12pt}
\end{table}

\subsubsection{Intersection Cache Hits} The intersection cache of our \textsc{E/I} operator is utilized more if the QVO extends (k--1)-matches to $a_k$ using adjacency lists with indices from $a_1...a_{k\text{--}2}$. 
Intersections that access the (k--1)$^{th}$ index cannot be reused
because $a_{k\text{--}1}$ is the result of an intersection performed in a previous \textsc{E/I} operator and will match to different vertex IDs. 
Instead, those accessing indices $a_1...a_{k-2}$ can potentially be reused.
We demonstrate that some plans perform significantly better than others only because they can utilize the intersection cache. Consider a variant of the diamond-X query in Figure~\ref{fig:symmetric-diamond-x}. One type of WCO plans for this query extend $u$$\rightarrow$$v$ edges to $(u,v,w)$ symmetric triangles by intersecting $u$'s backward and $v$'s forward adjacency lists. Then each triangle is extended to complete the query, intersecting again the forward and backward adjacency lists of one of the edges of the triangle.
There are two sub-groups of QVOs that fall under this type of plans: (i)
$a_2a_3a_1a_4$ and $a_2a_3a_4a_1$, which are equivalent plans due to symmetries in the query, so will perform exactly the same operations; and (ii) $a_1a_2a_3a_4$, $a_3a_1a_2a_4$, $a_3a_4a_2a_1$, and $a_4a_2a_3a_1$, which are also equivalent plans. Importantly, all of these plans cumulatively perform exactly the same intersections but those in group (i) and (ii) have different orders in which these intersections are performed, which lead to different intersection cache utilizations.

Table~\ref{tab:intersection-cache-effect} shows the performance of one representative plan from each sub-group: $a_2a_3a_1a_4$ and $a_1a_2a_3a_4$, on several graphs. The $a_2a_3a_1a_4$ plan is 4.4x faster on Epinions and 3x faster on Amazon. This is because when $a_2a_3a_1a_4$ extends $a_2a_3a_1$ triangles to complete the query, it will be accessing $a_2$ and $a_3$, so the first two indices in the triangles. For example if ($a_2$$=$$v_0$, $a_3$$=$$v_1$) extended to $t$ triangles $(v_0, v_1, v_2)$,...,$(v_0, v_1, v_{t+2})$, these partial matches will be fed into the next $\textsc{E/I}$ operator consecutively, and their extensions to $a_4$ will all require intersecting $v_0$ and $v_1$'s backward adjacency lists, so the cache would avoid $t$--$1$ intersections. Instead, the cache will not be utilized in the $a_1a_2a_3a_4$ plan. Our cache gives benefits similar to {\em factorization}~\cite{olteanu:size-bounds}. In factorized processing, the results of a query are represented as Cartesian products of independent components of the query.
In this case, matches of $a_1$ and $a_4$ are independent and can be done once for each match of $a_2a_3$.
A study of factorized processing is an interesting topic for future work.

\subsection{Cost Metric for WCO Plans}
\label{subsec:i-cost}
We introduce a new cost metric called {\em intersection cost} (i-cost), which we define as the size of adjacency lists that will be accessed and intersected by different WCO plans. Consider a WCO plan $\sigma$ that evaluates  sub-queries $Q_2$,...,$Q_{m}$, respectively, where $Q$$=$$Q_m$. Let $t$ be a (k--1)-match of $Q_{k\text{--}1}$ and suppose $t$ is extended to instances of $Q_{k}$ by intersecting a set of adjacency lists, described with  adjacency list descriptors $A_{k\text{--}1}$. 
Formally, i-cost of $\sigma$ is:
\begin{equation}
\label{eq:i-cost}
\sum_{Q_{k\text{--}1} \in Q_2...Q_{m-1}} \; \sum_{t \in Q_{k\text{--}1}} \; \sum_{\substack{(i, dir) \in A_{k\text{--}1} \\ \text{s.t. (i, dir) is accessed}}} |t[i].dir|
\end{equation}
We discuss how we estimate i-costs of plans in Section~\ref{sec:catalogue}. For now, note that Equation~\ref{eq:i-cost} captures the three effects of QVOs we identified:  (i) the $|t.dir|$ quantity captures the sizes of the adjacency lists in different directions; (ii) the second summation is over all intermediate matches, capturing the size of intermediate partial matches; and (iii) the last summation is over all adjacency lists that are accessed, so ignores the lists in the intersections that are cached. For the demonstrative experiments we presented in the previous section, we also  report the {\em actual} i-costs of different plans in Tables~\ref{table:direction-alds-experiment},~\ref{table:partial-matches}, and~\ref{tab:intersection-cache-effect}. The actual i-costs were computed in a profiled re-run of each experiment. Notice that in each experiment, i-costs of plans rank in the correct order of runtimes of plans.

There are alternative cost metrics from literature, such as the $C_{out}$~\cite{cluet:cout} and $C_{mm}$~\cite{leis:cmm} metrics, that would also do reasonably well in differentiating good and bad WCO plans.
However, these metrics capture only the effect of the number of intermediate match\-es. For example, they would not differentiate the plans in the asymmetric triangle query or the symmetric diamond-X query, i.e., the plans in Tables~\ref{table:direction-alds-experiment} and~\ref{tab:intersection-cache-effect} have the same actual $C_{out}$ and $C_{mm}$ costs.
\vspace{-7pt}
\section{Full Plan Space \& DP Optimizer}
\label{sec:full-plan-space}
In this section we describe our full plan space, which contain plans that include binary joins in addition to the \textsc{E/I} operator, the costs of these plans, and our dynamic programming optimizer.

\subsection{Hybrid Plans and HashJoin Operator}
\label{subsec:hybrid-plan-space}
In Section~\ref{sec:wco-plan-space}, we represented a WCO plan $\sigma$ as a chain, where each internal node $o_k$ had a single child labeled with $Q_k$, which was the projection of $Q$ onto the first $k$ query vertices in $\sigma$. 
A plan in our full plan space is a rooted tree as follows. Below, $Q_k$ refers to a projection of $Q$ onto an arbitrary set of $k$ query vertices.
\squishlist
\item Leaf nodes are labeled with a single query edge of $Q$.
\item Root is labeled with $Q$.
\item Each internal node $o_k$ is labeled with $Q_k$$=$$\{V_k, E_k\}$, with the {\em projection constraint} that $Q_k$ is a projection of $Q$ onto a subset of query vertices. $o_k$ has either one child or two children. If $o_k$ has one child $o_{k\text{--}1}$ with label $Q_{k\text{--}1}$$=$$\{V_{k\text{--}1}$, $E_{k\text{--}1}\}$, then $Q_{k\text{--}1}$
is a subgraph of $Q_k$  with one query vertex $q_v \in V_k$ and $q_v$'s incident edges in $E_k$ missing. This represents a WCO-style extension of partial matches of $Q_{k\text{--}1}$ by one query vertex to $Q_k$. If $o_k$ has two children $o_{c1}$ and $o_{c2}$ with labels $Q_{c1}$ and $Q_{c2}$, respectively, then $Q_k=Q_{c1} \cup Q_{c2}$ and $Q_k \neq Q_{c1}$ and $Q_k \neq Q_{c2}$. This represents a binary join of matches $Q_{c1}$ and $Q_{c2}$ to compute $Q_k$.
\squishend

As before, leaves map to \textsc{Scan} operator, an internal node $o_k$ with a single child maps to an \textsc{E/I} operator. If $o_k$ has two children, then it maps to a \textsc{Hash-Join} operator:

\noindent {\bf \textsc{Hash-Join}}: We use the classic hash join operator, which first creates a hash table of all of the tuples of $Q_{c1}$ on the common query vertices between $Q_{c1}$ and $Q_{c2}$. The table is then probed for each tuple of $Q_{c2}$.

Our plans are highly expressive and contain several classes of plans: (1) WCO plans from the previous section, in which each internal node has one child; (2) BJ plans, in which each node has two children and satisfy the projection constraint; and (3)  hybrid plans that satisfy the projection constraint. 
\iflong
We show in Appendix~\ref{app:eh-plan-space} that our hybrid plans contain EmptyHeaded's minimum-width GHD-based hybrid plans that satisfy the projection constraint. 
\else
In the longer version of this paper~\cite{mhedhbi:sqs-tech-report}, we show that our hybrid plans contain the minimum-width GHD-based hybrid plans that satisfy the projection constraint. 
\fi
For example the hybrid plan in Figure~\ref{fig:hybrid-plan-example} corresponds to a GHD for the diamond-X query with width 3/2. In addition, our plan space also contains hybrid plans that do not correspond to a GHD-based plan. Figure~\ref{fig:non-ghd-hybrid-plan-example} shows an example hybrid plan for the 6-cycle query that is not in EmptyHeaded's plan space. As we show in our evaluations, such plans can be very efficient for some queries.

The projection constraint prunes two classes of plans: 
\squishlist
\item[1.] Our plan space does not contain  BJ plans that first compute open triangles and then close them. Such BJ plans are in the plan spaces of existing optimizers, e.g., PostgreSQL, MySQL, and Neo4j. This is not a disadvantage because for each such plan, there is a more efficient WCO plan that computes triangles directly with an intersection of two already-sorted adjacency lists, avoiding the computation of open triangles. 
\item[2.] More generally, some of our hybrid plans contain the same query edge $a_i$$\rightarrow$$a_j$ in multiple parts of the join tree, which may look redundant because $a_i$$\rightarrow$$a_j$ is effectively joined multiple times. There can be alternative plans that  remove $a_i$$\rightarrow$$a_j$ from all but one of the sub-trees.
For example, consider the two hybrid plans $P_1$ and $P_2$ for the diamond-X query ($P_1$ is repeated from Figure~\ref{fig:hybrid-plan-example}). $P_2$ is not in our plan space because it does not satisfy the projection constraint because $a_2$$\rightarrow$$a_3$ is not in the right sub-tree.
Omitting such plans is also not a disadvantage because we duplicate $a_i$$\rightarrow$$a_j$ only if it closes cycles in a sub-tree, which  effectively is an additional filter that reduces the partial matches of the sub-tree. 
For example, on the Amazon graph, $P_1$ takes 14.2 seconds and $P_2$ 56.4 seconds.
\end{list}

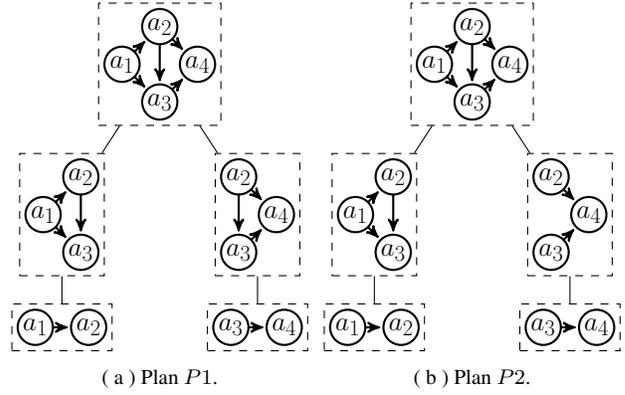
\begin{figure}
	\captionsetup{justification=centering}
	\centering
	\begin{subfigure}[b]{0.23\textwidth}
		\centering
		\begin{tikzpicture}[
		level 1/.style={sibling distance=2.6cm,level distance=2cm},
		level 2/.style={sibling distance=1cm,level distance=1.5cm},
		level 3/.style={sibling distance=1cm}]
		\node[draw,dashed,inner sep=\InnerSep] {
			\begin{tikzpicture}[scale=0.41,solid, transform shape,->,>=stealth', shorten >=1pt, auto,node distance=1.73cm, thick, main node/.style={circle,draw,font=\sffamily\Huge\bfseries}]
			\node[main node] (1) {$a_1$};
			\node[main node] (2) [above right of=1] {$a_2$};
			\node[main node] (3) [below right of=1] {$a_3$};
			\node[main node] (4) [below right of=2] {$a_4$};
			\path[every node/.style={font=\sffamily\small}]
			(1) edge (2) edge (3)
			(2) edge (4)
			(3) edge (4)
			(2) edge (3) ;
			\end{tikzpicture}
		}
		child {node[draw,dashed,inner sep=\InnerSep] {
				\begin{tikzpicture}[scale=0.41,solid, transform shape,->,>=stealth', shorten >=1pt, auto,node distance=1.73cm, thick, main node/.style={circle,draw,font=\sffamily\Huge\bfseries}]
				\node[main node] (1) {$a_1$};
				\node[main node] (2) [above right of=1] {$a_2$};
				\node[main node] (3) [below right of=1] {$a_3$};
				\path[every node/.style={font=\sffamily\small}]
				(1) edge (2) edge (3)
				(2) edge (3) ;
				\end{tikzpicture}
			}
			child {node[draw,dashed,inner sep=\InnerSep] {
					\begin{tikzpicture}[scale=0.41,solid, transform shape,->,>=stealth', shorten >=1pt, auto,node distance=1.73cm, thick, main node/.style={circle,draw,font=\sffamily\Huge\bfseries}]
					\node[main node] (1) {$a_1$};
					\node[main node] (2) [right of=1] {$a_2$};
					\path[every node/.style={font=\sffamily\small}]
					(1) edge (2) ;
					\end{tikzpicture}
			}}
		}
		child {node[draw,dashed,inner sep=\InnerSep] {
				\begin{tikzpicture}[scale=0.41,solid, transform shape,->,>=stealth', shorten >=1pt, auto,node distance=1.73cm, thick, main node/.style={circle,draw,font=\sffamily\Huge\bfseries}]
				\node[main node] (4) {$a_4$};
				\node[main node] (2) [above left of=4] {$a_2$};
				\node[main node] (3) [below left of=4] {$a_3$};
				\path[every node/.style={font=\sffamily\small}]
				(2) edge (3) edge (4)
				(3) edge (4);
				\end{tikzpicture}
			}
			child {node[draw,dashed,inner sep=\InnerSep] {
					\begin{tikzpicture}[scale=0.41,solid, transform shape,->,>=stealth', shorten >=1pt, auto,node distance=1.73cm, thick, main node/.style={circle,draw,font=\sffamily\Huge\bfseries}]
					\node[main node] (3) {$a_3$};
					\node[main node] (4) [right of=3] {$a_4$};
					\path[every node/.style={font=\sffamily\small}]
					(3) edge (4) ;
					\end{tikzpicture}
			}}
		}
		;
	\end{tikzpicture}
	\caption{Plan $P1$.}
	\label{fig:dec-share}
\end{subfigure}    
\begin{subfigure}[b]{0.23\textwidth}
	\centering
	\begin{tikzpicture}[
	level 1/.style={sibling distance=2.6cm,level distance=2cm},
	level 2/.style={sibling distance=1cm,level distance=1.5cm},
	level 3/.style={sibling distance=1cm}]
	\node[draw,dashed,inner sep=\InnerSep] {
		\begin{tikzpicture}[scale=0.41,solid, transform shape,->,>=stealth', shorten >=1pt, auto,node distance=1.73cm, thick, main node/.style={circle,draw,font=\sffamily\Huge\bfseries}]
		\node[main node] (1) {$a_1$};
		\node[main node] (2) [above right of=1] {$a_2$};
		\node[main node] (3) [below right of=1] {$a_3$};
		\node[main node] (4) [below right of=2] {$a_4$};
		\path[every node/.style={font=\sffamily\small}]
		(1) edge (2) edge (3)
		(2) edge (4)
		(3) edge (4)
		(2) edge (3) ;
		\end{tikzpicture}
	}
	child {node[draw,dashed,inner sep=\InnerSep] {
			\begin{tikzpicture}[scale=0.41,solid, transform shape,->,>=stealth', shorten >=1pt, auto,node distance=1.73cm, thick, main node/.style={circle,draw,font=\sffamily\Huge\bfseries}]
			\node[main node] (1) {$a_1$};
			\node[main node] (2) [above right of=1] {$a_2$};
			\node[main node] (3) [below right of=1] {$a_3$};
			\path[every node/.style={font=\sffamily\small}]
			(1) edge (2) edge (3)
			(2) edge (3) ;
			\end{tikzpicture}
		}
		child {node[draw,dashed,inner sep=\InnerSep] {
				\begin{tikzpicture}[scale=0.41,solid, transform shape,->,>=stealth', shorten >=1pt, auto,node distance=1.73cm, thick, main node/.style={circle,draw,font=\sffamily\Huge\bfseries}]
				\node[main node] (1) {$a_1$};
				\node[main node] (2) [right of=1] {$a_2$};
				\path[every node/.style={font=\sffamily\small}]
				(1) edge (2) ;
				\end{tikzpicture}
		}}
	}
	child {node[draw,dashed,inner sep=\InnerSep] {
			\begin{tikzpicture}[scale=0.41,solid, transform shape,->,>=stealth', shorten >=1pt, auto,node distance=1.73cm, thick, main node/.style={circle,draw,font=\sffamily\Huge\bfseries}]
			\node[main node] (4) {$a_4$};
			\node[main node] (2) [above left of=4] {$a_2$};
			\node[main node] (3) [below left of=4] {$a_3$};
			\path[every node/.style={font=\sffamily\small}]
			(2) edge (4)
			(3) edge (4);
			\end{tikzpicture}
		}
		child {node[draw,dashed,inner sep=\InnerSep] {
				\begin{tikzpicture}[scale=0.41,solid, transform shape,->,>=stealth', shorten >=1pt, auto,node distance=1.73cm, thick, main node/.style={circle,draw,font=\sffamily\Huge\bfseries}]
				\node[main node] (3) {$a_3$};
				\node[main node] (4) [right of=3] {$a_4$};
				\path[every node/.style={font=\sffamily\small}]
				(3) edge (4) ;
				\end{tikzpicture}
		}}
	}
	;
\end{tikzpicture}
\caption{Plan $P2$.}
\label{fig:dec-not-share}
\end{subfigure}
\vspace{-7pt}
\caption{Two plans: $P1$ shares a query edge and $P2$ does not.}
\label{fig:two-dec}
\vspace{-15pt}
\end{figure}

\subsection{Cost Metric for General Plans}
\label{subsec:full-plan-cost-metric}
A \textsc{Hash-Join} operator performs a very different computation than \textsc{E/I} operators, so the cost of \textsc{Hash-Join} needs to be normalized with i-cost. This is an approach taken by DBMSs to merge costs of multiple operators, e.g., a scan and a group-by, into a single cost metric.
Consider a \textsc{Hash-Join} operator $o_k$ that will join matches of $Q_{c1}$ and $Q_{c2}$ to compute $Q_k$. Suppose there are $n_{1}$ and $n_{2}$ instances of $Q_{c1}$ and $Q_{c2}$, respectively. Then $o_k$ will hash $n_{1}$ number of tuples into a table and probe this table $n_{2}$ times. 
We compute two weight constants $w_1$ and $w_2$ and calculate the cost of $o_k$ as $w_1n_1$ + $w_2n_2$ i-cost units. These weights can be hardcoded as done in the $C_{mm}$ cost metric~\cite{leis:cmm}, but we pick them empirically. In particular we run experiments in which we profile plans with \textsc{E/I} and \textsc{Hash-Join} operators and we log the (i-cost, time) pairs for the \textsc{E/I} operators, and the ($n_1$, $n_2$,  time) triples for the \textsc{Hash-Join} operators. The (i-cost, time) pairs allows us to convert time unit in the triples to  i-cost units. We then pick $w_1$ and $w_2$ that best fit these converted ($n_1$, $n_2$, i-cost) triples.

\subsection{Dynamic Programming Optimizer}
\label{subsubsec:dp-optimizer}
\iflong
Algorithm~\ref{alg:dp-optimizer} shows the pseudocode of our optimizer. 
\else
\fi
We next describe our optimizer, whose pseudocode is in the longer version of our paper~\cite{mhedhbi:sqs-tech-report}. Our optimizer takes as input a query $Q(V_Q, E_Q)$. We start by enumerating and computing the cost of all WCO plans. We discuss this step momentarily. We then initialize the cost of computing 2-vertex sub-queries of $Q$, so each query edge of $Q$, to the selectivity of the label on the edge. Then starting from $k=3$ up to $|V_Q|$, for each $k$-vertex sub-query $Q_k$ of $Q$, we find the lowest cost plan $P^*_{Q_k}$ to compute $Q_k$ in three different ways: 
\squishlist
\item[(i)] $P^*_{Q_k}$ is the lowest cost WCO plan that we enumerated\iflong  (line 5).\else. \fi
\item[(ii)] $P^*_{Q_k}$ extends the best plan $P^*_{Q_{k\text{--}1}}$ for a $Q_{k\text{--}1}$ by an \textsc{E/I} operator ($Q_{k\text{--}1}$ contains one fewer query vertex than $Q_k$)\iflong (lines 7-10).\else. \fi
\item[(iii)] $P^*_{Q_k}$ merges two best plans $P^*_{Q_{c1}}$ and $P^*_{Q_{c2}}$ for $Q_{c1}$ and $Q_{c2}$, respectively, with a \textsc{Hash-Join}\iflong (lines 12-15).\else. \fi
\squishend

The best plan for each $Q_k$ is stored in a {\em sub-query map}. We enumerate all WCO plans because the best WCO plan $P^*_{Q_{k}}$ for $Q_k$ is not necessarily an extension of the best WCO plan $P^*_{Q_{k\text{--}1}}$ for a $Q_{k\text{--}1}$ by one query vertex. That is because $P^*_{Q_{k}}$ may be extending a worse plan $P^{bad}_{Q_{k\text{--}1}}$ for $Q_{k\text{--}1}$ if the last extension has a good intersection cache utilization. 
Strictly speaking, this problem can arise when enumerating hybrid plans too,  if an \textsc{E/I} operator in case (ii) above follows a \textsc{Hash-Join}.  A full plan space enumeration would avoid this problem completely but we adopt dynamic programming to make our optimization time efficient, i.e., to make our optimizer efficient, we are potentially sacrificing picking the optimal plan in terms of estimated cost. However,  we verified that our optimizer returns the same plan as a full enumeration optimizer in all of our experiments. So at least for our experiments here, we have not sacrificed optimality. 

Finally, our optimizer omits plans that contain a \textsc{Hash-Join} that can be converted to an \textsc{E/I}. Consider the $a_1$$\rightarrow$$a_2$$\rightarrow$$a_3$ query. Instead of using a \textsc{Hash-Join} to materialize the $a_2$$\rightarrow$$a_3$ edges and then probe a scan of $a_1$$\rightarrow$$a_2$ edges, it is more efficient to use an \textsc{E/I} to extend $a_1$$\rightarrow$$a_2$ edges to $a_3$ using $a_2$'s forward adjacency list. 

\iflong
\renewcommand{\algorithmicrequire}{\textbf{Input:}}
\begin{algorithm} [t!]
\caption{DP Optimization Algorithm}
\label{alg:dp-optimizer}
\begin{algorithmic}[1]
	\Require $Q(V_Q, E_Q)$
	\State WCOP = enumerateAllWCOPlans($Q$) // {\em wco plans}	
	\State QPMap: {\em init each $a_i$$\xrightarrow{\text{$l_e$}}$$a_j$'s cost to the $\mu(l_e)$}
	\For{$k$ = 3, ..., $|V_Q|$}
	\For{$V_k \subseteq V$ s.t. $|V_k|$=k}
	\State $Q_k(V_k, E_k)$=$\Pi_{V_k}Q$; bestP = WCOP($Q_k$); minC = $\infty$
	\State {\em // Find best plan that extends to $Q_k$ by one query vertex}
	\For{$v_j \in V_k$ let $Q_{k\text{--}1}(V_{k\text{--}1}, E_{k\text{--}1}) = \Pi_{V_k\text{--}v_j}Q_k$}
	\State P = QPMap($Q_{k\text{--}1}$).extend($Q_k$);
	\If{cost(P) $<$ minC}
	\State bestPlan = P;
	\EndIf
	\EndFor
	\State {\em // Find best plan that generates $Q_i$ with a binary join}
	\For{$V_{c1}$,$V_{c2}$$\subset$$V_k$: $Q_{c1}$=$\Pi_{V_{c1}}$$Q_k$,$Q_{c2}$=$\Pi_{V_{c2}}$$Q_k$} \label{line:split}
	\State P = join(QPMap($Q_{c1}$), QPMap($Q_{c2}$));
	\If{cost(P) $<$ minC)}
	\State bestPlan = P;
	\EndIf
	\EndFor
	\EndFor
	\State QPMap($Q_k$) = bestPlan;
	\EndFor
	\State \Return QPMap(Q);
\end{algorithmic}
\end{algorithm}
\fi

\subsection{Plan Generation For Very Large Queries}
\label{subsec:large-query-optimization}
Our optimizer can take a very long time to generate a plan for large queries. For example, enumerating only the best WCO plan for a 20-clique requires inspecting 20! different QVOs, which would be prohibitive. To overcome this, we further prune plans for queries with more than 10 query vertices as follows: 
\squishlist
\item We avoid enumerating all WCO plans. Instead, WCO plans get enumerated in the DP part of the optimizer. Therefore, we possibly ignore good WCO plans that benefit from the intersection cache. 
\item At each iteration $k$, we keep only a subset (5 by default) $k$-vertex sub-queries of $Q$ with the lowest cost plans. So we store a subset of sub-queries in our sub-query map and enumerate only the $Q_k$ that can be generated from the sub-queries we stored previously in the map.
\squishend

\vspace{-10pt}
\section{Cost \& Cardinality Estimation}
\label{sec:catalogue}
To assign costs to the plans we enumerate, we need to estimate: (1) the cardinalities of the partial matches different plans generate; (2) the i-costs of extending a sub-query $Q_{k\text{--}1}$ to $Q_{k}$ by intersecting a set of adjacency lists in an \textsc{E/I} operator; and (3) the costs of \textsc{Hash-Join} operators. We focus on the setting where each subquery $Q_k$ has labels on the edges and the vertices. We use a data structure called the  {\em subgraph catalogue} to make the estimations. Table~\ref{table:ex-catalogue} shows an example catalogue. 

Each entry contains a key ($Q_{k\text{--}1}$, $A$, $a_k^{l_k}$), where $A$ is a set of (labeled) query edges and $a_k^{l_k}$ is a query vertex with label $l_k$. Let $Q_k$ be the subgraph that extends $Q_{k\text{--}1}$ with a query vertex labeled with $a_k^{l_k}$ and query edges in $A$. 
Each entry contains two estimates for extending a match of a sub-query $Q_{k\text{--}1}$ to $Q_k$ by intersecting a set of adjacency lists described by $A$:
	\squishlist
	\item[1.]$|$A$|$: Average sizes of the lists in $A$ that are intersected.
	\item[2.] $\mu$($Q_k$): Average number of $Q_k$ that will extend from one $Q_{k\text{--}1}$, i.e., the average number of vertices that: (i) are in the extension set of intersecting the adjacency lists $A$; and (ii) have label $l_k$.
	\squishend
In Table~\ref{table:ex-catalogue}, the query vertices of the input subgraph $Q_{k\text{--}1}$ are shown with canonicalized integers, e.g., 0, 1 or 2, instead of the non-canonicalized $a_i$ notation we used before.
Note that $Q_{k\text{--}1}$ can be extended to $Q_{k}$ using different $A$ with different i-costs. The fourth and fifth entries of Table~\ref{table:ex-catalogue}, which extend a single edge to an asymmetric triangle, demonstrate this possibility.

\begin{table}[t!]
\captionsetup{justification=centering}
	\centering
	{\small
		\begin{tabular}{lllll}
		\hline
		($Q_{k\text{--}1}$ & $A$ & $l_k$) & $|$$A$$|$ & $\mu$($Q_k$) \\
		\hline
		($1^{l_a}$$\xrightarrow{\text{$l_x$}}$$2^{l_b}$; & $L_1$$:$$2$$\xrightarrow{\text{$l_x$}}$; & $3^{l_a}$) & $|$$L_1$$|$$:$4.5& 4.5 \\
		($1^{l_a}$$\xrightarrow{\text{$l_x$}}$$2^{l_b}$; & $L_1$$:$$2$$\xrightarrow{\text{$l_x$}}$; & $3^{l_b}$) & $|$$L_1$$|$$:$4.5& 2.4 \\
				($1^{l_a}$$\xrightarrow{\text{$l_x$}}$$2^{l_b}$; & $L_1$$:$$2$$\xrightarrow{\text{$l_y$}}$; & $3^{l_a}$) & $|$$L_1$$|$$:$8.0& 3.2 \\

		($1^{l_a}$$\xrightarrow{\text{$l_x$}}$$2^{l_a}$; & 
		$L_1$$:$$1$$\xrightarrow{\text{$l_x$}}$,
		$L_2$$:$$2$$\xrightarrow{\text{$l_x$}}$; & $3^{l_a}$) & $|$$L_1$$|$$:$4.2, $|$$L_2$$|$$:$5.1& 1.5 \\
		($1^{l_a}$$\xrightarrow{\text{$l_x$}}$$2^{l_a}$; & 
		$L_1$$:$$1$$\xleftarrow{\text{$l_x$}}$,
		$L_2$$:$$2$$\xleftarrow{\text{$l_x$}}$; & $3^{l_a}$) & $|$$L_1$$|$$:$9.8, $|$$L_2$$|$$:$8.4& 1.5 \\		
		(...; & ...; & ...) & ... & ... \\
		\cline{1-5}
		\end{tabular}
	}
	\caption{A subgraph catalogue. A is a set of adjacency list descriptors; $\mu$ is selectivity.}
	\label{table:ex-catalogue}
\vspace{-10pt}
\end{table}
\vspace{-2pt}
\subsection{Catalogue Construction}
For each input $G$, we construct a catalogue containing all entries that extend an at most $h$-vertex subgraph to an ($h$+1)-vertex subgraph. By default we set $h$ to 3. When generating a catalogue entry for extending $Q_{k\text{--}1}$ to  $Q_{k}$, we do not find all instances of $Q_{k\text{--}1}$ and extend them to $Q_{k}$. Instead we first sample $Q_{k\text{--}1}$. We take a WCO plan that extends $Q_{k\text{--}1}$ to $Q_{k}$. We then sample $z$ random edges (1000 by default) uniformly at random from $G$ in the \textsc{Scan} operator.
The last \textsc{E/I} operator of the plan extends each partial match $t$ it receives to $Q_{k}$ by intersecting the adjacency lists in $A$. The operator measures the size of the adjacency lists in $A$ and the number of $Q_{k}$'s this computation produced. These measurements are averaged and stored in the catalogue as $|$A$|$ and $\mu$($Q_{k}$) columns.

\vspace{0.01pt}
\subsection{Cost Estimations}
\label{subsec:cost-estimation}
We use the catalogue to do three estimations as follows:

\noindent {\bf 1. Cardinality of $Q_k$}: To estimate the cardinality of $Q_k$, we pick a WCO plan $P$ that computes $Q_k$ through a sequence of ($Q_{j\text{--}1}$, $A_j$, $l_{j}$) extensions. The estimated cardinality of  $Q_k$ is the product of the $\mu$($A_j$) of the  ($Q_{j\text{--}1}$, $A_j$, $l_{j}$) entries in the catalogue.
If the catalogue contains entries with up to $h$-vertex subgraphs and $Q_k$ contains more than $h$ nodes, some of the entries we need for estimating the cardinality of $Q_k$ will be missing. Suppose for calculating the cardinality of $Q_k$, we need the $\mu$($A_x$) of an entry ($Q_{x\text{--}1}$, $A_x$, $l_{x}$) that is missing because $Q_{x\text{--}1}$ contains $x\text{--}1$ $>$ $h$ query vertices. Let $z$$=$($x$--$h$--1). In this case, we remove each $z$-size set of query vertices $a_{1}, ... a_{z}$ from $Q_{x\text{--}1}$ and $Q_{x}$, and the adjacency list descriptors from $A_x$ that include $1, ..., z$ in their indices. Let ($Q_{y\text{--}1}$, $A_y$, $l_{y}$) be the entry we get after a removal. We look at the $\mu$($A_y$) of ($Q_{y\text{--}1}$, $A_y$, $l_{y}$) in the catalogue. Out of all such $z$ set removals, we use the minimum $\mu$($A_y$) we find. 

Consider a missing entry for extending $Q_{k\text{--}1}$= $1$$\rightarrow$$2$$\rightarrow$$3$ by one query vertex to $4$ by intersecting three adjacency lists all pointing to $4$ from $1$, $2$, and $3$. For simplicity, let us ignore the labels on query vertices and edges. The resulting sub-query $Q_{k}$ will have two triangles: (i) an asymmetric triangle touching edge $1$$\rightarrow$$2$; and (ii) a symmetric triangle touching $2$$\rightarrow$$3$.  Suppose  entries in the catalogue indicate that an edge on average extends to 10 asymmetric triangles but to 0 symmetric triangles. We estimate that $Q_{k\text{--}1}$ will extend to zero $Q_{k}$ taking the minimum of our two estimates.

\noindent {\bf 2. I-cost of \textsc{E/I} operator:}  Consider an \textsc{E/I} operator $o_k$ extending $Q_{k\text{--}1}$ to $Q_{k}$ using adjacency lists $A$.  
We have two cases: 

\squishlist
\item No intersection cache: When $o_k$ will not utilize the intersection cache, we estimate i-cost of $o_k$ as:
\begin{equation}
\label{eq:cost-estimate}
\text{i-cost}(o_k) = \mu(Q_{k\text{--}1}) \times \sum_{L_i \in A} |L_i|
\end{equation}
Here, $\mu(Q_{k\text{--}1})$ is the estimated cardinality of $Q_{k\text{--}1}$, and $|L_i|$ is the average size of the adjacency list $L_i \in A$ that are logged in the catalogue for entry  ($Q_{k\text{--}1}$, $A$,  $a_k^{l_k}$) (i.e., the $|$$A$$|$ column).
\item Intersection cache utilization: If two or more of the adjacency list in $A$, say $L_i$ and $L_j$, access the vertices in a partial match $Q_j$ that is smaller than $Q_{k\text{--}1}$,
 then we multiply the estimated sizes of $L_i$ and $L_j$ with the estimated cardinality of $Q_{j}$ instead of $Q_{k\text{--}1}$. This is because we infer that $o_k$ will utilize the intersection cache for intersecting $L_i$ and $L_j$.
\squishend
Reasoning about utilization of intersection cache is critical in picking good plans. For example, recall our experiment from Table~\ref{tab:plan-spectrum-caching}  to demonstrate that the intersection cache broadly improves all plans for the diamond-X query. Our optimizer, which is ``cache-conscious'' picks $\sigma_2$ ($a_2a_3a_4a_1$). Instead, if we ignore the cache  and make our optimizer ``cache-oblivious'' by always estimating i-cost with Equation~\ref{eq:cost-estimate}, it picks the slower $\sigma_4$ ($a_1a_2a_3a_4$) plan. Similarly, our cache-conscious optimizer picks  $a_2a_3a_1a_4$ in our experiment from Table~\ref{tab:intersection-cache-effect}. Instead, the cache-oblivious optimizer assigns the same estimated i-cost to plans $a_2a_3a_1a_4$ and  $a_1a_2a_3a_4$, so cannot differentiate between these two plans and picks one arbitrarily.

\noindent {3. \bf Cost of \textsc{Hash-Join} operator:} Consider a \textsc{Hash-Join} operator  joining $Q_b$ and $Q_p$. The estimated cost of this operator is simply $w_1n_1$ + $w_2n_2$ (recall Section~\ref{subsec:full-plan-cost-metric}), where $n_1$ and $n_2$ are now the estimated cardinalities of $Q_b$ and $Q_p$, respectively.

\subsection{Limitations}
\label{subsec:card-limitations}
Similar to Markov tables~\cite{aboulnaga:markov} and MD- and Pattern-tree summaries~\cite{maduko:md-tree}, our catalogue is an estimation technique that is based on storing information about small size subgraphs and extending them to make estimates about larger subgraphs.  
We review these techniques in detail and discuss our differences in Section~\ref{sec:rw}. Here, we discuss several limitations that are inherent in such techniques. 

First, our estimates (both for i-cost and cardinalities) get worse as the size of the subgraphs for which we make estimates increase beyond $h$. Equivalently, as $h$ increases, our estimates for fixed-size large queries get better. 
At the same time, the size of the catalogue increases significantly as $h$ increases.
Similarly, the size of the catalogue increases as graphs get more heterogenous, i.e., contain more labels. Second, using larger sample sizes, i.e., larger $z$ values, increase the accuracy of our estimates but require more time to construct the catalogue.
Therefore $h$ and $z$ respectively trade off catalogue size and creation time with the accuracy of estimates.
We provide demonstrative experiments of these tradeoffs in Appendix~\ref{app:catalogue} for cardinality estimates.
\section{Adaptive WCO Plan Evaluation}
\label{sec:adaptive-plans}

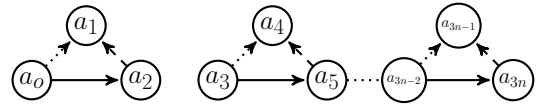
\begin{figure}
		\centering
	    \begin{tikzpicture}[scale=0.41, transform shape,->,>=stealth', shorten >=1pt, auto,node distance=2.5cm, thick, main node/.style={circle,draw,font=\sffamily\Huge\bfseries},minimum size=11mm]
	    \node[main node] (0) {$a_o$};
	    \node[main node] (1) [above right of=0] {$a_1$};
	    \node[main node] (2) [below right of=1] {$a_2$};
	    \node[main node] (3) [right of=2] {$a_3$};
        \node[main node] (4) [above right of=3] {$a_4$};
	    \node[main node] (5) [below right of=4] {$a_5$};
	    \node[main node] (6) [right of=5] {\Large{$a_{3n-2}$}};
        \node[main node] (7) [above right of=6] {\Large{$a_{3n-1}$}};
	    \node[main node] (8) [below right of=7] {\huge{$a_{3n}$}};
        \path[every node/.style={font=\sffamily\LARGE}]
	    (0) edge[dotted] (1) edge (2)
	    (2) edge [dashed](1)
        (3) edge[dotted] (4) edge (5)
	    (5) edge [dashed](4)
        (6) edge[dotted] (7) edge (8)
	    (8) edge [dashed](7);
		\path[-,every node/.style={font=\sffamily\LARGE}]
	    (5) edge[-,dotted] (6);
	    \end{tikzpicture}
	    \caption{Input graph for adaptive QVO example.}
	    \label{fig:3n-edge-graph}
	    \vspace{-7pt}
	\end{figure}
Recall that the $|A|$ and $\mu$ statistics stored in a catalogue entry ($Q_{k\text{--}1}$, $A$, $a_k^{l_k}$), are estimates of the adjacency list sizes (and selectivities) for matches of $Q_{k\text{--}1}$. These are  {\em estimates} based on {\em averages} over many sampled matches of $Q_{k\text{--}1}$. In practice, {\em actual} adjacency list sizes and selectivities of {\em individual} matches of $Q_{k\text{--}1}$ can be very different. Let us refer to parts of plans that are chains of one or more \textsc{E/I} operators as {\em WCO parts of plans}.  Consider a WCO part of a {\em fixed} plan $P$ that has a QVO $\sigma^*$ and extends partial matches of a sub-query $Q_i$ to matches of $Q_k$. Our optimizer picks $\sigma^*$ based on the estimates of the average statistics in the catalogue. Our adaptive evaluator updates our estimates for individual matches of $Q_{i}$ (and other sub-queries in this part of the plan)  based on actual statistics observed during evaluation and possibly changes $\sigma^*$ to another QVO for each individual match of $Q_i$.





\begin{example}
Consider the input graph $G$ shown in Figure~\ref{fig:3n-edge-graph}. $G$ contains 3n edges. 
Consider the diamond-X query and the WCO plan $P$ with $\sigma$$=$$a_2a_3a_4a_1$. Readers can verify that this plan will have an i-cost of 3n: 2n from extending solid edges, n from extending dotted edges, and 0 from extending dashed edges. 
 Now consider the following {\em adaptive} plan that picks $\sigma$ for the dotted and dashed edges as before but $\sigma'$$=$$a_2a_3a_1a_4$ for the solid edges. For the solid edges, $\sigma'$ incurs an i-cost of 0, reducing the i-cost to $n$. 
\end{example} 

\subsection{Adaptive Plans}
\begin{figure}
 \centering
 \captionsetup{justification=centering}
    \begin{tikzpicture}[grow=right,
    level 1/.style={sibling distance=2cm,level distance=1.95cm},
    level 2/.style={sibling distance=1cm,level distance=2.25cm},
    level 3/.style={sibling distance=1cm}]
    \node[draw,dashed,inner sep=\InnerSep] {
        \begin{tikzpicture}[scale=0.41,solid, transform shape,->,>=stealth', shorten >=1pt, auto,node distance=2cm, thick, main node/.style={circle,draw,font=\sffamily\Huge\bfseries}]
        \node[main node] (1) {$a_1$};
        \node[main node] (2) [right of=1] {$a_2$};
        \path[every node/.style={font=\sffamily\small}]
        (1) edge (2) ;
        \end{tikzpicture}
    }
    child {node[draw,dashed,inner sep=\InnerSep] {
        \begin{tikzpicture}[scale=0.41,solid, transform shape,->,>=stealth', shorten >=1pt, auto,node distance=2cm, thick, main node/.style={circle,draw,font=\sffamily\Huge\bfseries}]
        \node[main node] (1) {$a_1$};
        \node[main node] (3) [above right of=1] {$a_3$};
        \node[main node] (2) [below right of=3] {$a_2$};
        \path[every node/.style={font=\sffamily\small}]
        (1) edge (2) edge (3)
        (2) edge (3);
        \end{tikzpicture}
    }
      child {node[draw,dashed,inner sep=\InnerSep] {
              \begin{tikzpicture}[scale=0.41,solid, transform shape,->,>=stealth', shorten >=1pt, auto,node distance=2cm, thick, main node/.style={circle,draw,font=\sffamily\Huge\bfseries}]
        \node[main node] (1) {$a_1$};
        \node[main node] (2) [above right of=1] {$a_2$};
        \node[main node] (3) [below right of=1] {$a_3$};
        \node[main node] (4) [below right of=2] {$a_4$};
        \path[every node/.style={font=\sffamily\small}]
        (1) edge (2) edge (3)
        (2) edge (4)
        (3) edge (4)
        (2) edge (3) ;
        \end{tikzpicture}      
      }}
    }
    child {node[draw,dashed,inner sep=\InnerSep] {
        \begin{tikzpicture}[scale=0.41,solid, transform shape,->,>=stealth', shorten >=1pt, auto,node distance=2cm, thick, main node/.style={circle,draw,font=\sffamily\Huge\bfseries}]
        \node[main node] (1) {$a_1$};
        \node[main node] (2) [above right of=1] {$a_2$};
        \node[main node] (4) [below right of=2] {$a_4$};
        \path[every node/.style={font=\sffamily\small}]
        (1) edge (2)
        (2) edge (4);
        \end{tikzpicture}
    }
      child {node[draw,dashed,inner sep=\InnerSep] {
        \begin{tikzpicture}[scale=0.41,solid, transform shape,->,>=stealth', shorten >=1pt, auto,node distance=2cm, thick, main node/.style={circle,draw,font=\sffamily\Huge\bfseries}]
        \node[main node] (1) {$a_1$};
        \node[main node] (2) [above right of=1] {$a_2$};
        \node[main node] (3) [below right of=1] {$a_3$};
        \node[main node] (4) [below right of=2] {$a_4$};
        \path[every node/.style={font=\sffamily\small}]
        (1) edge (2) edge (3)
        (2) edge (4)
        (3) edge (4)
        (2) edge (3) ;
        \end{tikzpicture}
      }}
    }
    ;
\end{tikzpicture}
  	\caption{Example adaptive WCO plan (drawn horizontally).}
	\label{fig:adaptive-diamondx-example}
	\vspace{-10pt}
	
\end{figure}
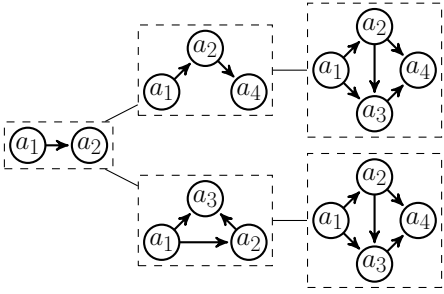

We optimize subgraph queries as follows. First, we get a {\em fixed} plan $P$ from our dynamic programming optimizer. If $P$ contains a chain of two or more \textsc{E/I} operators $o_i, o_{i+1} ..., o_{k}$, we replace it with an {\em adaptive} WCO plan. The adaptive plan extends the first partial matches $Q_i$ that $o_i$ takes as input in all possible (connected) ways to $Q_k$. In WCO plans $o_i$ is \textsc{Scan} and $Q_i$ is one query edge. Therefore in WCO plans, we fix the first two query vertices in a QVO and pick the rest adaptively. Figure~\ref{fig:adaptive-diamondx-example} shows the adaptive version of the fixed plan for the diamond-X query from Figure~\ref{fig:wco-plan-example}. We note that in addition to WCO plans, we adapt hybrid plans if they have a chain of two or more \textsc{E/I} operators. 

\subsection{Adaptive Operators}
\label{subsub:adaptive-ops}
Unlike the operators in fixed plans, our adaptive operators can feed their outputs to multiple operators. 
An adaptive operator $o_i$ is configured with a function $f$ that takes a partial match $t$ of  $Q_i$ and decides which of the next operators $t$ should be given. $f$ consists of two high-level steps: (1) For each possible $\sigma_j$ that can extend $Q_i$ to $Q_k$, $f$ re-evaluates the estimated i-cost of $\sigma_j$ by re-calculating the cost of plans using {\em updated cost estimates} (explained momentarily). $o_i$ gives $t$ to the next \textsc{E/I} operator of $\sigma_j^*$ that has the lowest re-calculated cost.
The cost of $\sigma_j$ is re-evaluated by changing the estimated adjacency list sizes that were used in cardinality and i-cost estimations with actual adjacency list sizes we obtain from $t$. 


\begin{example}
Consider the diamond-X query from Figure~\ref{fig:ldbj-plan-example} and suppose we have an adaptive plan in which the \textsc{Scan} operator matches edges to $a_2a_3$, so for each edge needs to decide whether to pick the ordering $\sigma_1:a_2a_3a_4a_1$ or $\sigma_2: a_2a_3a_1a_4$. Suppose the catalogue estimates the sizes of $|$$a_2$$\rightarrow$$|$ and $|$$a_3$$\rightarrow$$|$
as 100 and 2000, respectively. So we estimate the i-cost of  extending an $a_2a_3$ edge to $a_2a_3a_4$ as 2100. Suppose the selectivity $\mu_j$ of the number of triangles this intersection will generate is 10. Suppose \textsc{Scan} reads an edge $u$$\rightarrow$$v$ where $u$'s forward adjacency list size is 50 and $v$'s backward adjacency list size is 200. Then we update our i-cost estimate directly to 250 and $\mu_j$ to 10 $\times$ (50/100) $\times$ 200/2000=0.5.
\end{example}

As we show in our evaluations, adaptive QVO selection improves the performance of many WCO plans but more importantly guards our optimizer from picking bad QVOs. 
\section{System Implementation}
\label{sec:implementation}
We build our new techniques on top of Graphflow DBMS~\cite{kankanamge:graphflow}. Graphflow is a single machine, multi-threaded, main memory graph DBMS implemented in Java. The system supports a subset of the Cypher language~\cite{opencypher}.
We index both the forward and backward adjacency lists and store them in sorted vertex ID order. Adjacency lists are by default partitioned by the edge labels, {\em types} in Cypher jargon, and further by the labels of the destination vertices. With this partitioning, we can quickly access the edges of nodes matching a particular edge label and destination vertex label, allowing us to perform filters on labels very efficiently.
Our query plans follow a Volcano-style plan execution~\cite{graefe:volcano}.
Each plan $P$ has one final \textsc{Sink} operator, which connects to the final operators of all branches in $P$. The execution starts from the \textsc{Sink} operator and each operator asks for a tuple from one of its children until a \textsc{Scan} starts matching an edge. In adaptive parts of one-time plans, an operator $o_i$ may be called upon to provide a tuple from one of its parents, but due to adaptation, provide tuples to a different parent.

\iflong
We implemented a work-stealing-based technique to parallelize the evaluation of our plans. Let $w$ be the number of threads in the system. We give a copy of a plan $P$ to each worker and workers steal work from a single queue to start scanning ranges of edges in the \textsc{Scan} operators. Threads can perform extensions in the \textsc{E/I} operators without any coordination. Hash tables used  in \textsc{Hash-Join} operators are partitioned into $d$$>>$$w$ many hash table ranges. When constructing a hash table, workers grab locks to access each partition but setting $d$$>>$$w$ decreases the possibility of contention. Probing does not require coordination and is done independently.  If \textsc{Hash-Join}'s hash and probe children compute completely symmetric sub-queries, we compute that sub-query once, use it to construct the hash table, and then re-use it to probe.
\fi

\vspace{-8pt}
\section{Evaluation}
\label{sec:evaluation}

Our experiments aim to answer four questions: (1) How good are the plans our optimizer picks? (2) Which type of plans work better for which queries? (3) How much benefit do we get from adapting QVOs at runtime? (4) How do our plans and processing engine compare against EmptyHeaded (EH), which is the closest to our work and the most performant baseline we are aware of? 
\iflong
As part of our EH comparisons, we also tested the scalability of our single-threaded and parallel implementation on our largest graphs LiveJournal and Twitter. 
Finally, for completeness of our study, Appendix~\ref{app:cfl} compares our plans against CFL and Neo4j.
\else
As part of our EH comparisons, in the longer version of our paper~\cite{mhedhbi:sqs-tech-report} we tested the scalability of our single-threaded and parallel implementation on our largest graphs LiveJournal and Twitter. 
Finally, for completeness of our study,  Appendix~\ref{app:cfl} compares our plans against CFL, and the longer version of our paper~\cite{mhedhbi:sqs-tech-report} contains experiments against Neo4j.
\fi

\subsection{Setup}

\subsubsection{Hardware}
We use a single machine that has two Intel E5-2670 @2.6GHz CPUs and 512 GB of RAM. The machine has 16 physical cores and 32 logical cores. 
\iflong
Except our scalability experiments in Section \ref{subsub:scalability-experiments}, we use only one physical core. 
\else
Except for our scalability experiments in the long version of our paper, we use only one physical core. 
\fi
We set the maximum size of the JVM heap to 500 GB and keep JVM's default minimum size. We ran each experiment twice, one to warm-up the system and recorded measurements for the second run.

\subsubsection{Datasets}
\label{subsub:datasets}

\begin{table}[t!]
	\centering
	\begin{tabular}{llllllllll}\toprule
		Domain    & Name             & Nodes & Edges           \\
		\midrule
		Social    & Epinions (Ep)    &   76K &  509K    \\
		& LiveJournal (LJ) &  4.8M &   69M          \\
		& Twitter (Tw)     & 41.6M & 1.46B            \\
		Web       & BerkStan (BS)    &  685K &  7.6M  \\
		& Google (Go)      &  876K &  5.1M             \\
		Product   & Amazon (Am)      &  403K &  3.5M  \\
		\bottomrule
	\end{tabular}
	\vspace{-5pt}
	\caption{Datasets used.}
	\vspace{-12pt}
	\label{table:datasets-used}
\end{table}

\begin{figure}[t!]
	\centering
	\captionsetup{justification=centering}
	\begin{subfigure}[t]{0.08\textwidth}
        \centering
		\begin{tikzpicture}[scale=0.4, transform shape,->,>=stealth', shorten >=1pt, auto,node distance=1.95cm, thick, main node/.style={circle,draw,font=\sffamily\Huge\bfseries}]
			\node[main node] (1) {$a_1$};
			\node[main node] (2) [below right of=1] {$a_2$};
			\node[main node] (3) [below left of=1] {$a_3$};
			\path[every node/.style={font=\sffamily\small}]
			(1) edge (2) edge (3)
			(2) edge (3) ;
		\end{tikzpicture}
		\caption{Q1.}
		\label{fig:Q1}
	\end{subfigure}
	\begin{subfigure}[t]{0.092\textwidth}
        \centering
		\begin{tikzpicture}[scale=0.4, transform shape,->,>=stealth', shorten >=1pt, auto,node distance=1.95cm, thick, main node/.style={circle,draw,font=\sffamily\Huge\bfseries}]
			\node[main node] (1) {$a_1$};
			\node[main node] (2) [above right of=1] {$a_2$};
			\node[main node] (3) [below right of=1] {$a_3$};
			\node[main node] (4) [below right of=2] {$a_4$};
			\path[every node/.style={font=\sffamily\small}]
			(1) edge (2) edge (3)
			(2) edge (4)
			(3) edge (4) ;
		\end{tikzpicture}
		\caption{Q2.}
		\label{fig:Q2}
	\end{subfigure}
	\begin{subfigure}[t]{0.092\textwidth}
        \centering
		\begin{tikzpicture}[scale=0.4, transform shape,->,>=stealth', shorten >=1pt, auto,node distance=1.95cm, thick, main node/.style={circle,draw,font=\sffamily\Huge\bfseries}]
			\node[main node] (1) {$a_1$};
			\node[main node] (2) [above right of=1] {$a_2$};
			\node[main node] (3) [below right of=1] {$a_3$};
			\node[main node] (4) [below right of=2] {$a_4$};
			\path[every node/.style={font=\sffamily\small}]
			(1) edge (2) edge (3)
			(2) edge (4)
			(4) edge (3) ;
		\end{tikzpicture}
		\caption{Q3.}
		\label{fig:Q3}
	\end{subfigure}
	\begin{subfigure}[t]{0.092\textwidth}
        \centering
		\begin{tikzpicture}[scale=0.4, transform shape,->,>=stealth', shorten >=1pt, auto,node distance=1.95cm, thick, main node/.style={circle,draw,font=\sffamily\Huge\bfseries}]
			\node[main node] (1) {$a_1$};
			\node[main node] (2) [above right of=1] {$a_2$};
			\node[main node] (3) [below right of=1] {$a_3$};
			\node[main node] (4) [below right of=2] {$a_4$};
			\path[every node/.style={font=\sffamily\small}]
			(1) edge (2) edge (3)
			(2) edge (3) edge (4)
			(3) edge (4) ;
		\end{tikzpicture}
		\caption{Q4.}
		\label{fig:Q4}
	\end{subfigure}
	\begin{subfigure}[t]{0.12\textwidth}
        \centering
		\begin{tikzpicture}[scale=0.4, transform shape,->,>=stealth', shorten >=1pt, auto,node distance=1.95cm, thick, main node/.style={circle,draw,font=\sffamily\Huge\bfseries}]
			\node[main node] (1) {$a_1$};
			\node[main node] (2) [above right of=1] {$a_2$};
			\node[main node] (3) [below right of=1] {$a_3$};
			\node[main node] (4) [below right of=2] {$a_4$};
			\path[every node/.style={font=\sffamily\small}]
			(1) edge (2) edge (3)
			(2) edge (3) edge (4)
			(4) edge (3) ;
		\end{tikzpicture}
		\caption{Q5.}
		\label{fig:Q5}
	\end{subfigure}
	\begin{subfigure}[t]{0.09\textwidth}
        \centering
		\begin{tikzpicture}[scale=0.4, transform shape,->,>=stealth', shorten >=1pt, auto,node distance=1.95cm, thick, main node/.style={circle,draw,font=\sffamily\Huge\bfseries}]
			\node[main node] (1) {$a_1$};
			\node[main node] (2) [above right of=1] {$a_2$};
			\node[main node] (3) [below right of=1] {$a_3$};
			\node[main node] (4) [below right of=2] {$a_4$};
			\path[every node/.style={font=\sffamily\small}]
			(1) edge (2) edge (3) edge (4)
			(2) edge (3) edge (4)
			(3) edge (4) ;
		\end{tikzpicture}
		\caption{Q6.}
		\label{fig:Q6}
	\end{subfigure}
	\begin{subfigure}[t]{0.12\textwidth}
        \centering
		\begin{tikzpicture}[scale=0.4, transform shape,->,>=stealth', shorten >=1pt, auto,node distance=1.95cm, thick, main node/.style={circle,draw,font=\sffamily\Huge\bfseries}]
			\node[main node] (1) {$a_1$};
			\node[main node] (2) [above right of=1] {$a_2$};
			\node[main node] (3) [below right of=1] {$a_3$};
			\node[main node] (4) [right of=2] {$a_4$};
			\node[main node] (5) [right of=3] {$a_5$};
			\path[every node/.style={font=\sffamily\small}]
			(1) edge (2) edge (3) edge (4) edge (5)
			(2) edge (3) edge (4) edge (5)
			(3) edge (4) edge (5)
			(4) edge (5) ;
		\end{tikzpicture}
		\caption{Q7.}
		\label{fig:Q7}
	\end{subfigure}
	\begin{subfigure}[t]{0.12\textwidth}
        \centering
		\begin{tikzpicture}[scale=0.4, transform shape,->,>=stealth', shorten >=1pt, auto,node distance=1.95cm, thick, main node/.style={circle,draw,font=\sffamily\Huge\bfseries}]
			\node[main node] (1) {$a_1$};
			\node[main node] (3) [below right of=1] {$a_3$};
			\node[main node] (2) [below left of=3] {$a_2$};
			\node[main node] (4) [above right of=3] {$a_4$};
			\node[main node] (5) [below right of=3] {$a_5$};
			\path[every node/.style={font=\sffamily\small}]
			(1) edge (2) edge (3)
			(2) edge (3)
			(3) edge (4) edge (5)
			(4) edge (5) ;
		\end{tikzpicture}
		\caption{Q8.}
		\label{fig:Q8}
	\end{subfigure}\\ 
	\begin{subfigure}[t]{0.125\textwidth}
        \centering
		\begin{tikzpicture}[scale=0.4, transform shape,->,>=stealth', shorten >=1pt, auto,node distance=1.95cm, thick, main node/.style={circle,draw,font=\sffamily\Huge\bfseries}]
			\node[main node] (3) {$a_3$};
			\node[main node] (1) [above left of=3] {$a_1$};
			\node[main node] (4) [below left of=3] {$a_4$};
			\node[main node] (2) [above right of=3] {$a_2$};
			\node[main node] (5) [below right of=3] {$a_5$};
			\node[main node] (6) [below right of=2] {$a_6$};
			\path[every node/.style={font=\sffamily\small}]
			(1) edge (2)
			(2) edge (3)
			(3) edge (1) edge (4)
			(4) edge (5)
			(5) edge (3)
			(6) edge (2) edge (5) ;
		\end{tikzpicture}
		\caption{Q9.}
		\label{fig:Q9}
	\end{subfigure}
	\begin{subfigure}[t]{0.13\textwidth}
        \centering
		\begin{tikzpicture}[scale=0.4, transform shape,->,>=stealth', shorten >=1pt, auto,node distance=1.95cm, thick, main node/.style={circle,draw,font=\sffamily\Huge\bfseries}]
			\node[main node] (1) {$a_1$};
			\node[main node] (2) [above right of=1] {$a_2$};
			\node[main node] (3) [below right of=1] {$a_3$};
			\node[main node] (4) [below right of=2] {$a_4$};
			\node[main node] (5) [above right of=4] {$a_5$};
			\node[main node] (6) [below right of=4] {$a_6$};
			\path[every node/.style={font=\sffamily\small}]
			(1) edge (2) edge (3)
			(2) edge (4)
			(3) edge (4)
			(4) edge (5) edge (6)
			(5) edge (6) ;
		\end{tikzpicture}
		\caption{Q10.}
		\label{fig:Q10}
	\end{subfigure}
	\begin{subfigure}[t]{0.12\textwidth}
        \centering
		\begin{tikzpicture}[scale=0.4, transform shape,->,>=stealth', shorten >=1pt, auto,node 	distance=1.95cm, thick, main node/.style={circle,draw,font=\sffamily\Huge\bfseries}]
			\node[main node] (1) {$a_1$};
			\node[main node] (2) [above right of=1] {$a_2$};
			\node[main node] (3) [below right of=1] {$a_3$};
			\node[main node] (4) [above left of=1] {$a_4$};
			\node[main node] (5) [below left of=1] {$a_5$};
			\path[every node/.style={font=\sffamily\small}]
			(1) edge (2) edge (3) edge (4)
	        (5) edge (1) ;
		\end{tikzpicture}
		\caption{Q11.}
		\label{fig:Q11}
	\end{subfigure}
    \begin{subfigure}[t]{0.1\textwidth}
	    \centering
	    \begin{tikzpicture}[scale=0.4, transform shape,->,>=stealth', shorten >=1pt, auto,node distance=1.95cm, thick, main node/.style={circle,draw,font=\sffamily\Huge\bfseries}]
			\node[main node] (1) {$a_1$};
			\node[main node] (2) [right of=1] {$a_2$};
			\node[main node] (3) [below of=2] {$a_3$};
			\node[main node] (4) [below of=3] {$a_4$};
			\node[main node] (5) [left of=4] {$a_5$};
	        \node[main node] (6) [above  of=5] {$a_6$};
			\path[every node/.style={font=\sffamily\small}]
			(1) edge (2) 
	        (2) edge (3) 
			(3) edge (4) 
			(4) edge (5) 
			(5) edge (6)
	        (6) edge (1);
		\end{tikzpicture}
		\caption{Q12.}
		\label{fig:Q12}
	\end{subfigure}
    \begin{subfigure}[t]{0.1\textwidth}
    	\centering
	    \begin{tikzpicture}[scale=0.4, transform shape,->,>=stealth', shorten >=1pt, auto,node distance=1.95cm, thick, main node/.style={circle,draw,font=\sffamily\Huge\bfseries}]
			\node[main node] (1) {$a_1$};
			\node[main node] (2) [right of=1] {$a_2$};
			\node[main node] (3) [below of=2] {$a_3$};
			\node[main node] (4) [below of=3] {$a_4$};
			\node[main node] (5) [left of=4] {$a_5$};
			\node[main node] (6) [above  of=5] {$a_6$};
			\path[every node/.style={font=\sffamily\small}]
			(1) edge (2) 
	        (2) edge (3)
	        (3) edge (4)
			(4) edge (5) 
			(5) edge (6) ;
		\end{tikzpicture}
		\caption{Q13.}
		\label{fig:Q13}
	\end{subfigure}
	\begin{subfigure}[t]{0.18\textwidth}
		\centering
		\begin{tikzpicture}[scale=0.4, transform shape,->,>=stealth', shorten >=1pt, auto,node distance=2.25cm, thick, main node/.style={circle,draw,font=\sffamily\Huge\bfseries}]
			\node[main node] (1) {$a_1$};
			\node[main node] (7) [below right of=1] {$a_7$};
			\node[main node] (6) [right of=7] {$a_6$};
			\node[main node] (5) [above right of=6] {$a_5$};
			\node[main node] (2) [above of=1] {$a_2$};
			\node[main node] (4) [above of=5] {$a_4$};
			\node[main node] (3) [above=10pt] at ($(2)!0.5!(4)$) {$a_3$};
			\path[every node/.style={font=\sffamily\small}]
			(1) edge (2) edge (3) edge (4) edge (5) edge (6) edge (7)
			(2) edge (3) edge (4) edge (5) edge (6) edge (7)
			(3) edge (4) edge (5) edge (6) edge (7)
			(4) edge (5) edge (6) edge (7)
			(5) edge (6) edge (7)
			(6) edge (7) ;
		\end{tikzpicture}
		\caption{Q14.}
		\label{fig:Q14}
	\end{subfigure}
	\vspace{-5pt}
	\caption{Subgraph queries used for evaluations.}
	\vspace{-12pt}
	\label{fig:queries-used}
\end{figure}
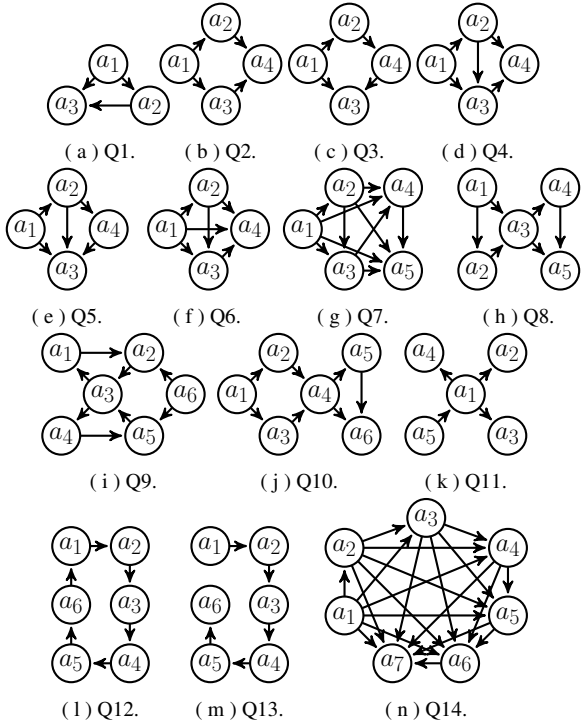

The datasets we use are in Table \ref{table:datasets-used}.\footnote{We obtained the graphs from reference \cite{snapnets} except for the Twitter graph, which we obtained from reference \cite{Kwak10www}.} Our datasets differ in several structural properties: (i) size; (2) how skewed their forward and backward adjacency lists distribution is; and (3) average clustering coefficients, which is a measure of the cyclicity of the graph, specifically the amount of cliques in it. The datasets also come from a variety of application domains: social networks, the web, and product co-purchasing. Each dataset's catalogue was generated with $z$=1000 and $h$=3 except for Twitter, where we set $h$=2.

\subsubsection{Queries}
For the experiments in this section, we used the 14 queries shown in Figure \ref{fig:queries-used}, which contain both acyclic and cyclic queries with dense and sparse connectivity with up to 7 query vertices and 21 query edges. In our experiments, we consider both labeled and unlabeled queries. Our datasets and queries are not labeled by default and we label them randomly. We use the notation $QJ_i$ to refer to evaluating the subgraph query $QJ$ on a dataset for which we randomly generate a label $l$ on each edge, where $l\in \{l_1, l_2, \ldots, l_i\}$. For example, evaluating $Q3_2$ on Amazon indicates randomly adding one of two possible labels to each data edge in Amazon and query edge on $Q3$. If a query was unlabeled we simply write it as $QJ$.

\subsection{Plan Suitability For Different Queries and Optimizer Evaluation}
\label{subsubsec:suitability}

\iflong
\begin{figure*}[htp!]
	\centering
	\captionsetup{justification=centering}
    \begin{subfigure}[b]{0.100\textwidth}
		\centering
		\includegraphics[scale=0.48]{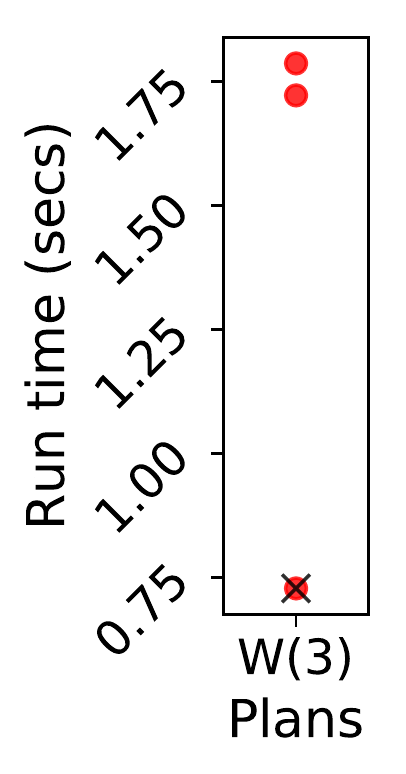}
		\caption{Q1, Am.}
		\label{fig:suit-q1-amazon}
	\end{subfigure}
    \begin{subfigure}[b]{0.100\textwidth}
		\centering
		\includegraphics[scale=0.48]{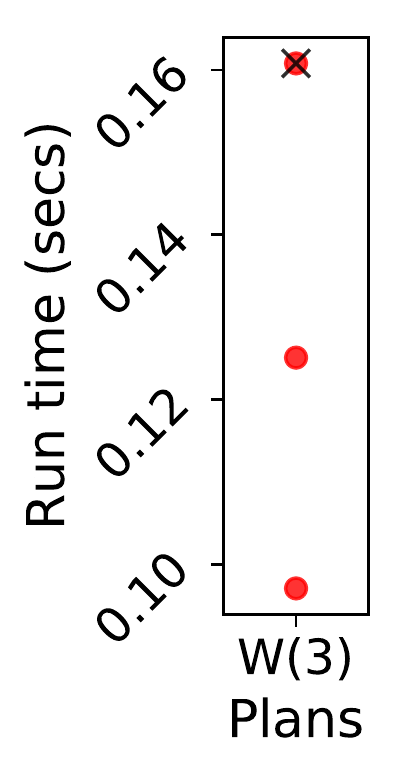}
		\caption{Q1$_3$, Ep.}
		\label{fig:suit-q1-epinions_3}
	\end{subfigure}
    \begin{subfigure}[b]{0.100\textwidth}
		\centering
		\includegraphics[scale=0.48]{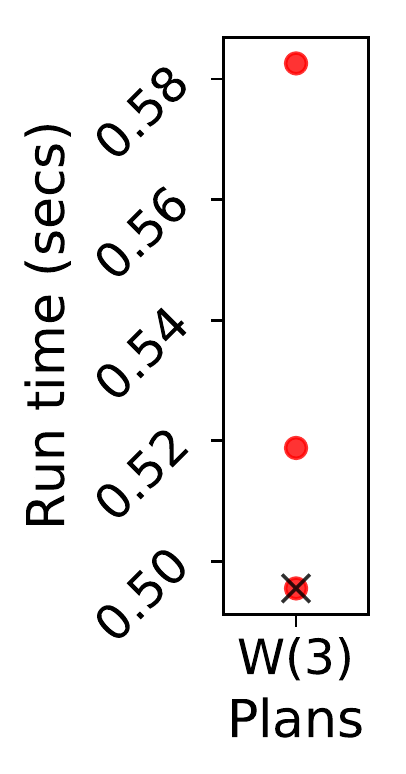}
		\caption{Q1$_5$, Go.}
		\label{fig:suit-q1-google_5}
	\end{subfigure}
    \begin{subfigure}[b]{0.100\textwidth}
		\centering
		\includegraphics[scale=0.48]{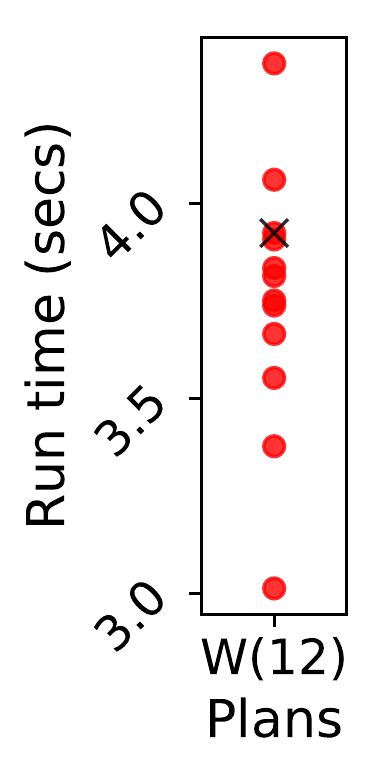}
		\caption{Q6, Am.}
		\label{fig:suit-q6-amazon}
	\end{subfigure}
	\begin{subfigure}[b]{0.100\textwidth}
		\centering
		\includegraphics[scale=0.48]{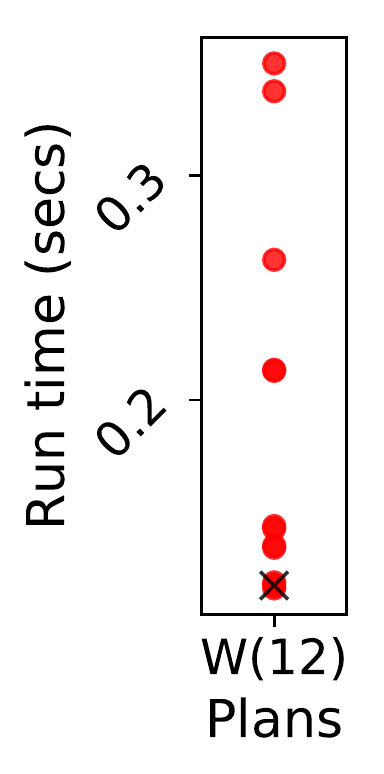}
		\caption{Q6$_3$, Ep.}
		\label{fig:suit-q6-epinions_3}
	\end{subfigure}
	\begin{subfigure}[b]{0.100\textwidth}
		\centering
		\includegraphics[scale=0.48]{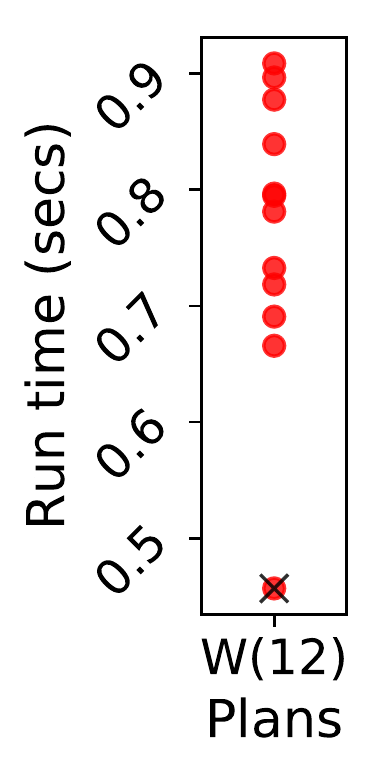}
		\caption{Q6$_5$, Go.}
		\label{fig:suit-q6-google_5}
	\end{subfigure}
	\begin{subfigure}[b]{0.100\textwidth}
		\centering
		\includegraphics[scale=0.48]{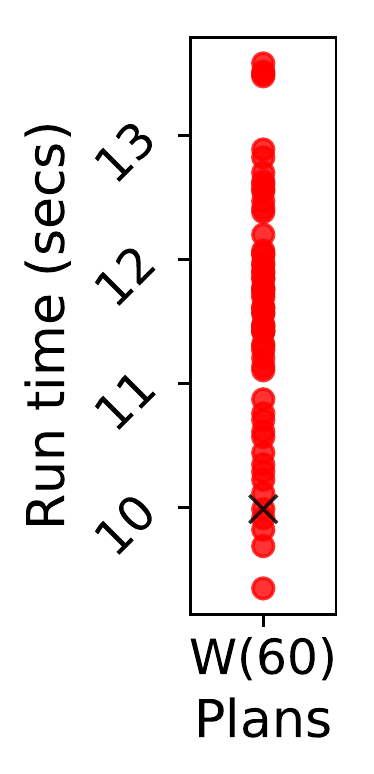}
		\caption{Q7, Am.}
		\label{fig:suit-q7-amazon}
	\end{subfigure}
	\begin{subfigure}[b]{0.100\textwidth}
		\centering
		\includegraphics[scale=0.48]{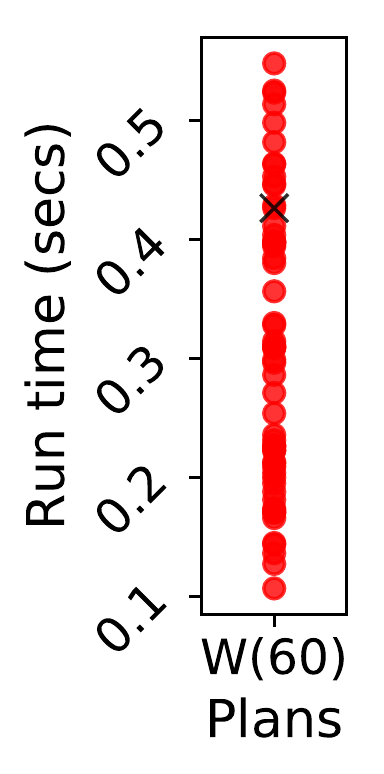}
		\caption{Q7$_3$, Ep.}
		\label{fig:suit-q7-epinions_3}
	\end{subfigure}
	\begin{subfigure}[b]{0.100\textwidth}
		\centering
		\includegraphics[scale=0.48]{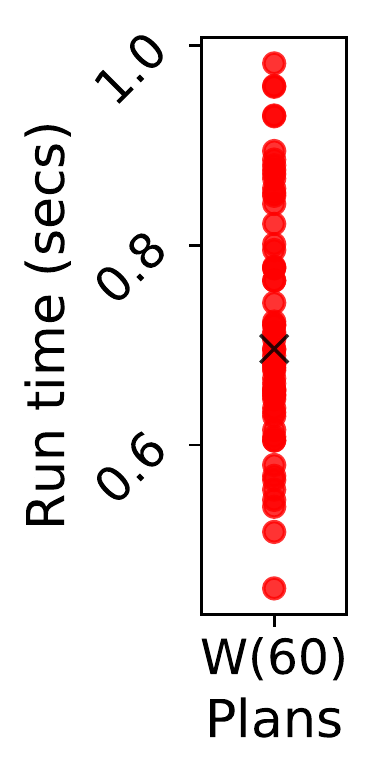}
		\caption{Q7$_5$, Go.}
		\label{fig:suit-q7-google_5}
	\end{subfigure}
    \begin{subfigure}[b]{0.150\textwidth}
		\centering
		\includegraphics[scale=0.48]{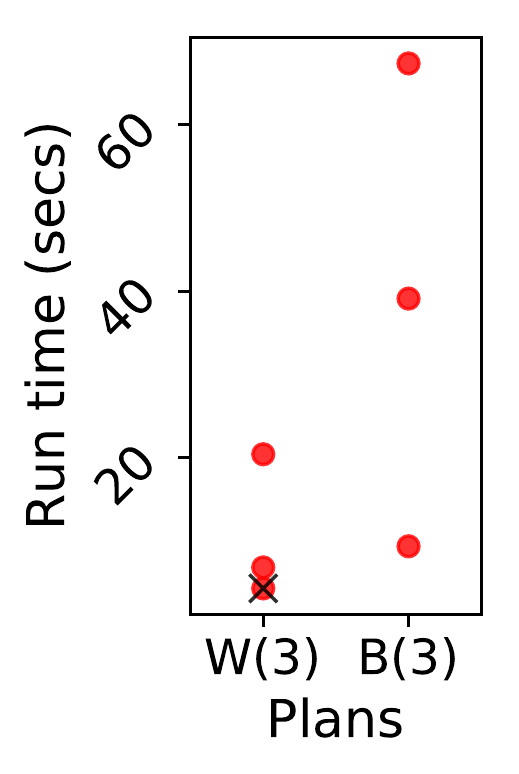}
		\caption{Q2, Am.}
		\label{fig:suit-q2-amazon}
	\end{subfigure}
	\begin{subfigure}[b]{0.150\textwidth}
		\centering
		\includegraphics[scale=0.48]{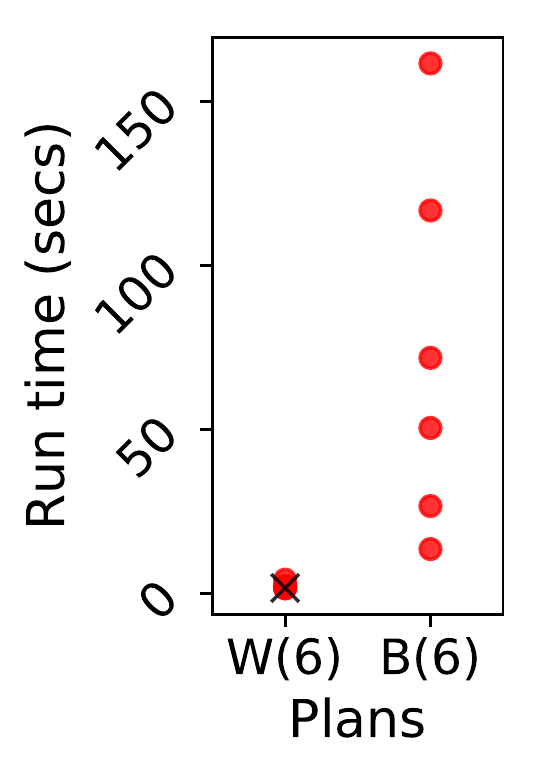}
		\caption{Q2$_3$, Ep.}
		\label{fig:suit-q2-epinions_3}
	\end{subfigure}
	\begin{subfigure}[b]{0.150\textwidth}
		\centering
		\includegraphics[scale=0.48]{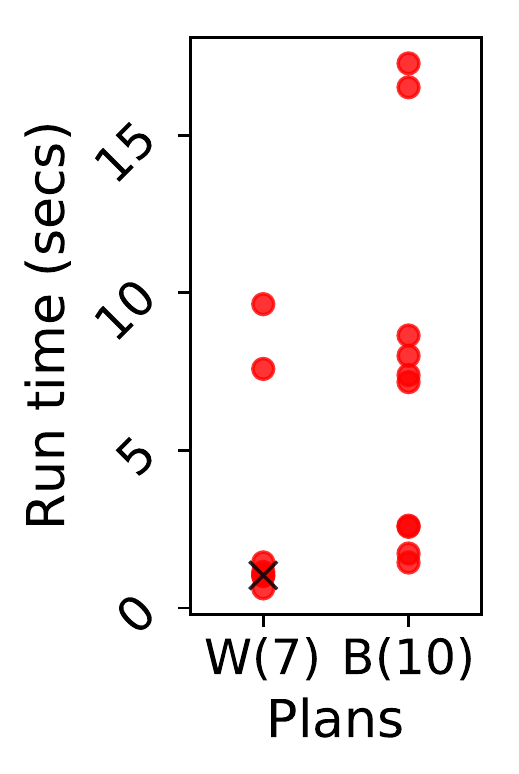}
		\caption{Q2$_5$, Go.}
		\label{fig:suit-q2-google_5}
	\end{subfigure}
    \begin{subfigure}[b]{0.150\textwidth}
		\centering
		\includegraphics[scale=0.48]{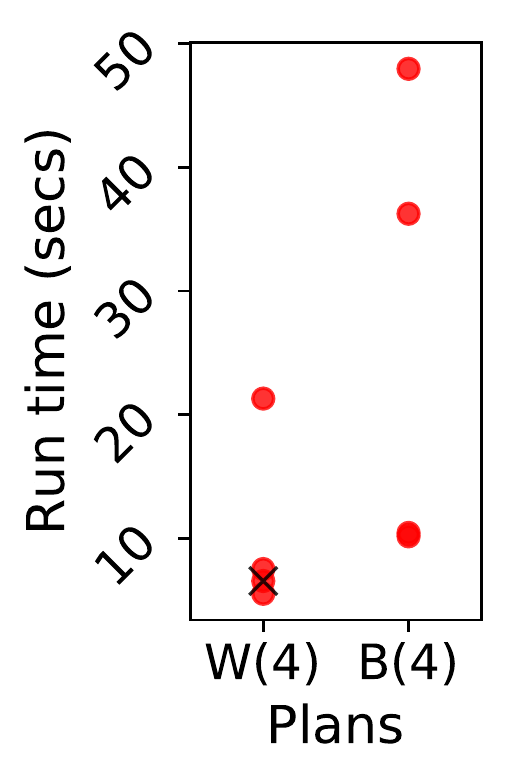}
		\caption{Q3, Am.}
		\label{fig:suit-q3-amazon}
	\end{subfigure}
	\begin{subfigure}[b]{0.150\textwidth}
		\centering
		\includegraphics[scale=0.48]{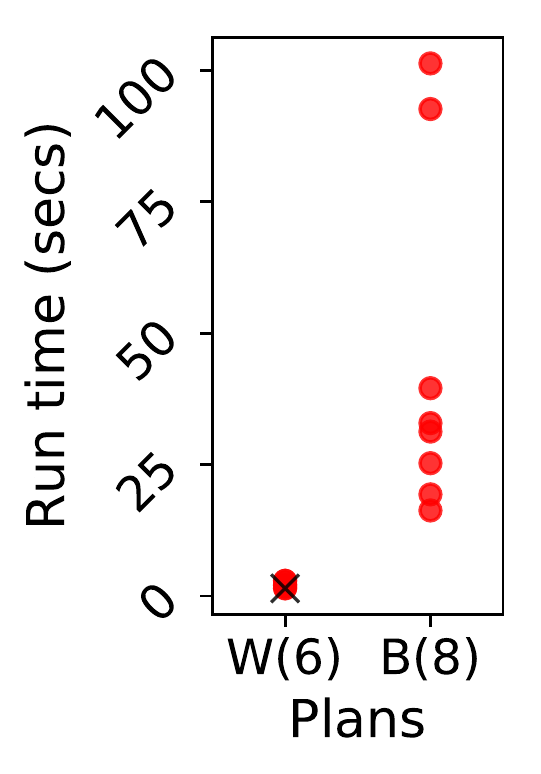}
		\caption{Q3$_3$, Ep.}
		\label{fig:suit-q3-epinions_3}
	\end{subfigure}
	\begin{subfigure}[b]{0.150\textwidth}
		\centering
		\includegraphics[scale=0.48]{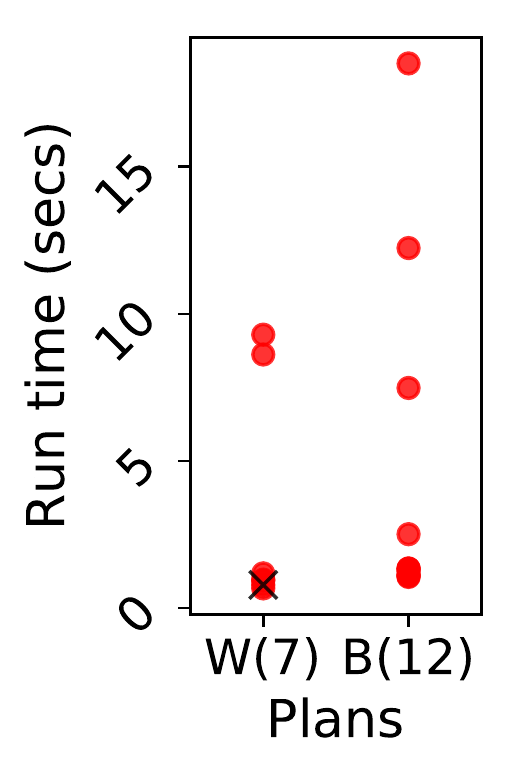}
		\caption{Q3$_5$, Go.}
		\label{fig:suit-q3-google_5}
	\end{subfigure}
    \begin{subfigure}[b]{0.150\textwidth}
	\centering
		\includegraphics[scale=0.48]{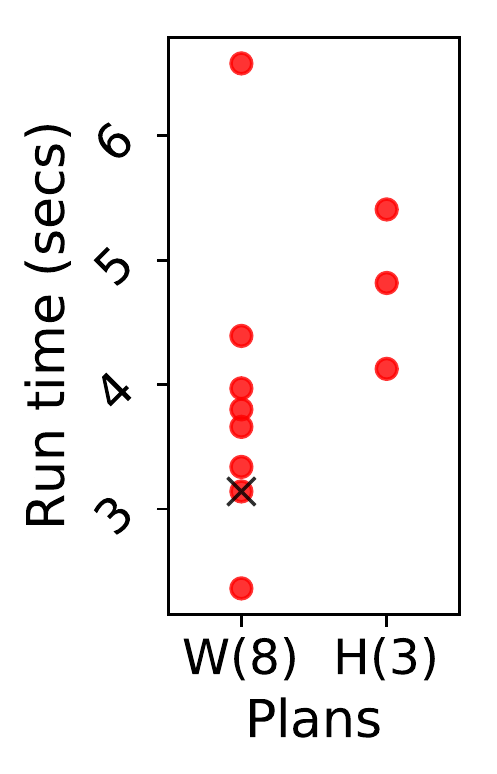}
		\caption{Q4, Am.}
		\label{fig:suit-q4-amazon}
	\end{subfigure}
	\begin{subfigure}[b]{0.150\textwidth}
		\centering
		\includegraphics[scale=0.48]{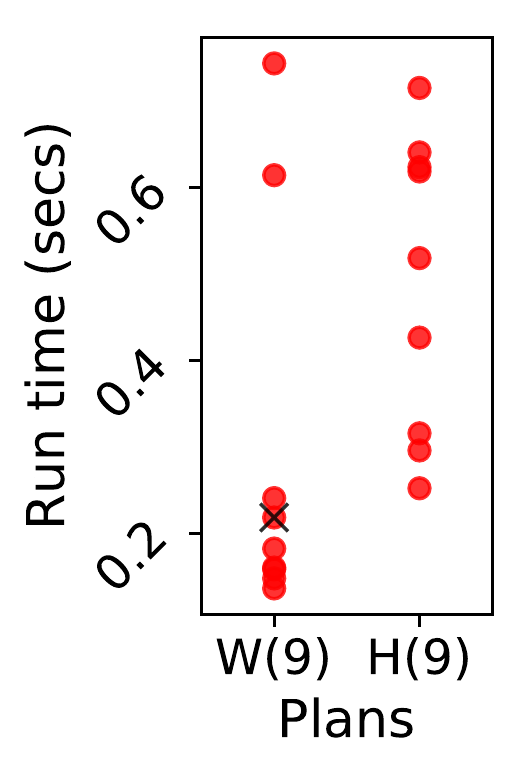}
		\caption{Q4$_3$, Ep.}
		\label{fig:suit-q4-epinions_3}
	\end{subfigure}
	\begin{subfigure}[b]{0.150\textwidth}
		\centering
		\includegraphics[scale=0.48]{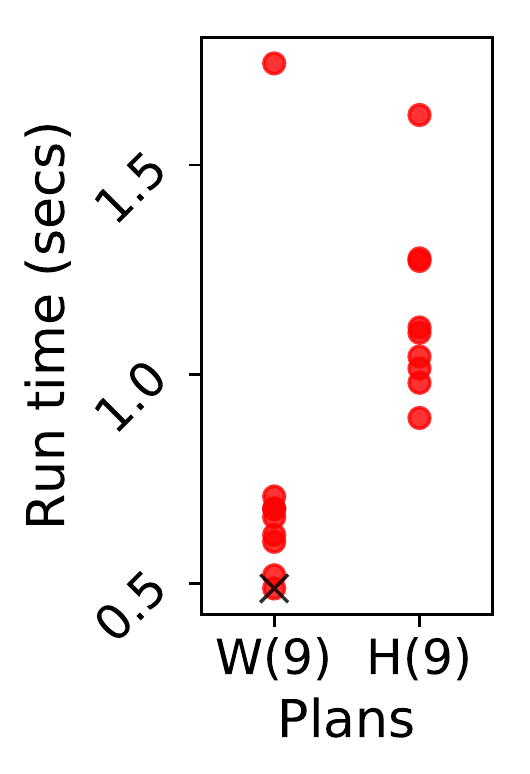}
		\caption{Q4$_5$, Go.}
		\label{fig:suit-q4-google_5}
	\end{subfigure}
    \begin{subfigure}[b]{0.150\textwidth}
		\centering
		\includegraphics[scale=0.48]{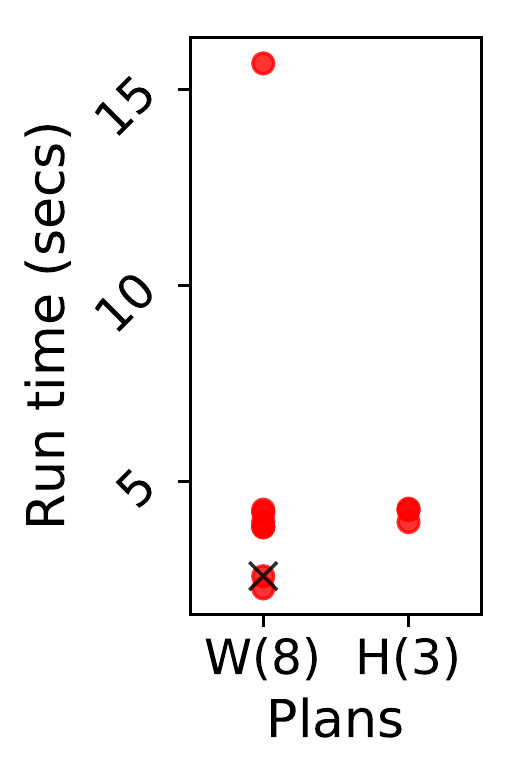}
		\caption{Q5, Am.}
		\label{fig:suit-q5-amazon}
	\end{subfigure}
	\begin{subfigure}[b]{0.150\textwidth}
		\centering
		\includegraphics[scale=0.48]{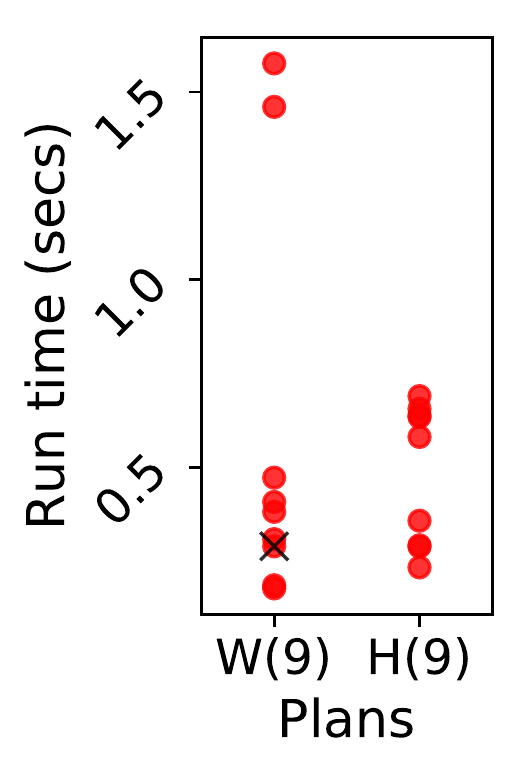}
		\caption{Q5$_3$, Ep.}
		\label{fig:suit-q5-epinions_3}
	\end{subfigure}
	\begin{subfigure}[b]{0.150\textwidth}
		\centering
		\includegraphics[scale=0.48]{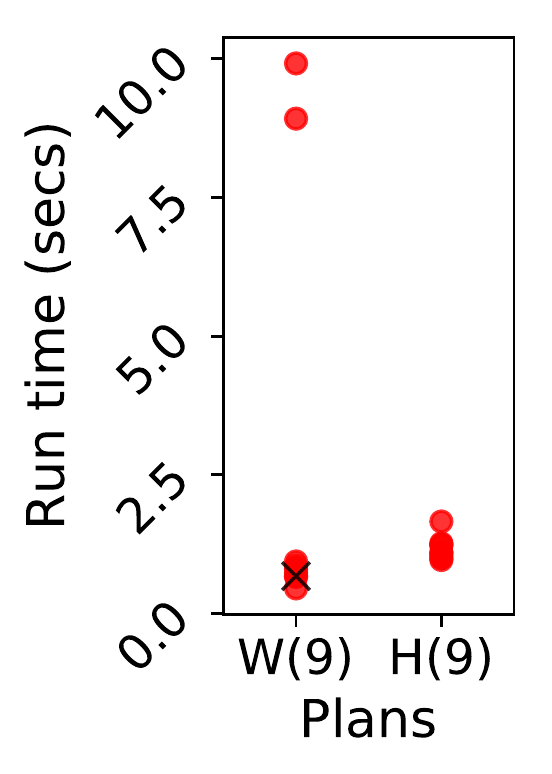}
		\caption{Q5$_5$, Go.}
		\label{fig:suit-q5-google_5}
	\end{subfigure}
	\begin{subfigure}[b]{0.150\textwidth}
		\centering
		\includegraphics[scale=0.48]{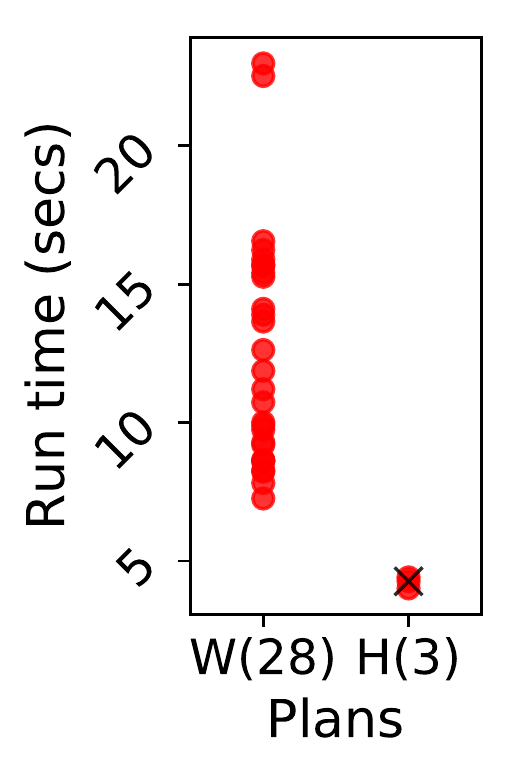}
		\caption{Q8, Am.}
		\label{fig:suit-q8-amazon}
	\end{subfigure}
	\begin{subfigure}[b]{0.150\textwidth}
		\centering
		\includegraphics[scale=0.48]{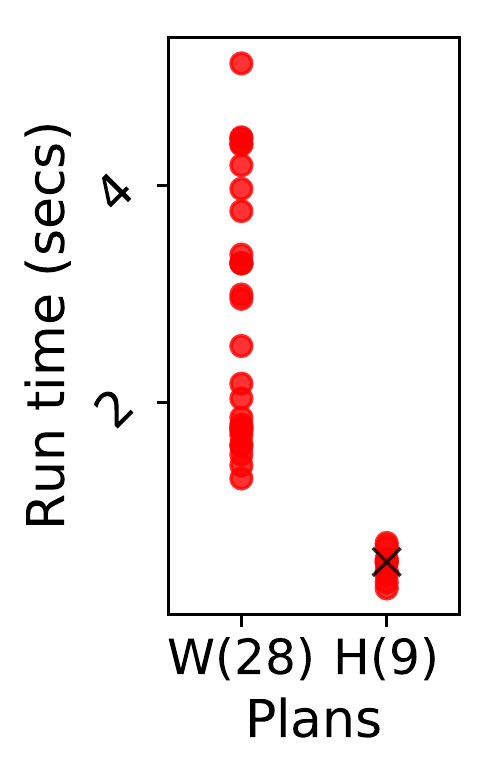}
		\caption{Q8$_3$, Ep.}
		\label{fig:suit-q8-epinions_3}
	\end{subfigure}
	\begin{subfigure}[b]{0.150\textwidth}
		\centering
		\includegraphics[scale=0.48]{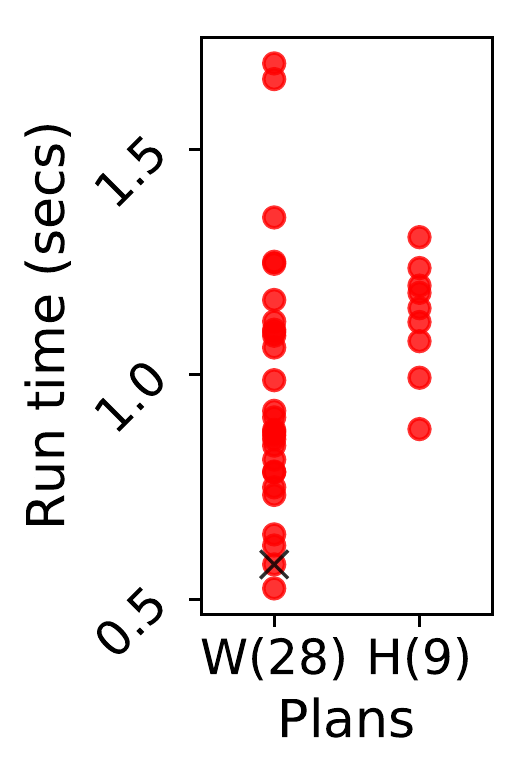}
		\caption{Q8$_5$, Go.}
		\label{fig:suit-q8-google_5}
	\end{subfigure}
	\begin{subfigure}[b]{0.150\textwidth}
		\centering
		\includegraphics[scale=0.48]{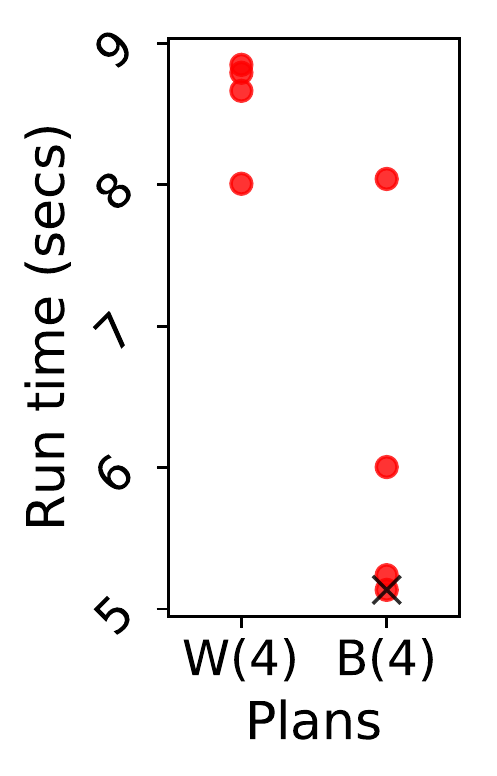}
		\caption{Q11, Am.}
		\label{fig:suit-q11-amazon}
	\end{subfigure}
	\begin{subfigure}[b]{0.150\textwidth}
		\centering
		\includegraphics[scale=0.48]{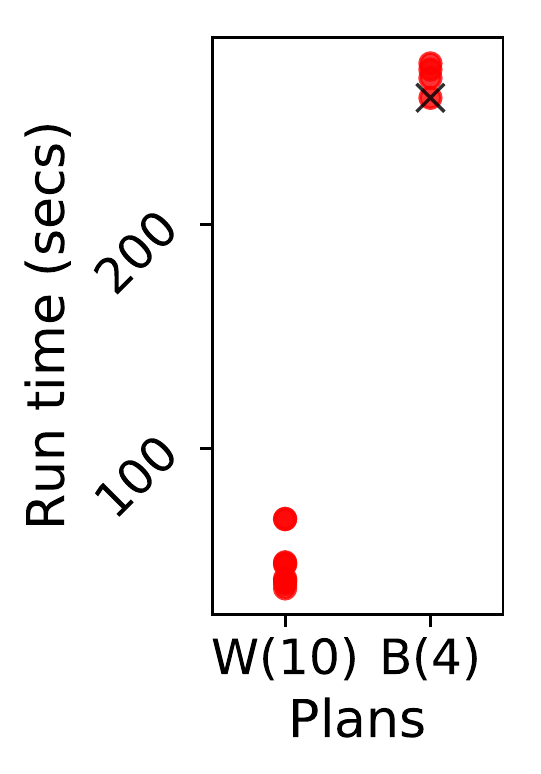}
		\caption{Q11$_3$, Ep.}
		\label{fig:suit-q11-epinions_3}
	\end{subfigure}
	\begin{subfigure}[b]{0.150\textwidth}
		\centering
		\includegraphics[scale=0.48]{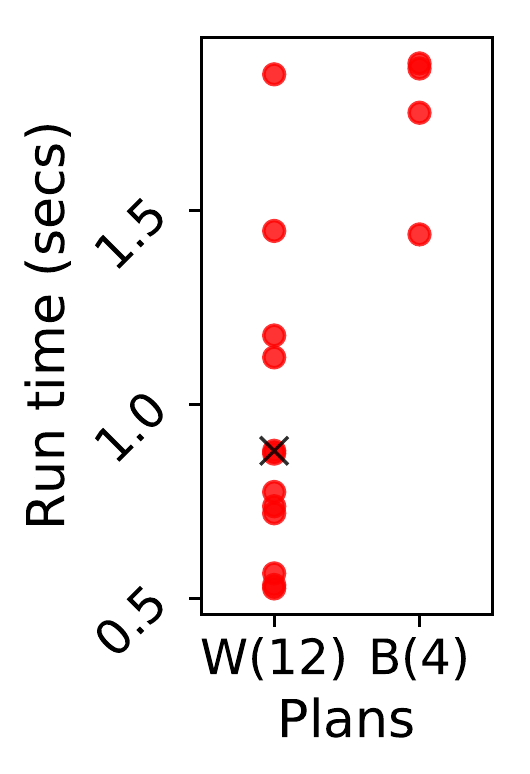}
		\caption{Q11$_3$, Go.}
		\label{fig:suit-q11-google_5}
	\end{subfigure}
	\begin{subfigure}[b]{0.155\textwidth}
		\centering
		\includegraphics[scale=0.48]{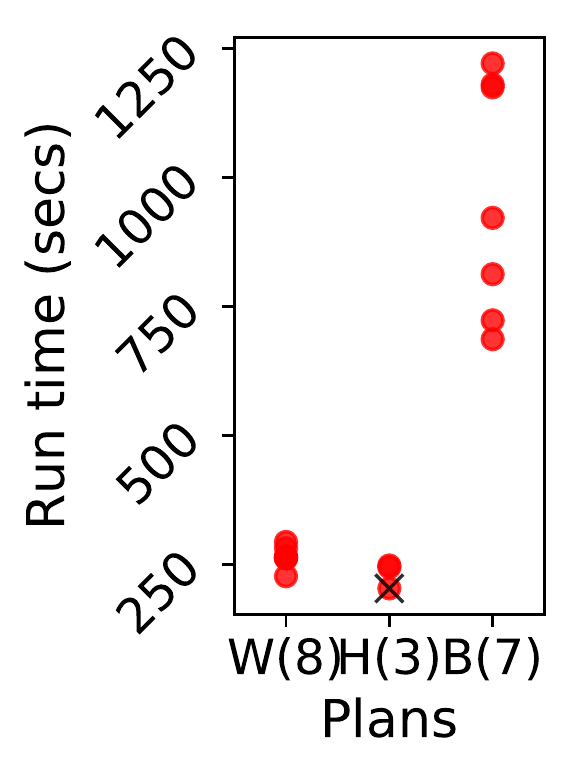}
		\caption{Q12, Am.}
		\label{fig:suit-q12-amazon}
	\end{subfigure}
	\begin{subfigure}[b]{0.19\textwidth}
		\centering
		\includegraphics[scale=0.48]{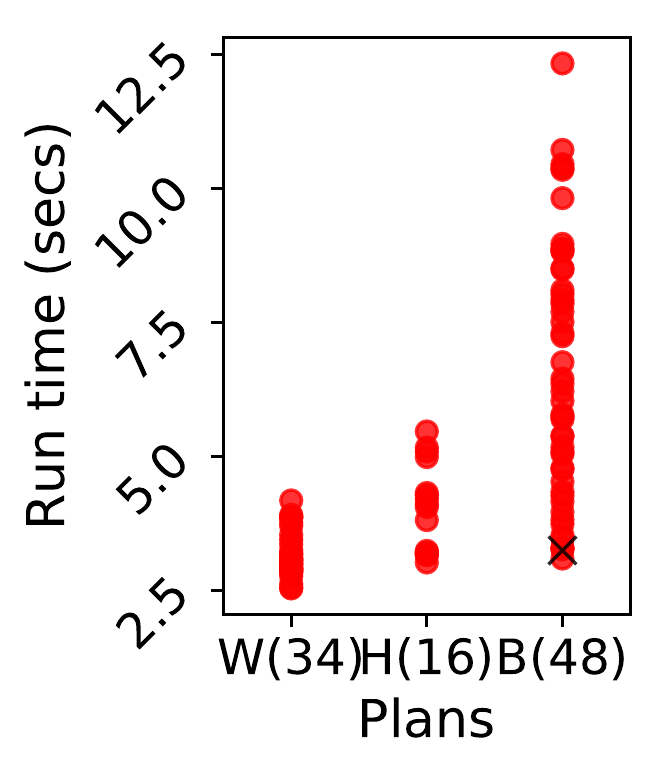}
		\caption{Q12$_5$, Go.}
		\label{fig:suit-q12-google_5}
	\end{subfigure}
	\begin{subfigure}[b]{0.155\textwidth}
		\centering
		\includegraphics[scale=0.48]{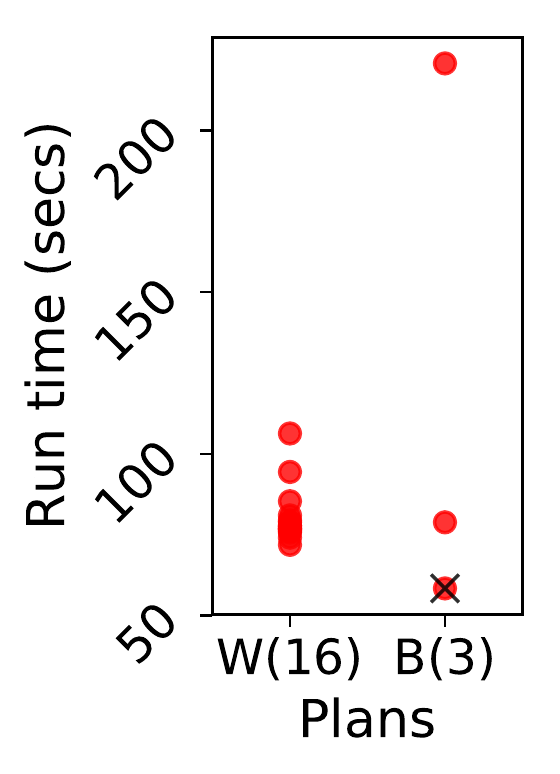}
		\caption{Q13, Am.}
		\label{fig:suit-q13-amazon}
	\end{subfigure}
	\begin{subfigure}[b]{0.155\textwidth}
		\centering
		\includegraphics[scale=0.48]{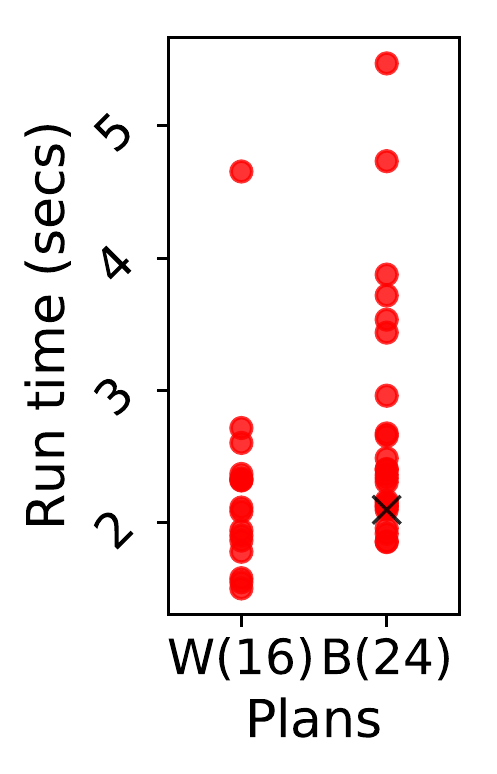}
		\caption{Q13$_5$, Go.}
		\label{fig:suit-q13-google_5}
	\end{subfigure}
	\caption{Plan spectrum charts for suitability and goodness of DP optimizer experiments.}
	\label{fig:suitability-and-goodness}
\end{figure*}

\else
\begin{figure*}[htp!]
	\centering
	\captionsetup{justification=centering}
    \begin{subfigure}[b]{0.16\textwidth}
		\centering
		\includegraphics[scale=0.5]{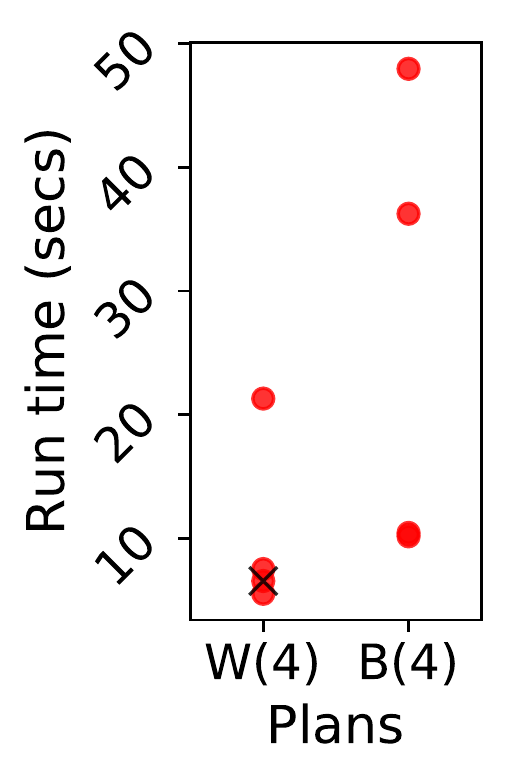}
		\caption{Q3, Amazon.}
		\label{fig:suit-q3-amazon}
	\end{subfigure}
	\begin{subfigure}[b]{0.16\textwidth}
		\centering
		\includegraphics[scale=0.5]{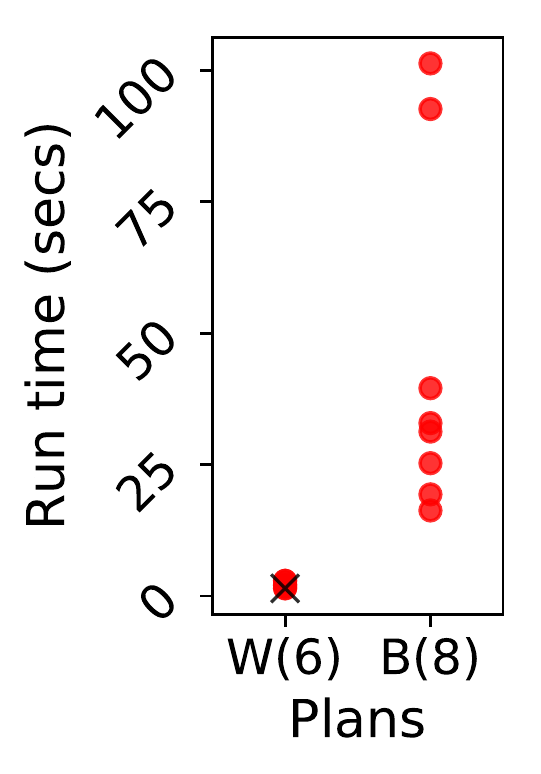}
		\caption{Q3$_3$, Epinions.}
		\label{fig:suit-q3-epinions_3}
	\end{subfigure}
    \begin{subfigure}[b]{0.16\textwidth}
		\centering
		\includegraphics[scale=0.5]{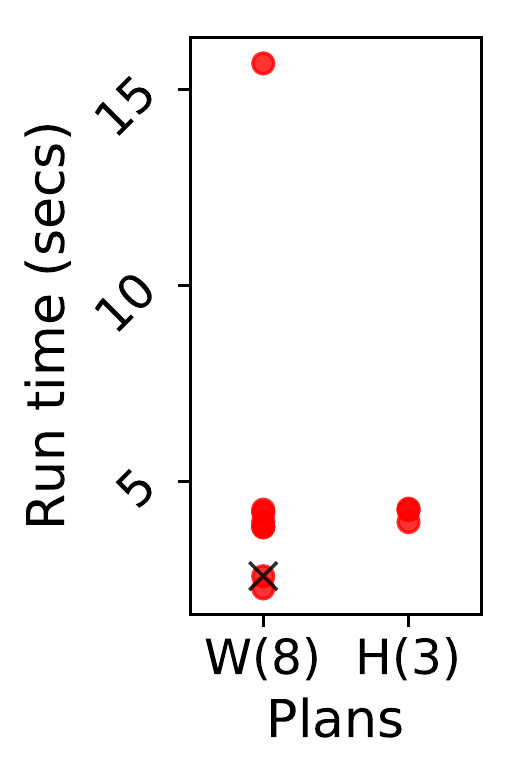}
		\caption{Q5, Amazon.}
		\label{fig:suit-q5-amazon}
	\end{subfigure}
	\begin{subfigure}[b]{0.16\textwidth}
		\centering
		\includegraphics[scale=0.5]{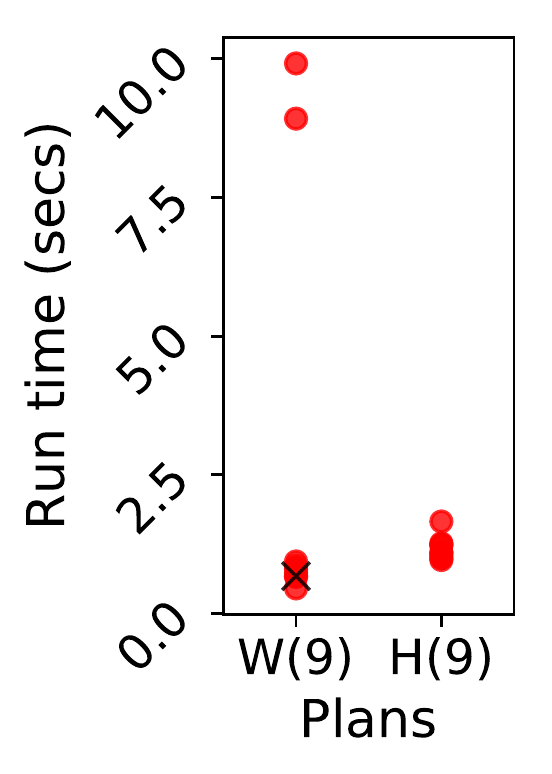}
		\caption{Q5$_5$, Google.}
		\label{fig:suit-q5-google_5}
	\end{subfigure}
	\begin{subfigure}[b]{0.16\textwidth}
		\centering
		\includegraphics[scale=0.5]{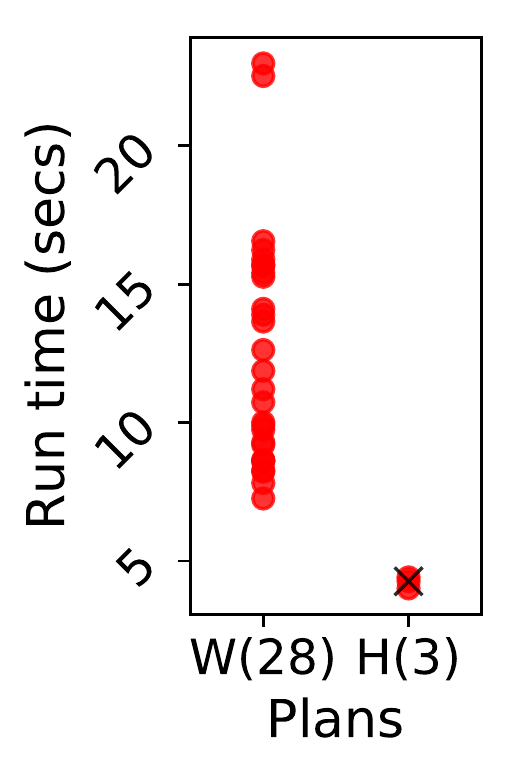}
		\caption{Q8, Amazon.}
		\label{fig:suit-q8-amazon}
	\end{subfigure}
	\begin{subfigure}[b]{0.16\textwidth}
		\centering
		\includegraphics[scale=0.5]{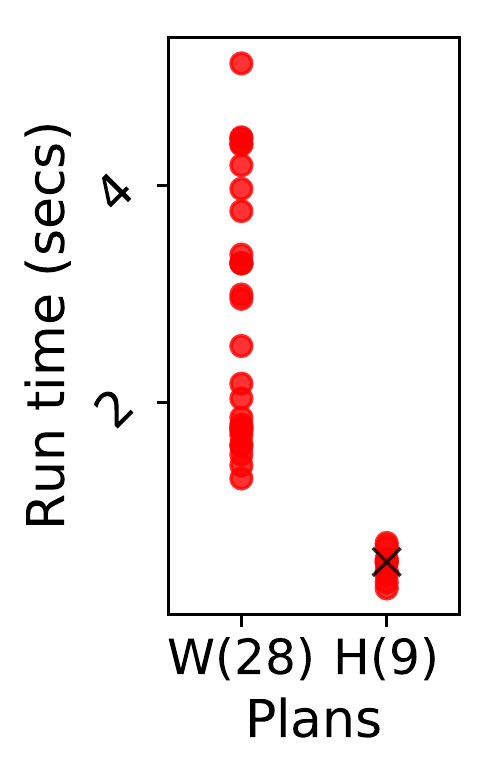}
		\caption{Q8$_3$, Epinions.}
		\label{fig:suit-q8-epinions_3}
	\end{subfigure}
	\begin{subfigure}[b]{0.16\textwidth}
		\centering
		\includegraphics[scale=0.5]{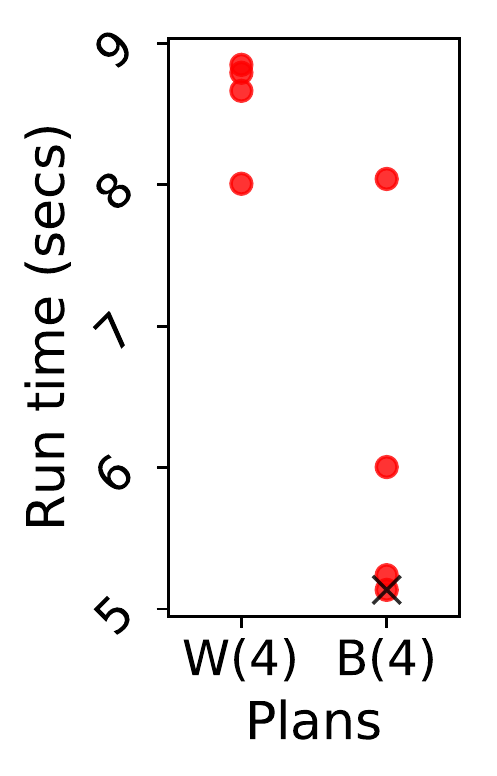}
		\caption{Q11, Amazon.}
		\label{fig:suit-q11-amazon}
	\end{subfigure}
	\begin{subfigure}[b]{0.145\textwidth}
		\centering
		\includegraphics[scale=0.48]{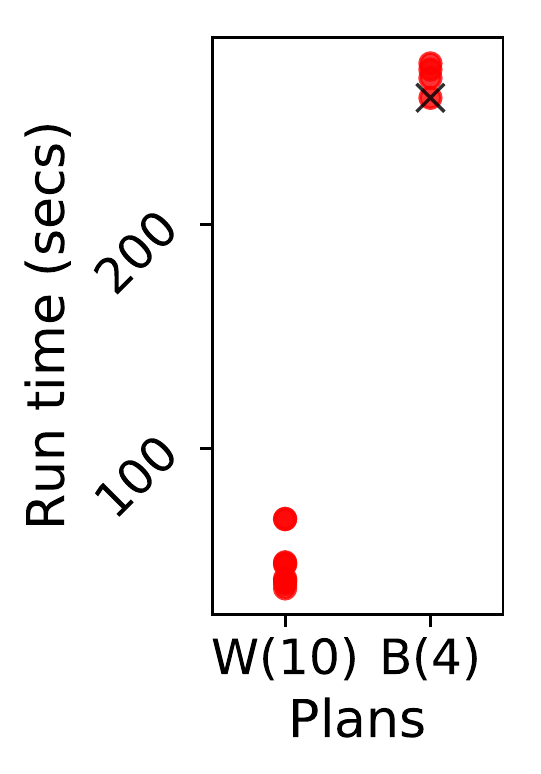}
		\caption{Q11$_3$, Epinions.}
		\label{fig:suit-q11-epinions_3}
	\end{subfigure}
	\begin{subfigure}[b]{0.155\textwidth}
		\centering
		\includegraphics[scale=0.5]{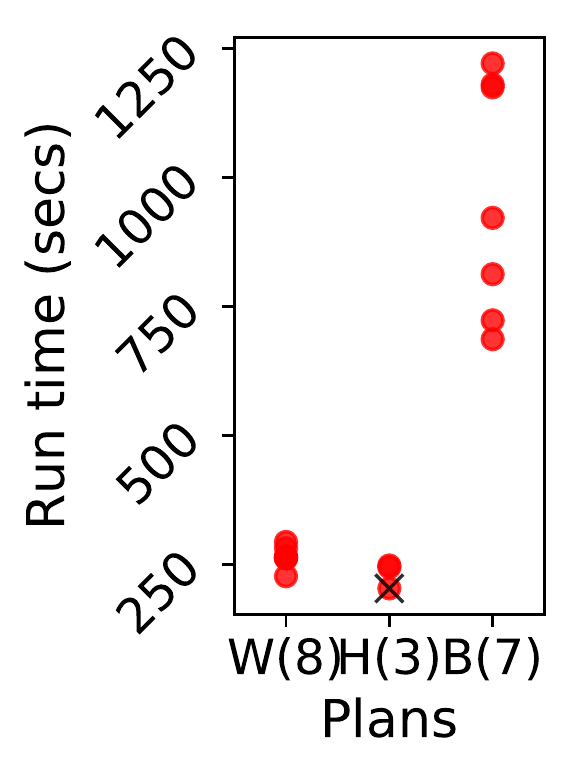}
		\caption{Q12, Amazon.}
		\label{fig:suit-q12-amazon}
	\end{subfigure}
	\begin{subfigure}[b]{0.19\textwidth}
		\centering
		\includegraphics[scale=0.5]{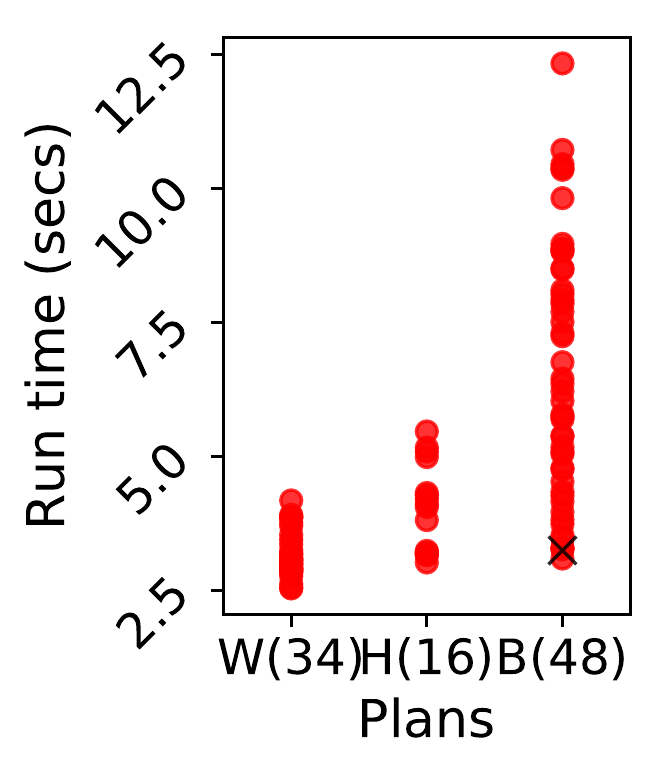}
		\caption{Q12$_5$, Google.}
		\label{fig:suit-q12-google_5}
	\end{subfigure}
	\begin{subfigure}[b]{0.155\textwidth}
		\centering
		\includegraphics[scale=0.5]{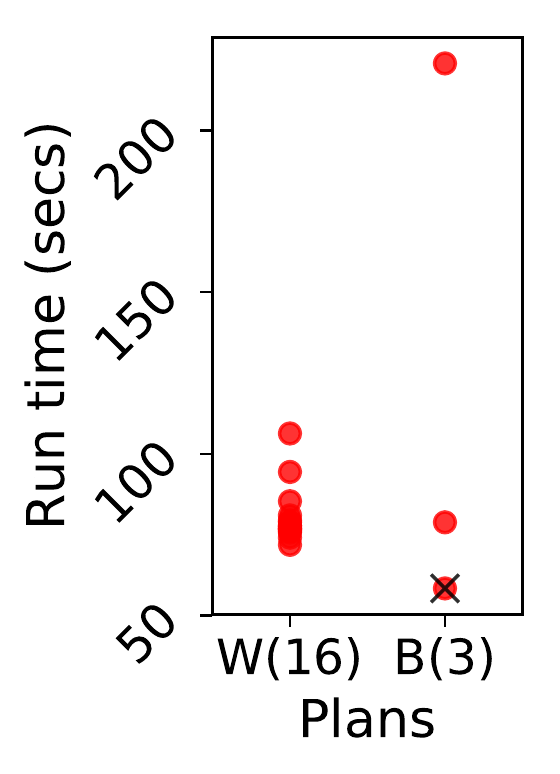}
		\caption{Q13, Amazon.}
		\label{fig:suit-q13-amazon}
	\end{subfigure}
	\begin{subfigure}[b]{0.155\textwidth}
		\centering
		\includegraphics[scale=0.5]{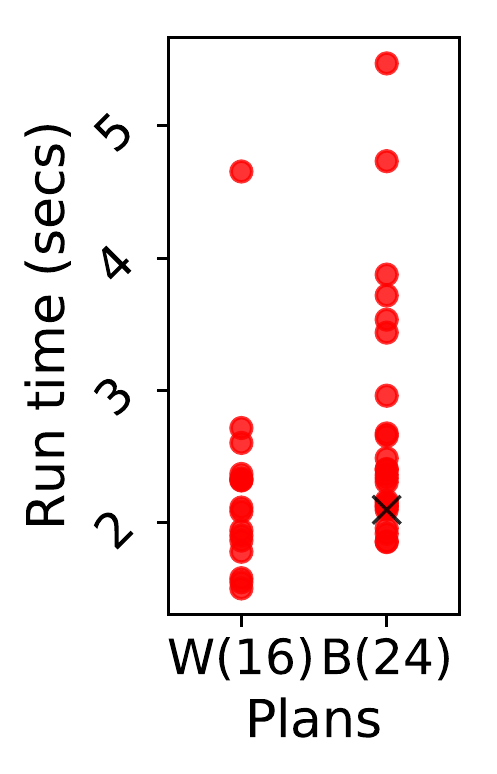}
		\caption{Q13$_5$, Google.}
		\label{fig:suit-q13-google_5}
	\end{subfigure}
	\caption{Plan spectrum charts for suitability and goodnesss of DP optimizer experiments.}
    \vspace{-12pt}
	\label{fig:suitability-and-goodness}
\end{figure*}
\fi

In order to evaluate how good are the plans our optimizer generates, we compare the plans we pick against all other possible plans in a query's plan spectrum. This also allows us to study which types of plans are suitable for which queries. We generated plan spectrums of queries $Q1$-$Q8$ and $Q11$-$Q13$ on Amazon without labels, Epinions with 3 labels, and Google with 5 labels. The spectrums of $Q12$ and $Q13$ on Epinions took a prohibitively long time to generate and are omitted. 
\iflong
All of our spectrums are shown in Figure~\ref{fig:suitability-and-goodness}.
\else
Due to space constraints we present 12 of the 31 spectrums in Figure~\ref{fig:suitability-and-goodness}.
In particular we omit spectrums of $Q1$, $Q6$, $Q7$, which are cliques and for which our optimizer only generates WCO plans, and $Q2$ and $Q4$, which are similar to $Q3$ and $Q5$, respectively. 
The omitted spectrums are in the longer version of our paper~\cite{mhedhbi:sqs-tech-report}. 
\fi
Each circle in Figure~\ref{fig:suitability-and-goodness} is the runtime of a plan and $\times$ is the plan our optimizer picks.

We first observe that different types of plans are more suitable for different queries. The main structural properties of a query that govern which types of plans will perform well are how large and how cyclic the query is. For clique-like densely cyclic queries, such as $Q5$, and small sparsely-cyclic queries, such as $Q3$, best plans are WCO.
On acyclic queries, such as $Q11$ and $Q13$, BJ plans are best on some datasets and WCO plans on others. On acyclic queries WCO plans are equivalent to left deep BJ plans, which are worse than bushy BJ plans on some datasets.
Finally, hybrid plans are best plans for queries that contain small cyclic structure that do no share edges, such as $Q8$.

Our most interesting query is $Q12$, which is a 6-cycle query. $Q12$ can be evaluated efficiently with both WCO and hybrid plans (and reasonably well with some BJ plans). The hybrid plans first perform binary joins to compute 4-paths, and then extend 4-paths into 6-cycles with an intersection. Figure~\ref{fig:non-ghd-hybrid-plan-example} from Section~\ref{sec:introduction} shows an example of such hybrid plans. These plans do not correspond to the GHDs in EH's plan space. On the Amazon graph, one of these hybrid plans is optimal and our optimizer picks that plan.
On  Google graph our optimizer picks an efficient BJ plan although the optimal plan is WCO. 

Our optimizer's plans were broadly optimal or very close to optimal across our experiments. Specifically, our optimizer's plan was optimal in 15 of our 31 spectrums, was within 1.4x of the optimal in 21 spectrum and within 2x in 28 spectrums. In 2 of the 3 cases we were more than 2x of the optimal, the absolute runtime difference was in sub-seconds. There was only one experiment in which our plan was not close to the optimal plan, which is shown in Figure~\ref{fig:suit-q11-epinions_3}. Observe that our optimizer picks different types of plans across different types of queries. In addition, as we demonstrated with $Q12$ above, we can pick different plans for the same query on different data sets ($Q8$ and $Q13$ are other examples).

Although we do not study query optimization time in this paper, our optimizer generated a plan within  331ms in all of our experiments except for Q7$_5$ on Google which took ~1.4 secs.

\subsection{Adaptive WCO Plan Evaluation}
\label{subsubsec:adaptive-evaluation}

In order to understand the benefits we get by adaptively picking QVOs, we studied the spectrums of WCO plans of $Q2$, $Q3$, $Q4$, $Q5$, and $Q6$, and hybrid plans for $Q10$ on Epinions, Amazon and Google graphs. These are the queries in which our DP optimizer's fixed plans contained a chain of two or more \textsc{E/I} operators (so we could adapt them). The spectrum of $Q10$ on Epinions took a prohibitively long time to generate and is omitted.
\iflong
Figure~\ref{fig:adaptive-spectrums} shows the 17 spectrums we generated.
\else
Due to space constraints, Figure~\ref{fig:adaptive-spectrums} shows 12 of the 17 spectrums we generated, for one of the graphs for each query. The rest of our spectrums are presented in the longer version of our paper~\cite{mhedhbi:sqs-tech-report}.
\fi
In the case of $Q2$, $Q3$, and $Q4$, selecting QVOs adaptively overall improves the performance of every fixed plan. For example, the fixed plan our DP optimizer picks for $Q3$ on Epinions improves by 1.2x but other plans improve by up to 1.6x. $Q10$'s spectrum for hybrid plans are similar to $Q3$ and $Q4$'s. Each hybrid plan of $Q10$ computes the diamonds on the left and triangles on the right and joins on $a_4$. Here, we can adaptively compute the diamonds (but not the triangles). Each fixed hybrid plan improves by adapting and some improve by up to 2.1x.
On $Q5$ most plans' runtimes remain similar but one WCO plan improves by 4.3x. The main benefit of adapting  is that it makes our optimizer more robust against picking bad QVOs. Specifically, the deviation between the best and worst  plans are smaller in adaptive plans than fixed plans.

The only exception to these observations is $Q6$, where several plan's performance gets worse, although the deviation between good and bad plans still become smaller. We observed that for cliques, the overheads of adaptively picking QVOs is higher than other queries. This is because: (i) cost re-evaluation accesses many actual adjacency list sizes, so the overheads are high; and (ii) the QVOs of cliques have similar behaviors: each one extends edges to triangles, then four cliques, etc.), and the benefits are low.

\iflong
\begin{figure*}[t!]
\centering
\captionsetup{justification=centering}
	\begin{subfigure}[b]{0.14\textwidth}
		\centering
		\includegraphics[scale=0.5]{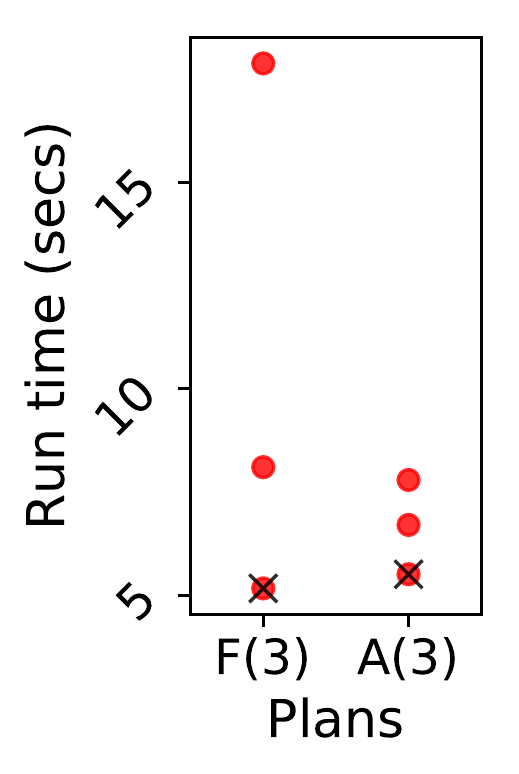}
		\caption{Q2, Amazon.}
		\label{fig:adaptive-q2-amazon}
	\end{subfigure}
\begin{subfigure}[b]{0.14\textwidth}
	\centering
	\includegraphics[scale=0.5]{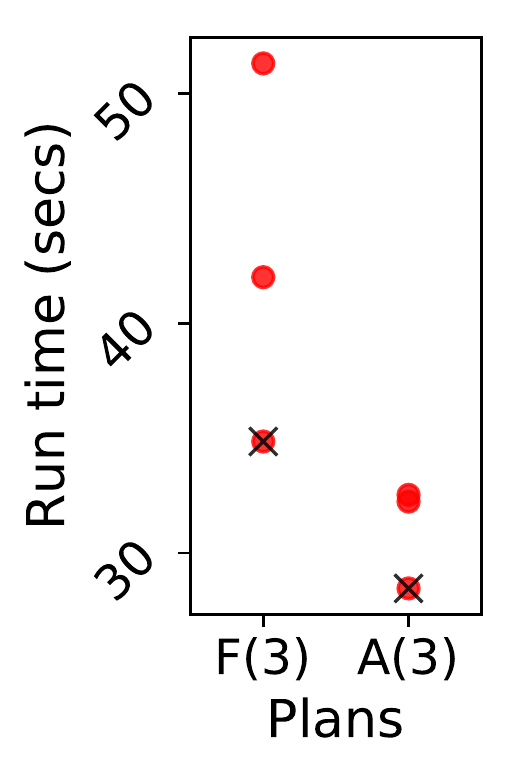}
	\caption{Q2, Epinions.}
	\label{fig:adaptive-q2-epinions}
\end{subfigure}
\begin{subfigure}[b]{0.14\textwidth}
	\centering
	\includegraphics[scale=0.5]{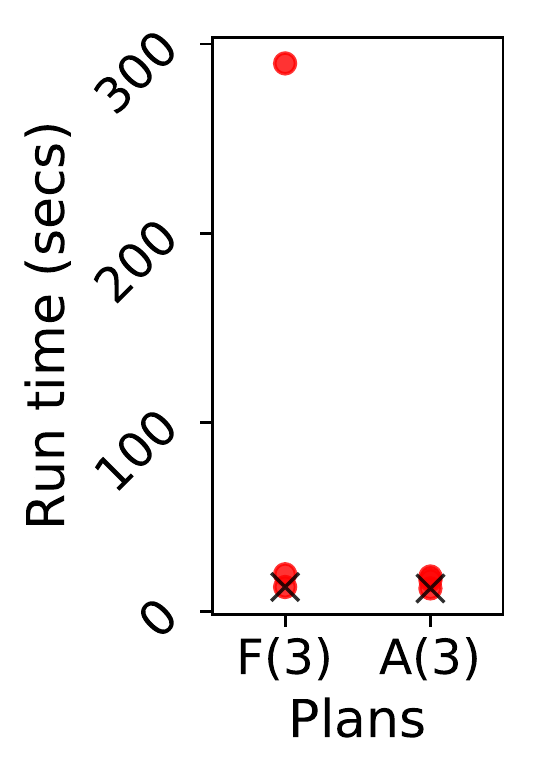}
	\caption{Q2, Google.}
	\label{fig:adaptive-q2-google}
\end{subfigure}
	\begin{subfigure}[b]{0.14\textwidth}
		\centering
		\includegraphics[scale=0.5]{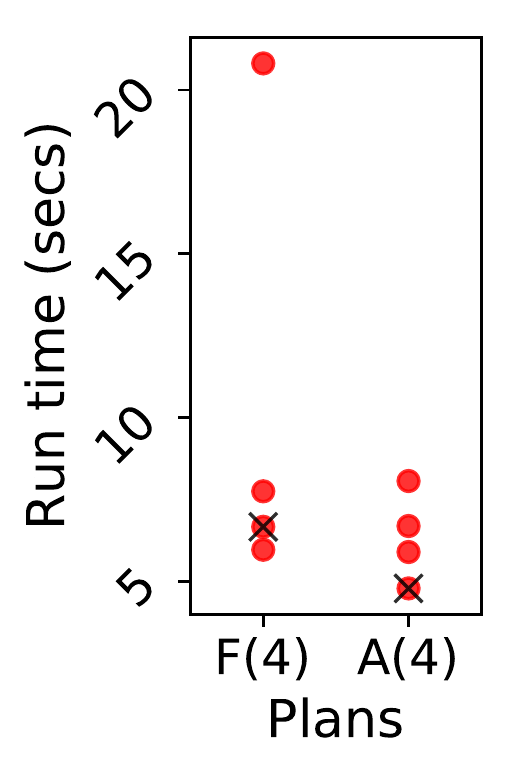}
		\caption{Q3, Amazon.}
		\label{fig:adaptive-q3-amazon}
	\end{subfigure}
\begin{subfigure}[b]{0.14\textwidth}
	\centering
	\includegraphics[scale=0.5]{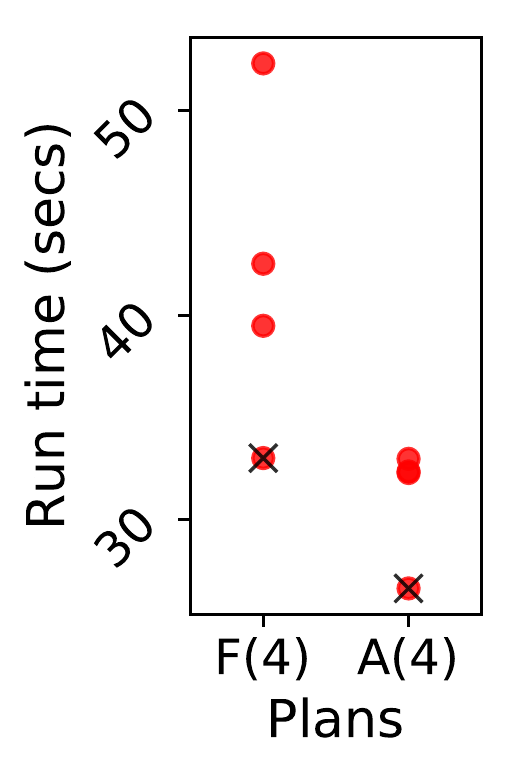}
	\caption{Q3, Epinions.}
	\label{fig:adaptive-q3-epinions}
\end{subfigure}
\begin{subfigure}[b]{0.14\textwidth}
	\centering
	\includegraphics[scale=0.5]{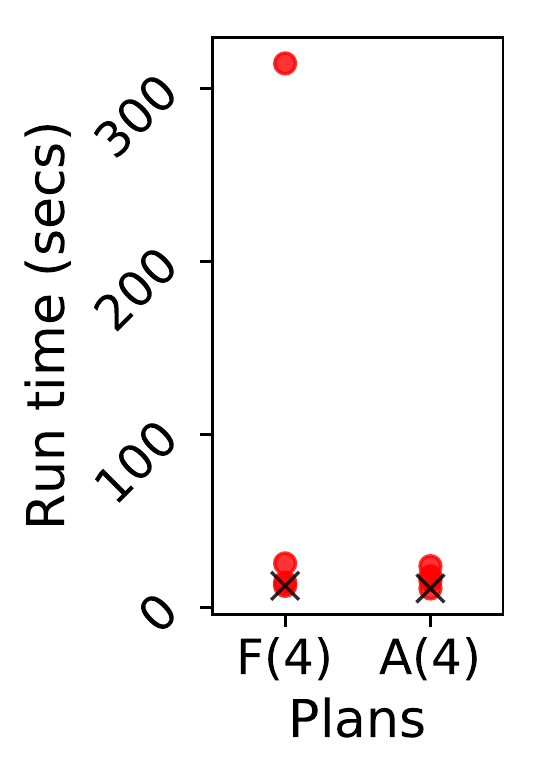}
	\caption{Q3, Google.}
	\label{fig:adaptive-q3-google}
\end{subfigure}
\begin{subfigure}[b]{0.14\textwidth}
	\centering
	\includegraphics[scale=0.5]{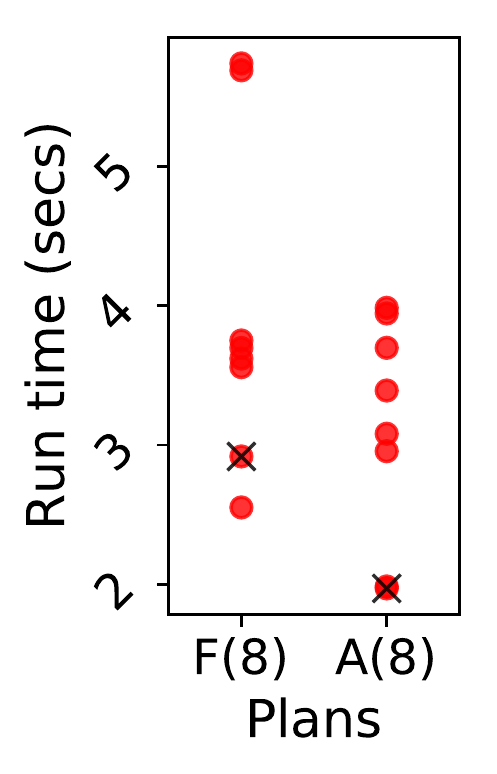}
	\caption{Q4, Amazon.}
	\label{fig:adaptive-q4-amazon}
\end{subfigure}
\begin{subfigure}[b]{0.14\textwidth}
	\centering
	\includegraphics[scale=0.5]{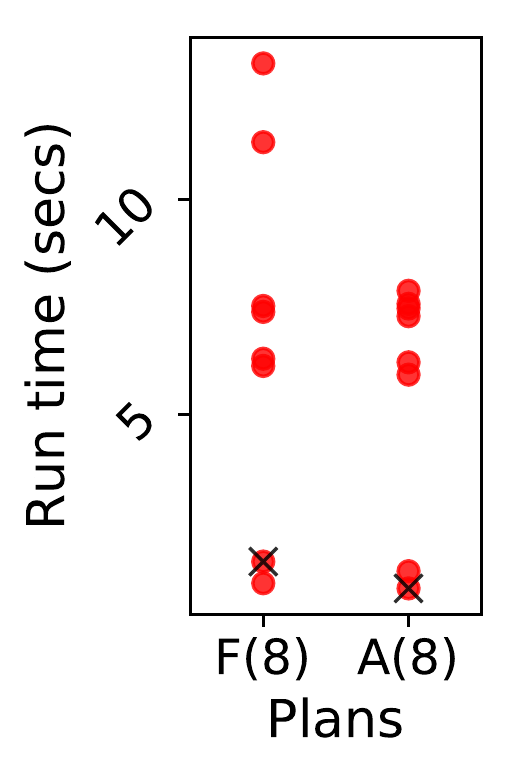}
	\caption{Q4, Epinions.}
	\label{fig:adaptive-q4-epinions}
\end{subfigure}
\begin{subfigure}[b]{0.14\textwidth}
	\centering
	\includegraphics[scale=0.5]{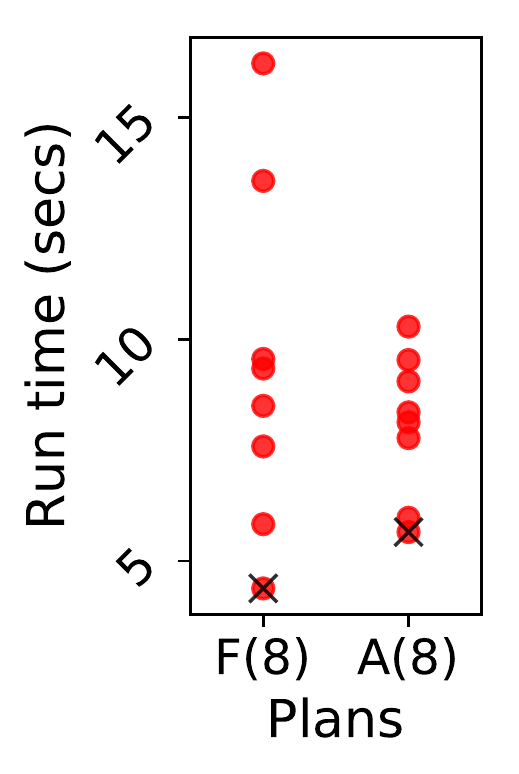}
	\caption{Q4, Google.}
	\label{fig:adaptive-q4-google}
\end{subfigure}
\begin{subfigure}[b]{0.14\textwidth}
	\centering
	\includegraphics[scale=0.5]{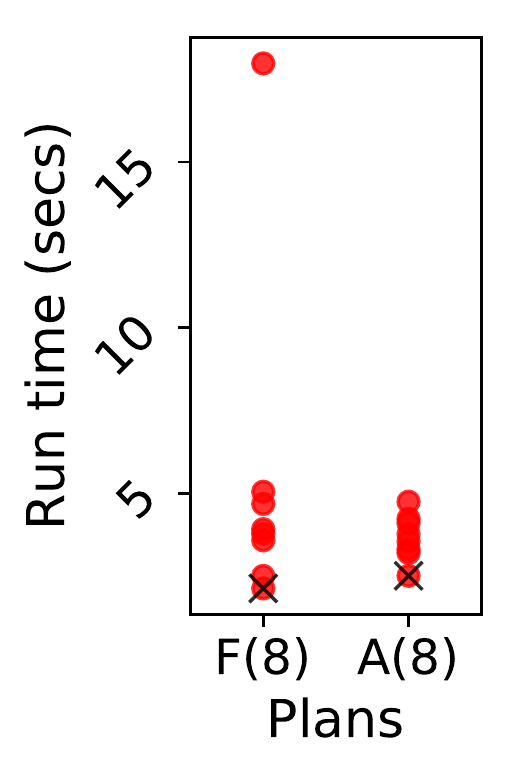}
	\caption{Q5, Amazon.}
	\label{fig:adaptive-q5-amazon}
\end{subfigure}
\begin{subfigure}[b]{0.14\textwidth}
	\centering
	\includegraphics[scale=0.5]{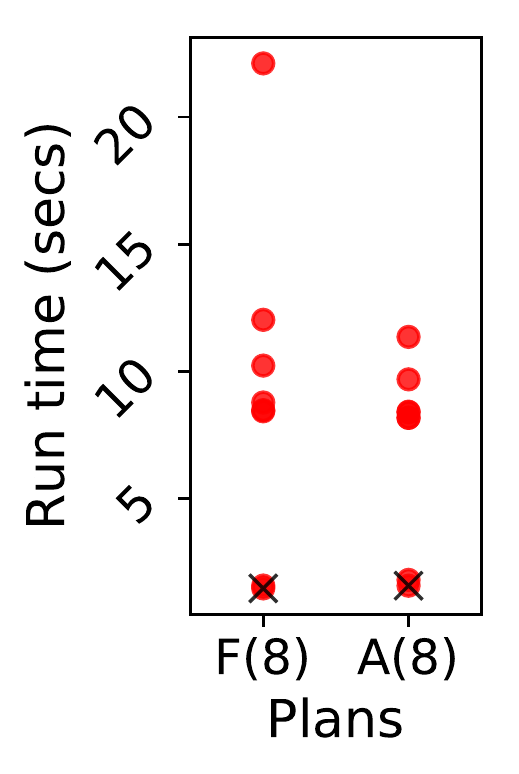}
	\caption{Q5, Epinions.}
	\label{fig:adaptive-q5-epinions}
\end{subfigure}
\begin{subfigure}[b]{0.14\textwidth}
	\centering
	\includegraphics[scale=0.5]{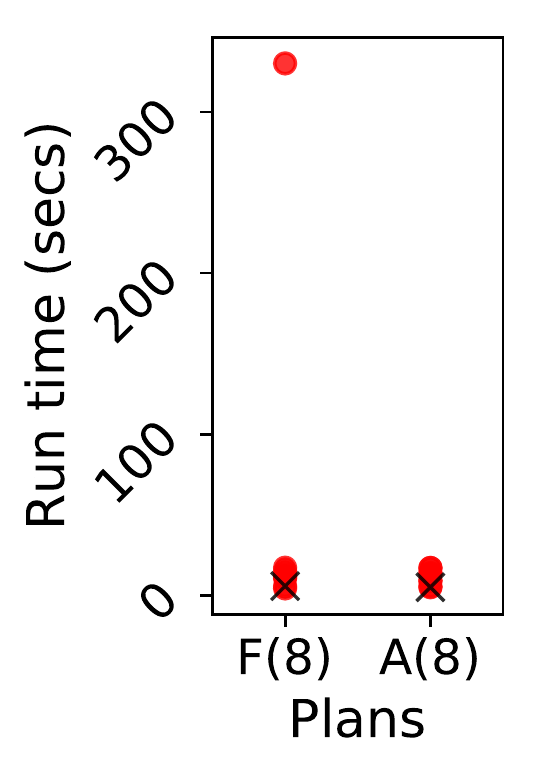}
	\caption{Q5, Google.}
	\label{fig:adaptive-q5-google}
\end{subfigure}
\begin{subfigure}[b]{0.14\textwidth}
	\centering
	\includegraphics[scale=0.5]{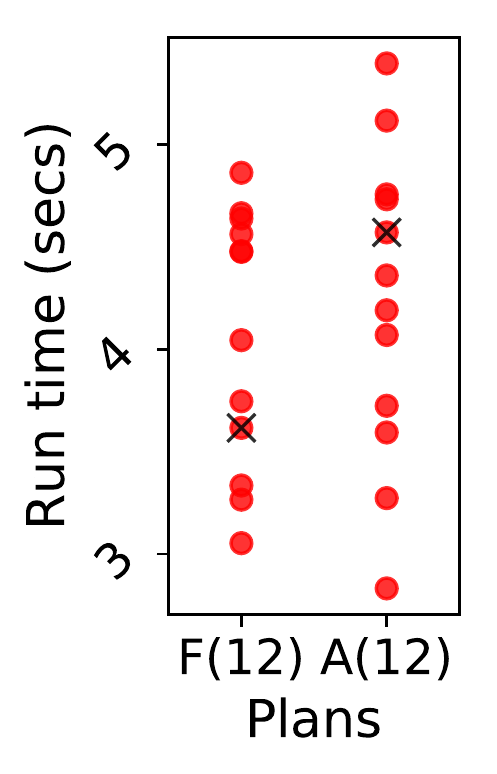}
	\caption{Q6, Amazon.}
	\label{fig:adaptive-q6-amazon}
\end{subfigure}
\begin{subfigure}[b]{0.15\textwidth}
	\centering
	\includegraphics[scale=0.5]{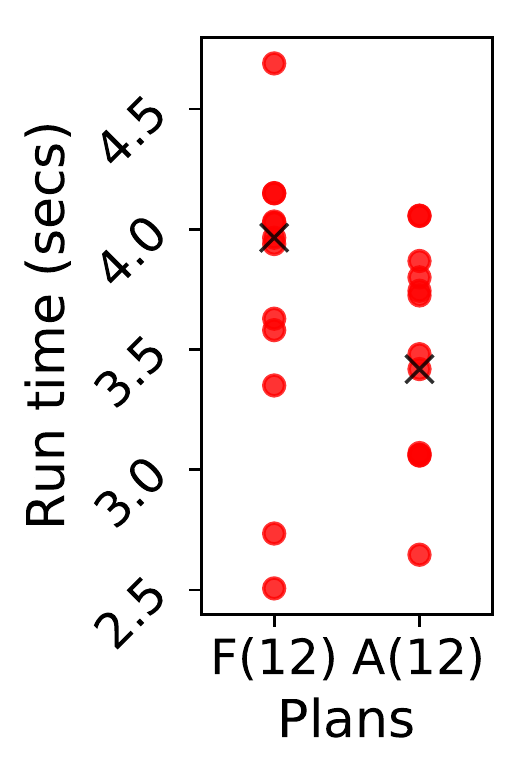}
	\caption{Q6, Epinions.}
	\label{fig:adaptive-q6-epinions}
\end{subfigure}
\begin{subfigure}[b]{0.14\textwidth}
	\centering
	\includegraphics[scale=0.5]{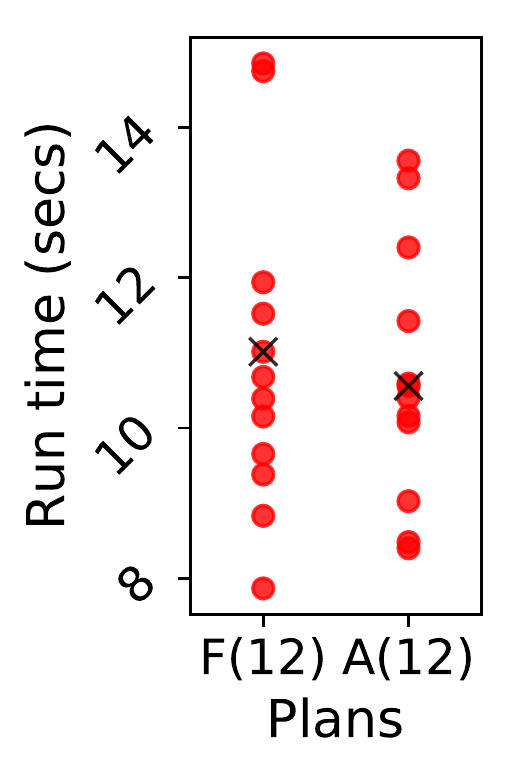}
	\caption{Q6, Google.}
	\label{fig:adaptive-q6-google}
\end{subfigure}
\begin{subfigure}[b]{0.14\textwidth}
	\centering
	\includegraphics[scale=0.5]{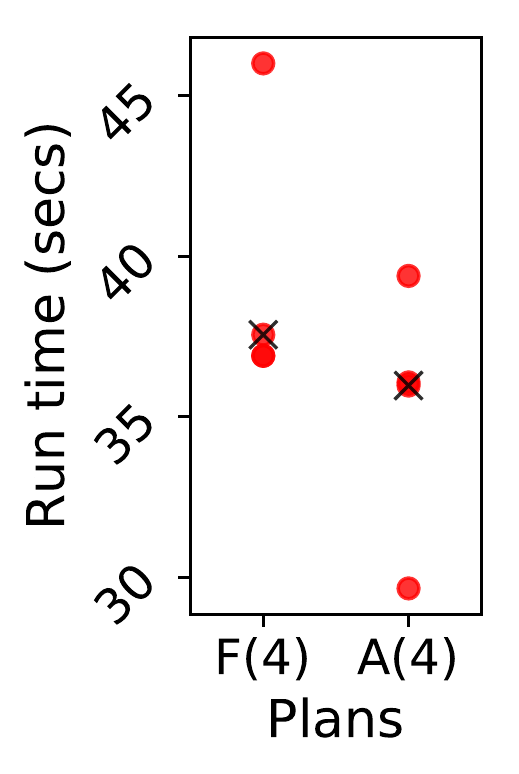}
	\caption{Q10, Amazon.}
	\label{fig:adaptive-q10-amazon}
\end{subfigure}
\begin{subfigure}[b]{0.14\textwidth}
	\centering
	\includegraphics[scale=0.5]{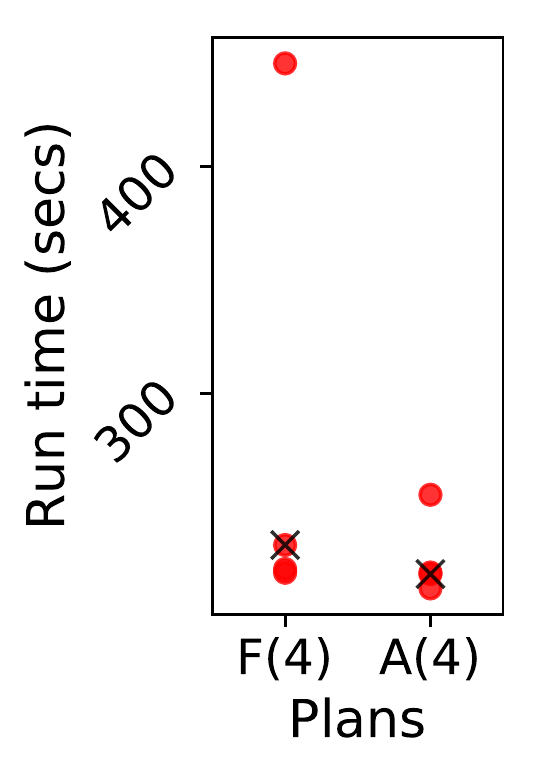}
	\caption{Q10, Google.}
	\label{fig:adaptive-q10-google}
\end{subfigure}
\caption{Adaptive plan spectrums.}
\vspace{-15pt}
\label{fig:adaptive-spectrums}
\end{figure*}
\else
\begin{figure*}[htp]
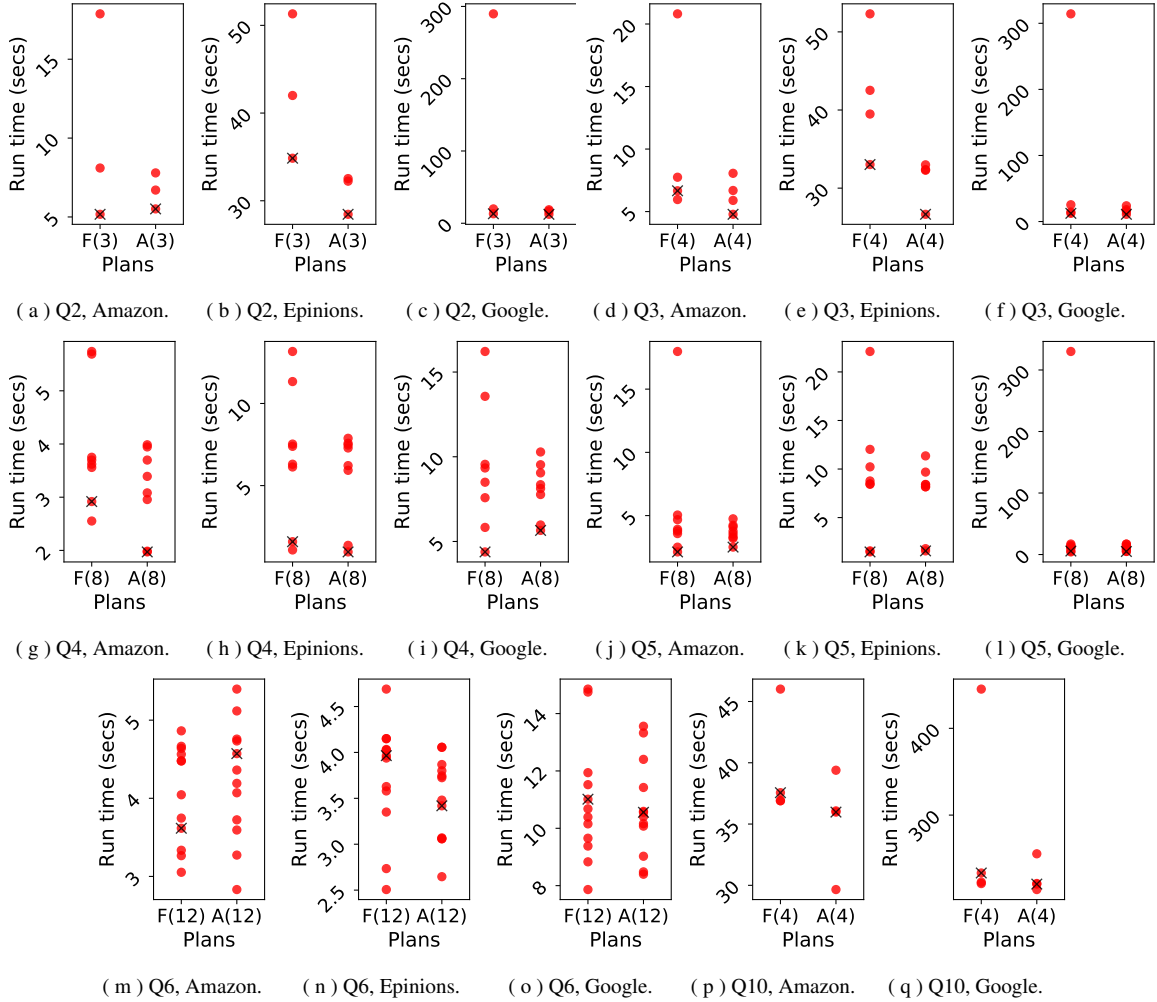

\centering
\captionsetup{justification=centering}
\begin{subfigure}[b]{0.14\textwidth}
	\centering
	\includegraphics[scale=0.5]{adaptive-q2-epinions}
	\caption{Q2, Epinions.}
	\label{fig:adaptive-q2-epinions}
\end{subfigure}
\begin{subfigure}[b]{0.14\textwidth}
	\centering
	\includegraphics[scale=0.5]{adaptive-q2-google}
	\caption{Q2, Google.}
	\label{fig:adaptive-q2-google}
\end{subfigure}
\begin{subfigure}[b]{0.14\textwidth}
	\centering
	\includegraphics[scale=0.5]{adaptive-q3-epinions}
	\caption{Q3, Epinions.}
	\label{fig:adaptive-q3-epinions}
\end{subfigure}
\begin{subfigure}[b]{0.14\textwidth}
	\centering
	\includegraphics[scale=0.5]{adaptive-q3-google}
	\caption{Q3, Google.}
	\label{fig:adaptive-q3-google}
\end{subfigure}
\begin{subfigure}[b]{0.14\textwidth}
	\centering
	\includegraphics[scale=0.5]{adaptive-q4-amazon}
	\caption{Q4, Amazon.}
	\label{fig:adaptive-q4-amazon}
\end{subfigure}
\begin{subfigure}[b]{0.14\textwidth}
	\centering
	\includegraphics[scale=0.5]{adaptive-q4-epinions}
	\caption{Q4, Epinions.}
	\label{fig:adaptive-q4-epinions}
\end{subfigure}
\begin{subfigure}[b]{0.14\textwidth}
	\centering
	\includegraphics[scale=0.5]{adaptive-q5-amazon}
	\caption{Q5, Amazon.}
	\label{fig:adaptive-q5-amazon}
\end{subfigure}
\begin{subfigure}[b]{0.14\textwidth}
	\centering
	\includegraphics[scale=0.5]{adaptive-q5-epinions}
	\caption{Q5, Epinions.}
	\label{fig:adaptive-q5-epinions}
\end{subfigure}
\begin{subfigure}[b]{0.14\textwidth}
	\centering
	\includegraphics[scale=0.5]{adaptive-q6-amazon}
	\caption{Q6, Amazon.}
	\label{fig:adaptive-q6-amazon}
\end{subfigure}
\begin{subfigure}[b]{0.15\textwidth}
	\centering
	\includegraphics[scale=0.5]{adaptive-q6-epinions}
	\caption{Q6, Epinions.}
	\label{fig:adaptive-q6-epinions}
\end{subfigure}
\begin{subfigure}[b]{0.14\textwidth}
	\centering
	\includegraphics[scale=0.5]{adaptive-q10-amazon}
	\caption{Q10, Amazon.}
	\label{fig:adaptive-q10-amazon}
\end{subfigure}
\begin{subfigure}[b]{0.14\textwidth}
	\centering
	\includegraphics[scale=0.5]{adaptive-q10-google}
	\caption{Q10, Google.}
	\label{fig:adaptive-q10-google}
\end{subfigure}
\caption{Adaptive plan spectrums.}
\vspace{-15pt}
\label{fig:adaptive-spectrums}
\end{figure*}
\fi

\subsection{EmptyHeaded (EH) Comparisons}
\label{subsub:eh-comparisons}

EH is one of the most efficient systems for one-time subgraph queries and its plans are the closest to ours.
Recall from Section~\ref{sec:introduction} that EH has a cost-based optimizer that picks a GHD with the minimum width, i.e., EH picks a GHD with the lowest AGM bound across all of its sub-queries. This allows EH to often (but not always) pick good decompositions. 
However: (1) EH does not optimize the choice of QVOs for computing its sub-queries; and (2) EH cannot pick plans that have intersections after a binary join, as such plans do not correspond to GHDs. In particular, the QVO EH picks for a query $Q$ is the lexicographic order of the variables used for query vertices when a user issues the query. EH's only heuristic is that QVOs of two sub-queries that are joined start with query vertices on which the join will happen.
Therefore by issuing the same query with different variables, users can make EH pick a good or a bad ordering. This shortcoming has the advantage though that by making EH pick good QVOs, we can show that our orderings also improve EH. The important point is that  EH does not optimize for QVOs. We therefore report EH's performance with both ``bad'' variables (\texttt{EH-b}) and ``good'' variables (\texttt{EH-g}). For good orderings we use the ordering that Graphflow picks.  For bad orderings, we generated the spectrum of plans in EH (explained momentarily) and picked the worst-performing ordering for the GHD EH picks.
For our experiments we ran $Q3$, $Q5$, $Q7$, $Q8$, $Q9$, $Q12$, and $Q13$ on  Amazon, Google, and Epinions. We first explain how we generated EH spectrums and then present our results.

\subsubsection{EH Spectrums}
\label{subsub:eh-spectrum}

\begin{figure}[t!]
	\centering
	\captionsetup{justification=centering}
	\begin{subfigure}[b]{0.145\textwidth}
		\centering
		\includegraphics[scale=0.48]{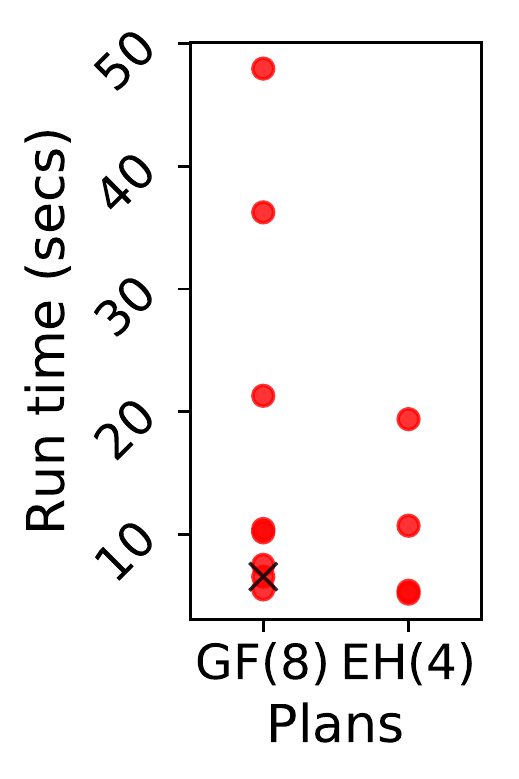}
		\caption{$Q3$, Amazon.}
		\label{fig:gf-eh-q3-amazon}
	\end{subfigure}
	\begin{subfigure}[b]{0.145\textwidth}
		\centering
		\includegraphics[scale=0.48]{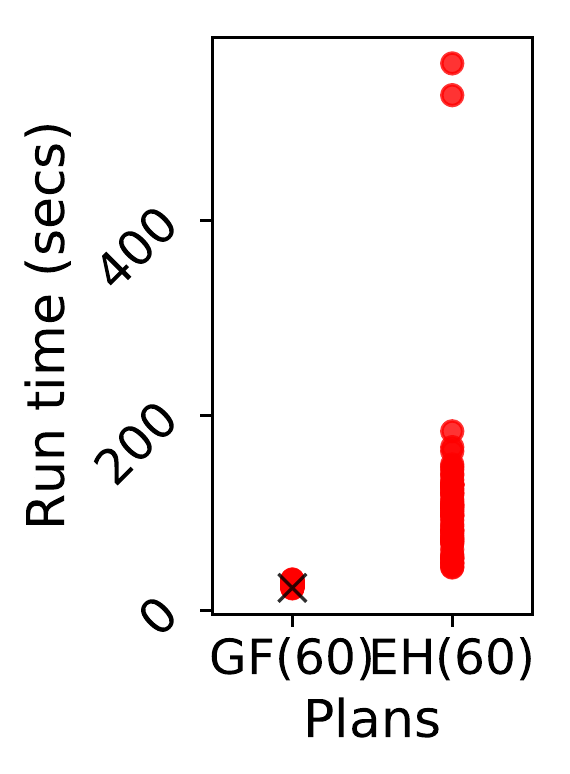}
		\caption{$Q7$, Epinions.}
		\label{fig:gf-eh-q7-google}
	\end{subfigure}
	\begin{subfigure}[b]{0.145\textwidth}
		\centering
		\includegraphics[scale=0.48]{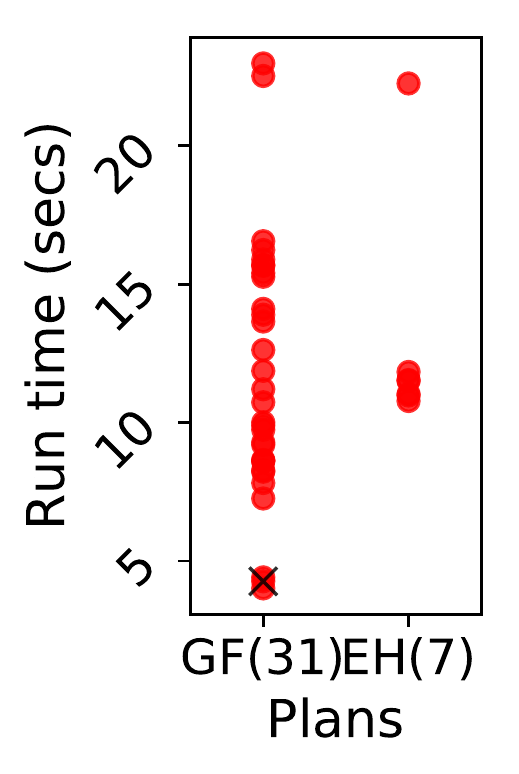}
		\caption{$Q8$, Amazon.}
		\label{fig:gf-eh-q8-amazon}
	\end{subfigure}
	\vspace{-5pt}
	\caption{Plan spectrum charts for EmpthHeaded (EH).}
	\label{fig:spectrum-gf-eh}
	\vspace{-15pt}
\end{figure}

Given a query, EH's query planner enumerates a set of minimum width GHDs and picks one of these GHDs.
To define the plan spectrum of EH, we took all of these GHDs, and by rewriting the query with all possible different variables, we generate all possible QVOs of the sub-queries of the GHD that EH considers.
Figure~\ref{fig:spectrum-gf-eh} shows a sample of the spectrums for Q3 and Q7 on Amazon and for Q8 on Epinions along with Graphflow's plan spectrum (including WCO, BJ, and hybrid plans) for comparison. For Q9, Q12, and Q13 we could not generate spectrums as every EH plan took more than our 30 minutes time limit. For Q7, both Graphflow and EH generate only WCO plans. For Q8, EH generates two GHDs (two triangles joined on $a_3$) whose different QVOs give 4 different plans for a total of 8. One of the plans in the spectrum is omitted as it had memory issues. We note that out of these queries, Q8 and Q9 were the only queries for which EH generated two different decompositions (ignoring the QVOs of sub-queries). For Q but neither decomposition under any QVO ran within our time limit  on our datasets.

\begin{table*}[t!]
	\centering
	\captionsetup{justification=centering}
	\begin{tabular}{ m{1cm} m{0.7cm}   
			m{0.56cm}
			m{0.56cm} m{0.56cm} m{0.56cm} m{0.56cm} m{0.56cm} m{0.56cm}  
			m{0.56cm} m{0.56cm} m{0.56cm} m{0.56cm} m{0.56cm} m{0.56cm}
			m{0.56cm} m{0.56cm}} 
		\toprule
		& & Q1 & Q3 & Q3$_2$ & Q5 & Q5$_2$ & Q7 & Q7$_2$ & Q8 & Q8$_2$ & Q9 & Q9$_2$ & Q12 & Q12$_2$ & Q13 & Q13$_2$ \\
		\midrule
		Amazon &
		\texttt{EH-b}\newline
		\texttt{EH-g}\newline
		\texttt{GF}
		& 1.0 \newline 0.6 \newline  0.6
		& 19.0 \newline 5.4 \newline  5.5
		& 3.4 \newline 1.3 \newline 2.1
		& 47.1 \newline 3.3 \newline  1.9
		& 9.2 \newline 1.5 \newline 0.8
		& 91.4 \newline 21.2 \newline 9.53
		& 11.6 \newline 1.7 \newline 0.9
		& 22.2 \newline 10.6 \newline 5.1
		& 1.8 \newline 1.4 \newline 2.0
		& $Mm$ \newline $Mm$ \newline 24.7
		& $Mm$ \newline $Mm$\newline 2.4
		& $Mm$ \newline $Mm$ \newline 209.2
		& $Mm$ \newline $Mm$ \newline 14.8
		& $Mm$ \newline $Mm$ \newline 48.0
		& $Mm$ \newline $Mm$ \newline 11.25
		\\
		\midrule
		Google &
		\texttt{EH-b}\newline
		\texttt{EH-g}\newline
		\texttt{GF}
		& 1.9 \newline 1.4 \newline 2.6
		& 444.5 \newline 12.0 \newline  14.0
		& 42.6 \newline 2.1 \newline 4.0
		& 401.1 \newline 11.3 \newline 5.9
		& 77.6 \newline 2.3 \newline 2.1
		& 1.04K \newline 107.3 \newline 48.8
		& 23.4 \newline 4.8 \newline 3.3
		& 66.6 \newline 35.8 \newline 17.0
		& 16.0 \newline 3 \newline 4.5
		& $TL$ \newline $TL$ \newline 236.2
		& $TL$ \newline $TL$ \newline 6.9
		& $Mm$ \newline $Mm$ \newline 510.6
		& $Mm$ \newline $Mm$ \newline 73.8
		& $Mm$ \newline $Mm$ \newline 1.44K
		& $Mm$ \newline $Mm$ \newline 70.1
		\\
		\midrule
		Epinions &  
		\texttt{EH-b}\newline
		\texttt{EH-g}\newline
		\texttt{GF}
		& 0.5 \newline 0.2 \newline  0.4
		& 42.7 \newline 26.6 \newline  28.1
		& 6.5 \newline 1.7 \newline 4.6
		& 64.5 \newline 3.5 \newline 1.5
		& 11.4 \newline 0.9 \newline 0.6
		& 560.7 \newline 45.7 \newline 23.7
		& 2.9 \newline 0.8 \newline 1.2
		& 1.01K \newline 117.2 \newline 37.5
		& 22.0 \newline 7.0 \newline 5.4
		& $TL$ \newline $TL$ \newline 865.3
		& $TL$ \newline $TL$ \newline 26.1
		& $Mm$ \newline $Mm$ \newline $TL$
		& $Mm$ \newline $Mm$ \newline $TL$
		& $TL$ \newline $TL$ \newline $TL$
		& $TL$ \newline $TL$ \newline $TL$
		\\
		\bottomrule
	\end{tabular}
	\caption{Runtime in secs of Graphflow plans (\texttt{GF}) and EmptyHeaded with good orderings (\texttt{EH-g}) and bad orderings (\texttt{EH-b}).\\ $TL$ indicates the query did not finish in 30 mins. $Mm$ indicates the system ran out of memory.}
	\vspace{-15pt}
	\label{tab:eh-comparisons}
\end{table*}

\subsubsection{Graphflow vs EH Comparisons}
\label{subsub:gf-vs-eh}

We ran our queries on Graphflow with adapting off. To compare, we ran EH's plan with good and bad QVOs for $Q3$, $Q5$, $Q7$, $Q8$  (recall no EH plan ran within our time limit for $Q9$, $Q12$, and $Q13$).
We repeated the experiments once with no labels and once with two labels. 
Table~\ref{tab:eh-comparisons} shows our results. Except for $Q1$ on Google and $Q8_2$ on Amazon where the difference is only 500ms and 200ms, respectively. Graphflow is always faster than \texttt{EH-b}, where the runtime is as high as 68x in one instance. 
The most performance difference is on $Q5$ and Google, for which both our system and EH uses a WCO plan. When we force EH to pick our good QVOs, on smaller size queries EH can be more efficient than our plans. For example, although Graphflow is 32x faster than \texttt{EH-b} on $Q3$ Google, it is 1.2x slower than \texttt{EH-g}. Importantly \texttt{EH-g} is always faster than  \texttt{EH-b}, showing that our QVOs improve runtimes consistently in a completely independent system that implements WCO-style processing.

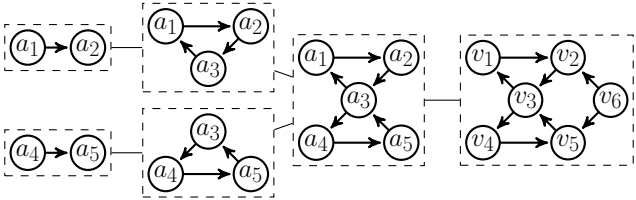
\begin{figure}[t!]
\captionsetup{justification=centering}
	\centering
    \begin{tikzpicture}[grow=left,
    level 1/.style={sibling distance=1.4cm,level distance=2.5cm},
    level 2/.style={sibling distance=1.4cm,level distance=2cm},
    level 3/.style={sibling distance=1.4cm,level distance=2cm}]
    \node[draw,dashed,inner sep=\InnerSep] {
		\begin{tikzpicture}[scale=0.4,solid, transform shape,->,>=stealth', shorten >=1pt, auto,node distance=2cm, thick, main node/.style={circle,draw,font=\sffamily\Huge\bfseries}]
			\node[main node] (3) {$v_3$};
			\node[main node] (1) [above left of=3] {$v_1$};
			\node[main node] (4) [below left of=3] {$v_4$};
			\node[main node] (2) [above right of=3] {$v_2$};
			\node[main node] (5) [below right of=3] {$v_5$};
			\node[main node] (6) [below right of=2] {$v_6$};
			\path[every node/.style={font=\sffamily\small}]
			(1) edge (2)
			(2) edge (3)
			(3) edge (1) edge (4)
			(4) edge (5)
			(5) edge (3)
			(6) edge (2) edge (5) ;
		\end{tikzpicture}
    }
    child {node[draw,dashed,inner sep=\InnerSep] {
		\begin{tikzpicture}[scale=0.4,solid, transform shape,->,>=stealth', shorten >=1pt, auto,node distance=2cm, thick, main node/.style={circle,draw,font=\sffamily\Huge\bfseries}]
		\node[main node] (3) {$a_3$};
		\node[main node] (1) [above left of=3] {$a_1$};
		\node[main node] (4) [below left of=3] {$a_4$};
		\node[main node] (2) [above right of=3] {$a_2$};
		\node[main node] (5) [below right of=3] {$a_5$};
		\path[every node/.style={font=\sffamily\small}]
		(1) edge (2)
		(2) edge (3)
		(3) edge (1) edge (4)
		(4) edge (5)
		(5) edge (3) ;
		\end{tikzpicture}
     }
    child {node[draw,dashed,inner sep=\InnerSep] {
        \begin{tikzpicture}[scale=0.4,solid, transform shape,->,>=stealth', shorten >=1pt, auto,node distance=2cm, thick, main node/.style={circle,draw,font=\sffamily\Huge\bfseries}]
        \node[main node] (1) {$a_1$};
        \node[main node] (3) [below right of=1] {$a_3$};
        \node[main node] (2) [above right of=3] {$a_2$};
        \path[every node/.style={font=\sffamily\small}]
        (1) edge (2)
        (2) edge (3)
        (3) edge (1) ;
        \end{tikzpicture}
    }
      child {node[draw,dashed,inner sep=\InnerSep] {
        \begin{tikzpicture}[scale=0.4,solid, transform shape,->,>=stealth', shorten >=1pt, auto,node distance=2cm, thick, main node/.style={circle,draw,font=\sffamily\Huge\bfseries}]
        \node[main node] (1) {$a_1$};
        \node[main node] (2) [right of=1] {$a_2$};
        \path[every node/.style={font=\sffamily\small}]
        (1) edge (2) ;
        \end{tikzpicture}
      }}
    }
    child {node[draw,dashed,inner sep=\InnerSep] {
        \begin{tikzpicture}[scale=0.4,solid, transform shape,->,>=stealth', shorten >=1pt, auto,node distance=2cm, thick, main node/.style={circle,draw,font=\sffamily\Huge\bfseries}]
        \node[main node] (4) {$a_4$};
        \node[main node] (3) [above right of=4] {$a_3$};
        \node[main node] (5) [below right of=3] {$a_5$};
        \path[every node/.style={font=\sffamily\small}]
        (3) edge (4)
        (4) edge (5)
        (5) edge (3) ;
        \end{tikzpicture}
    }
      child {node[draw,dashed,inner sep=\InnerSep] {
        \begin{tikzpicture}[scale=0.4,solid, transform shape,->,>=stealth', shorten >=1pt, auto,node distance=2cm, thick, main node/.style={circle,draw,font=\sffamily\Huge\bfseries}]
        \node[main node] (4) {$a_4$};
        \node[main node] (5) [right of=4] {$a_5$};
        \path[every node/.style={font=\sffamily\small}]
        (4) edge (5) ;
        \end{tikzpicture}
      }}
    }
    }
	;
\end{tikzpicture}
        \caption{Plan (drawn horizontally) with seamless mixing of intersections and binary joins on Q9.}
        \vspace{-15pt}
		\label{fig:seamless-mixing}
\end{figure}

We next discuss $Q9$, which demonstrates again the benefit we get by seamlessly mixing intersections with binary joins. Figure~\ref{fig:seamless-mixing} shows the plan our optimizer picks on $Q9$ on all of our datasets. Our plan separately computes two triangles, joins them, and finally performs a 2-way intersection. This execution does not correspond to the GHD-based plans of EH, so is not in the plan space of EH. Instead, EH considers two GHDs for this query but neither of them finished within our time limit.

\iflong
\subsection{Scalability Experiments}
\label{subsub:scalability-experiments}

We next demonstrate the scalability of Graphflow on larger datasets and linear scalability across physical cores. We evaluated $Q1$ on LiveJournal and Twitter, $Q2$ on LiveJournal, and $Q14$, which is a very difficult 7-clique query, on Google. We repeated each query with 1, 2, 4, 8, 16, and 32 cores, except we use 8, 16, and 32 cores on the Twitter graph. Figure~\ref{fig:scalability-plots} shows our results. Our plans scale linearly until 16 cores with a slight slow down when moving 32 cores which uses all system resources. For example, going from 1 core to 16 cores, our runtime is reduced by 13x for $Q1$ on LiveJournal, 16x for $Q2$ on LiveJournal and 12.3x for $Q14$ on Google.

\begin{figure}[htp]
	\centering
	\captionsetup{justification=centering}
	\begin{subfigure}[b]{0.235\textwidth}
		\centering
		\includegraphics[scale=0.45]{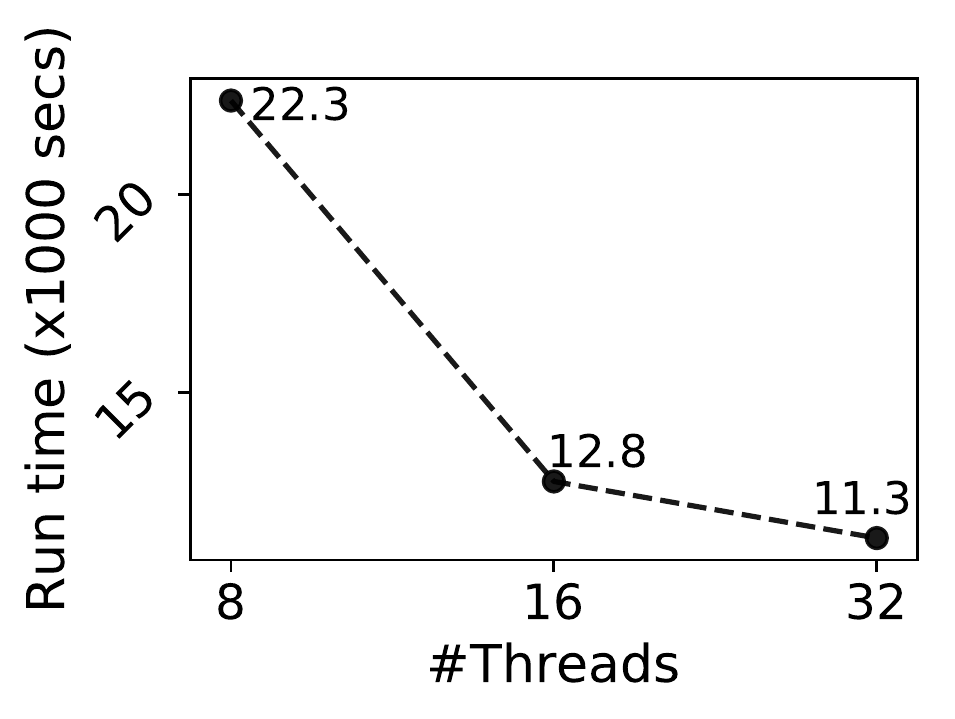}
		\caption{Q1, Twitter.}
		\label{fig:scale-q1-twitter}
	\end{subfigure}
	\begin{subfigure}[b]{0.235\textwidth}
		\centering
		\includegraphics[scale=0.45]{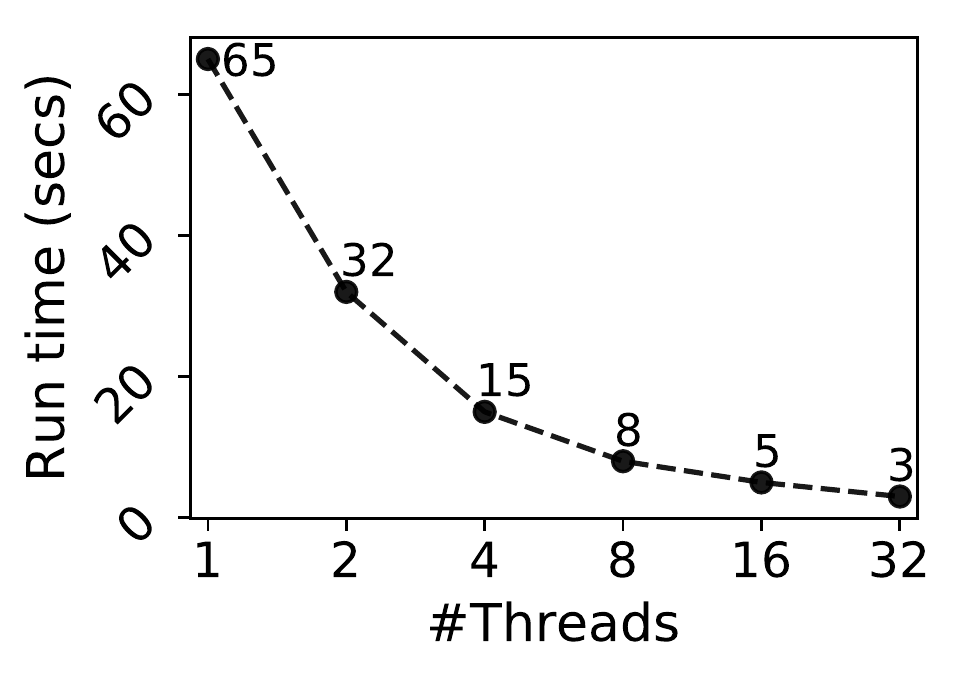}
		\caption{Q1, LiveJournal.}
		\label{fig:scale-q1-livejournal}
	\end{subfigure}
	\begin{subfigure}[b]{0.235\textwidth}
		\centering
		\includegraphics[scale=0.45]{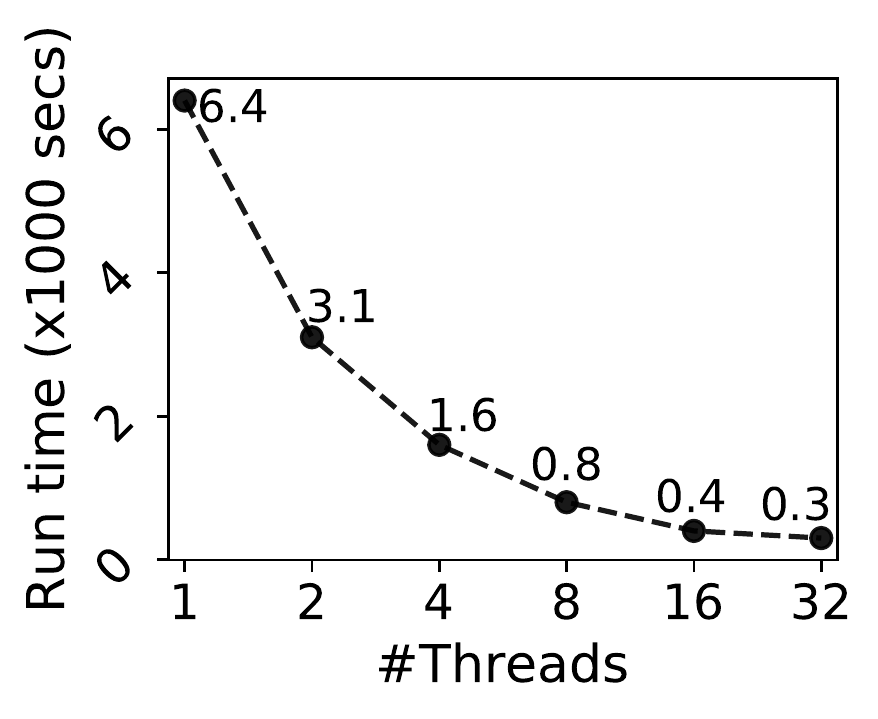}
		\caption{Q2, LiveJournal.}
		\label{fig:scale-q2-livejournal}
	\end{subfigure}
	\begin{subfigure}[b]{0.235\textwidth}
		\centering
		\includegraphics[scale=0.45]{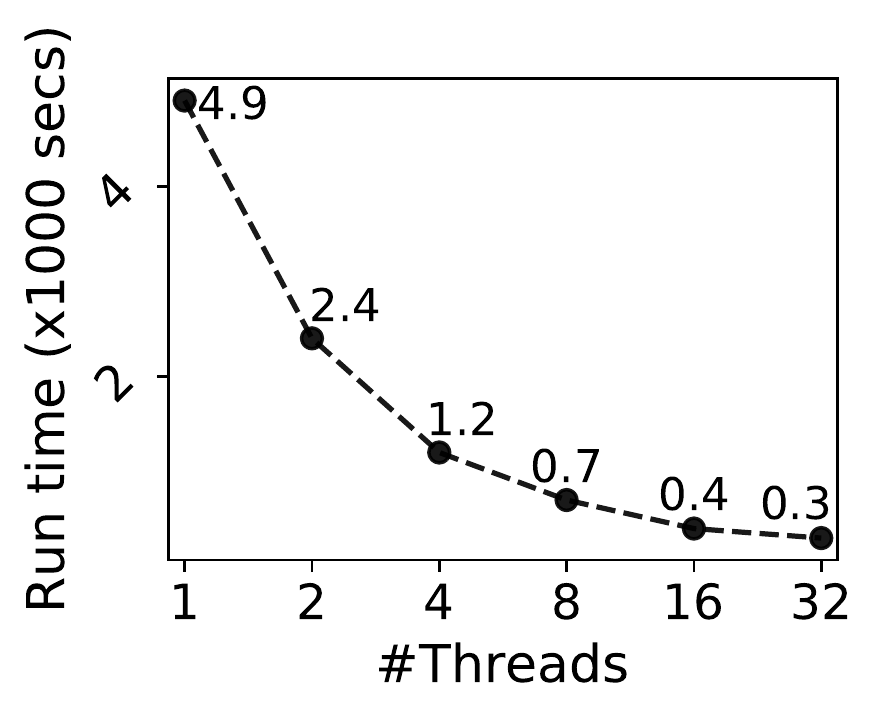}
		\caption{Q14, Google.}
		\label{fig:scale-q14-google}
	\end{subfigure}
	\caption{Scalability experiments.}
	\vspace{-7pt}
	\label{fig:scalability-plots}
	\vspace{-14pt}
\end{figure}
\fi
\vspace{-10pt}
\section{Related Work}
\label{sec:rw}

We review related work in WCO join algorithms, subgraph query evaluation algorithms, and cardinality estimation techniques related to our catalogue. For join and subgraph query evaluation, we focus on serial algorithms and single node systems. Several distributed solutions have been developed in the context of graph data processing~\cite{shao:psgl, lai:seed}, RDF engines~\cite{abdelaziz:spartex, zeng:trinity}, or multiway joins of relational tables~\cite{afrati:mrjoins, ammar:bigjoin, koutris:survey}. We do not review this literature here in detail.
There is also a rich body of work on adaptive query processing in relational systems for which we refer readers to reference~\cite{deshpande:adaptive}.
 
\noindent {\bf WCO Join Algorithms} Prior to GJ, there were two other WCO join algorithms  called NPRR~\cite{ngo:nprr} and Leapfrog TrieJoin (LFTJ)~\cite{veldhuizen:lftj}. Similar to GJ, these algorithms also perform attribute-at-a-time join processing using intersections.
The only reference that studies QVOs in these algorithms is reference~\cite{chu:tributary}, which studies picking the QVO for LFTJ algorithm in the context of multiway relational joins.
The reference picks the QVO based on the distinct values in the attribute of the relations. In subgraph query context, this heuristic ignores the structure of the query, e.g., whether the query is cyclic or not, and effectively orders the query vertices based on the selectivity of the labels on them. For example, this heuristic becomes degenerate if the query vertices do not have labels.


\noindent {\bf Subgraph Query Evaluation Algorithms:} Many of the earlier subgraph isomorphism algorithms are based on Ullmann's branch and bound or backtracking method~\cite{ullman:subgraph}. The algorithm conceptually performs a query-vertex-at-a-time matching using an arbitrary QVO. 
This algorithm has been improved with different techniques to pick better QVOs and filter partial matches, often focusing on queries with labels~\cite{cordella:vf, cordella:vf2, shang:si}. Turbo$_{ISO}$, for example, proposes to merge similar query vertices (same label and neighbours) to minimize the number of partial matches and perform the Cartesian product to expand the matches at the end. 
CFL~\cite{cfl}  decomposes the query into a dense subgraph and a forest, and process the dense subgraph first to reduce the number of partial matches. CFL also uses an index called {\em compact path index (CPI)} which estimates the number of matches for each root-to-leaf query path in the query and is used to enumerate the matches as well. We compare our approach to CFL in Appendix~\ref{app:cfl}.
A systematic comparison of our approach against these approaches is beyond the scope of this paper. 
Our approach is specifically designed to be directly implementable on any DBMS that adopts a cost-based optimizer and decomposable operator-based query plans. In contrast, these algorithms do not seem easy to decompose into a set of database operators. Studying how these algorithms can be turned into database plans is an interesting area of research. 

Another group of algorithms index different structures in input graphs, such as frequent paths, trees, or triangles, to speed up query evaluation~\cite{zhao:index, yan:index}.  Such approaches can be complementary to our approach. For example, reference~\cite{ammar:bigjoin} in the distributed setting demonstrated how to speed up GJ-based WCO plans by indexing triangles in the graph.

\noindent {\bf Cardinality Estimation using Small-size Graph Patterns:}
Our catalogue is closely related to Markov tables~\cite{aboulnaga:markov}, and MD- and Pattern-tree summaries from reference~\cite{maduko:md-tree}. Similar to our catalogue, both of these techniques store information about small-size subgraphs to make cardinality estimates for larger subgraphs. Markov tables were introduced to estimate cardinalities of paths in XML trees and store exact cardinalities of small size paths to estimate longer paths. MD- and Pattern-tree techniques store exact cardinalities of small-size acyclic patterns, and are used to estimate the cardinalities of larger subgraphs (acyclic and cyclic) in general graphs. These techniques are limited to cardinality estimation and store only acyclic patterns. In contrast, our catalogue stores information about acyclic and cyclic patterns and is used for both cardinality and i-cost estimation. In addition to selectivity ($\mu$) estimates that are used for cardinality estimation, we store information about the sizes of the adjacency lists (the $|A|$ values), which allows our optimizer to differentiate between WCO plans that generate the same number of intermediate results, so have same cardinality estimates, but incur different i-costs. Storing cyclic patterns in the catalogue allow us to make accurate estimates for cyclic queries.


\vspace{-7pt}
\section{Conclusions}
\label{sec:conclusion}

\noindent
We described a cost-based dynamic programming optimizer that enumerates a plan space that contains WCO plans, BJ plans, and a large class of hybrid plans. Our i-cost metric captures the several runtime effects of QVOs we identified through extensive experiments. Our optimizer generates novel hybrid plans that seamlessly mix intersections with binary joins, which  are not in the plan space of prior optimizers for  subgraph queries.
Our approach has several limitations which give us directions for future work. 
First, our optimizer can benefit from more advanced cardinality and i-cost estimators, such as those based on sampling outputs or machine learning. Second, for very large queries, currently our optimizer enumerates a limited part of our plan space. Studying faster plan enumeration methods, similar to those discussed in~\cite{Neumann:2018}, is an important future work direction. Finally, existing literature on subgraph matching has several optimizations, such as factorization~\cite{olteanu:size-bounds} or postponing the Cartesian product optimization~\cite{cfl}, for evaluating identifying and evaluating independent components of a query separately. We believe these are efficient optimizations that can be integrated into our optimizer.

\small{
\bibliographystyle{abbrv}
\bibliography{references}
}
\balance
\appendix

\iflong
\section{Subsumed EH Plans}
\label{app:eh-plan-space}

We show that our plan space contains the EH's GHD-based plans that satisfy the projection constraint. For details on GHDs how EH picks GHDs we refer the reader to reference~\cite{aberger:eh}.  Briefly, a GHD $D$ of $Q$ is a decomposition of $Q$ where each node $i$ is labeled with a sub-query $Q_i$ of $Q$. The interpretation of a GHD $D$ as a join plan is as follows: each sub-query is evaluated using Generic Join first and materialized into an intermediate table. Then, starting from the leaves, each table is joined into its parent in an arbitrary order. So a GHD can easily be turned into a join plan $T$ in our notation (from Section~\ref{sec:full-plan-space}) by ``expanding'' each sub-query $Q_i$ into a WCO (sub-) plan according to the $\sigma$ that EH picks for $Q_i$ and adding intermediate nodes in $T$ that are the results of the joins that EH performs.
Given $Q$, EH picks the GHD $D^*$ for $Q$ as follows. First, EH loops over each GHD $D$ of $Q$, and computes the worst-case size of the subqueries, which are computed by the AGM bounds of these queries (i.e., the minimum {\em fractional edge covers} of sub-queries; see~\cite{atserias:agm}). The maximum size of the subqueries is the width of GHD and the GHD with the minimum width is picked. This effectively implies that one of these GHDs satisfy our projection constraint. This is because adding a missing query edge to $Q_i(V_i, E_i)$ can only decrease its fractional edge cover.
To see this consider $Q_i'(V_i, E_i')$, which contains $V'$ but also any missing query edge in $E_i$. Any fractional edge cover for $Q_i$ is a fractional edge cover for $Q_i'$ (by giving weight 0 to $E_i' \minus E_i$ in the cover), so the minimum fractional edge cover of $Q_i'$ is at most that for $Q_i$, proving that $D^*$ is in our plan space. 

We verified that for every query from Figure~\ref{fig:queries-used}, the plans EH picks satisfy the projection constraint. However, there are minimum-width GHDs that do not satisfy this constraint. For example, for Q10, EH finds two minimum-width GHDs: (i) one that joins a diamond and a triangle (width 2); and (ii) one that joins a three path ($a_2a_1a_3a_4$) joined with a triangle with an extended edge (also width 2). The first GHD satisfies the projection constraint, while the second one does not. EH (arbitrarily) picks the first GHD. As we argued in Section~\ref{subsec:hybrid-plan-space}, satisfying the projection constraint is not a disadvantage, as it makes the plans generate fewer intermediate tuples. For example, on a Gnutella peer-to-peer graph~\cite{snap} (neither GHD finished in a reasonable amount of time on our datasets from Table~\ref{table:datasets-used}), the first GHD for Q10 takes around 150ms, while the second one does not finish within 30 minutes. 

\fi

\balance
\section{Catalogue Experiments}
\label{app:catalogue}

We present preliminary experiments to show two tradeoffs: (1) the space vs estimation quality tradeoff that parameter $h$ determines; and (2) construction time vs estimation quality tradeoff that parameter $z$ determines. For estimation quality we evaluate cardinality estimation and omit the estimation of adjacency list sizes, i.e., the $|A|$ column, that we use in our i-cost estimates.
We first generated all 5-vertex size unlabeled queries. This gives us 535 queries. For each query, we assign labels at random given the number of labels in the dataset (we consider Amazon with 1 label, Google with 3 labels). Then for each dataset, we construct two sets of catalogues: (1) we fix $z$ to 1000, and construct a catalogue with $h$=$2$, $h$=$3$, and $h$=$4$ and record the number of entries in the catalogue; (2) we fix $h$ to 3 and construct a catalogue with $z$=$100$, $z$=$500$, $z$=$1000$, and $z$=$5000$ and record the construction time. Then, for each labeled query $Q$, we first compute its actual cardinality, $|Q_{true}|$, and record the estimated cardinality of $Q$, $Q_{est}$ for each catalogue we constructed. Using these estimation we record the q-error of the estimation, which is max($|$Q$_{est}$$|$ / $|$Q$_{true}$$|$, $|$Q$_{true}$$|$ / $|$Q$_{est}$$|$). This is an error metric used in prior work cardinality estimation~\cite{leis:cardinality} that is at least 1, where 1 indicates completely accurate estimation. As a very basic baseline, we also compared our catalogues to the cardinality estimator of PostgreSQL. For each dataset, we created an Edge relation $E$(from, to). We create two composite indexes on the table on (from, to) and (to, from) which are equivalent to our forward and backward adjacency lists. We collected stats on each table through the ANALYZE command. We obtain PostgreSQL's estimate by writing each query in an equivalent SQL select-join query and running EXPLAIN on the SQL query.

Our results are shown in Tables \ref{table:q-error-z} and \ref{table:q-error-h} as cumulative distributions as follows: for different q-error bounds $\tau$, we show the number of queries  that a particular catalogue estimated with q-error at most $\tau$. As expected, larger $h$ and larger $z$ values lead to less q-error, while respectively yielding larger catalogue sizes and longer construction times,. The biggest q-error differences are obtained when moving from $h$=$3$ to $h$=$4$ and $z$=$100$ to $z$=$500$. There are a few exception $\tau$ values when the larger h or z values lead to very minor decreases in the number of queries within the $\tau$ bound but the trend holds broadly.

\begin{table}[H]
	\captionsetup{justification=centering}
	\centering
	\begin{tabular}{ m{0.2cm} m{0.7cm} m{0.8cm} m{0.5cm} m{0.5cm} m{0.5cm} m{0.6cm} m{0.7cm} m{0.7cm} }\toprule
		& $z$ & time(s) & $\leq$2 & $\leq$3 & $\leq$3 & $\leq$5 & $\leq$10 & $>$20 \\
		\midrule
		
		Am & 
		
		100 
		
		500 
		
		1,000
		
		5,000  & 
		
		0.1 
		
		0.3 
		
		0.5
		
		1.5 &
		
		318
		
		384
		
		383
		
		384 &
		
		445
		
		486
		
		481
		
		475 &
		
		510
		
		520
		
		519
		
		518 &
		
		526
		
		527
		
		529
		
		529 &
		
		529
		
		530
		
		532
		
		532 &
		
		535
		
		535
		
		535
		
		535\\\midrule
		
		Go$_3$ & 
		
		100
		
		500
		
		1,000
		
		5,000  & 
		
		3.1
		
		9.3
		
		17.0
		
		66.1 &
		
		166
		
		214
		
		222
		
		219 &
		
		276
		
		310
		
		315
		
		322 &
		
		356
		
		371
		
		371
		
		373 &
		
		415
		
		430
		
		430
		
		432 &
		
		461
		
		477
		
		475
		
		473 &
		
		535
		
		535
		
		535
		
		535 \\
		
		\bottomrule
	\end{tabular}
	\vspace{-5pt}
	\caption{Q-error and construction time for different $z$ values.}
	\label{table:q-error-z}
	\vspace{-8pt}
\end{table}

\begin{table}[H]
	\captionsetup{justification=centering}
	\centering
	\begin{tabular}{ m{0.2cm} m{0.2cm} m{0.2cm} m{0.7cm} m{0.5cm} m{0.5cm} m{0.5cm} m{0.6cm} m{0.7cm} m{0.7cm} }\toprule
		& & h & $|$entries$|$ & $\leq$2 & $\leq$3 & $\leq$3 & $\leq$5 & $\leq$10 & $>$20 \\
		\midrule
		
		Am & 
		
		\texttt{GF} &
		
		2 
		
		3
		
		4 &
		
		8
		
		138
		
		2858 &
		
		348
		
		381
		
		498 &
		
		464
		
		482
		
		510 &	
		
		512
		
		512
		
		518 &
		
		523
		
		524
		
		524 &
		
		527
		
		527
		
		527 &
		
		535
		
		535
		
		535 \\
		
		& \texttt{PG} & - & - & 15 & 15 & 23 & 23 & 25 & 535 \\\midrule
		
		Go$_3$ & 
		
		\texttt{GF} &
		
		2
		
		3
		
		4 &
		
		144
		
		20.3K
		
		11.9M &
		
		181
		
		222
		
		441 &
		
		289
		
		315
		
		497 &
		
		375
		
		371
		
		515 &
		
		447
		
		430
		
		524 &
		
		492
		
		475
		
		529 &
		
		535
		
		535
		
		535 \\
		
		& \texttt{PG} & - & - & 0 & 0 & 0 & 0 & 0 & 535 \\
		
		\bottomrule
	\end{tabular}
	\vspace{-5pt}
	\caption{Q-error and catalogue size (MB) for different $h$ values.}
	\label{table:q-error-h}
\end{table}
\vspace{-5pt}

\iflong
\section{CFL Comparison}
\label{app:cfl}
\else
\section{CFL Comparison}
\label{app:cfl}
\vspace{1pt}
\fi

\begin{table}[H]
	\captionsetup{justification=centering}
	\centering
	\begin{tabular}{ m{0.25cm} m{0.4cm} m{1.78cm} m{1.95cm} m{2.2cm} } \toprule
		$|$T$|$ & & \textbf{Q10s} &  \textbf{Q15s} &  \textbf{Q20s} \\
		
		\midrule
		
		$10^5$ & \texttt{GF} 		
		
		\texttt{CFL} & 7.3
		
		9.3 (\textbf{1.2x}) & 6.0
		
		17.5 (\textbf{2.9x}) & 5.5
		
		40.5 (\textbf{7.3x}) \\\midrule
		
		$10^8$ & \texttt{GF} 		
		
		\texttt{CFL} & 625.6
		
		4,818.9 (\textbf{7.7x}) & 665.5
		
		5,898.1 (\textbf{8.8x}) & 797.2
		
		7,104.1 (\textbf{8.9x}) \\\midrule
		
		& & \textbf{Q10d} &  \textbf{Q15d} &  \textbf{Q20d} \\
		\midrule
		
		$10^5$ & \texttt{GF} 		
		
		\texttt{CFL} & 29.2 (\textbf{2.2x})
		
		13.2 & 99.8
		
		389.9 (\textbf{3.9x}) & 142.0
		
		1,140.7 (\textbf{8.0x}) \\\midrule
		
		$10^8$ & \texttt{GF} 		
		
		\texttt{CFL} & 1,159.6
		
		7,974.3 (\textbf{6.8x}) & 1,906.2
		
		11,656.2 (\textbf{6.1x}) & 1,556.9
		
		19,135.7 (\textbf{12.2x})\\\bottomrule
	\end{tabular}
	\caption{Average run-time (secs) of Graphflow (\texttt{GF}) and \texttt{CFL}.}
	\label{table:cfl-vs-gf}
\end{table}

CFL~\cite{cfl} is an efficient algorithm in literature that can evaluate labeled subgraph queries as in our setting. The main optimization of CFL is what is referred to as ``postponing Cartesian products'' in the query. Essentially, these are (conditionally) independent parts of the query that can be matched separately and appear as Cartesian products in the output. CFL decomposes a query into a dense {\em core} and a {\em forest}. Broadly, the algorithm first matches the core, where fewer matches are expected and there is less chance of independence between the parts. Then the forest is matched. In both parts, any detected Cartesian products are postponed and evaluated independently. This reduces the number of intermediate results the algorithm generates. CFL also builds an index called CPI, which is used to quickly enumerate matches of paths in the query during evaluation.
We follow the setting from the evaluation section of reference~\cite{cfl}. We obtained the CFL code and 6 different query sets used in reference~\cite{cfl} from the authors. Each query set contains 100 randomly generated queries that are either sparse (average query vertex degree $\leq$ 3) or dense  (average query vertex degree $>$ 3). We used three sparse query sets Q10s, Q15, and Q20s containing queries with 10, 15, and 20 query vertices, respectively. Similarly, we used three dense query sets Q10d, Q15d, and Q20d. To be close to their setup, we use the \texttt{human} dataset from the original CFL paper. The dataset contains $86282$ edges, $4674$ vertices, $44$ distinct labels. We report the average run-time per query for each query set when we limit the output to $10^5$ and $10^8$ matches as done in reference~\cite{cfl}.
Table~\ref{table:cfl-vs-gf} compares the runtime of Graphflow and CFL on the 6 query sets. Except for one of our experiments, on Q10d with 10$^5$ output size limit, Graphflow's runtimes are faster (between 1.2x to 12.2x) than CFL. We note that although our runtime results are faster than CFL on average, readers should not interpret these results as one approach being superior to another. For example, we think the postponing of Cartesian products optimization and a CPI index are good techniques and can improve our approach.  However, one major advantage of our approach is that we do flat tuple-based processing using standard database operators, so our techniques can easily be integrated into existing graph databases. It is less clear how to decompose CFL-style processing into database operators.

\iflong
\section{Neo4j Comparison}
\label{app:neo4j}
\vspace{4pt}

Neo4j is perhaps the most popular graph DBMS that uses BJ plans to evaluate queries. We used Neo4j v.3.1.0. Our runtime results are significantly faster (up to 837x) as show in Table~\ref{table:neo4j-vs-gf} and we expect similar results against systems using BJ plans. Aside from our advantage of using WCO plans with good QVOs, our implementation has several advantages against Neo4j: (1) Graphflow is a prototype system, which is inherently more efficient as it supports fewer features; and (2) instead of Neo4j's linked lists storing Java objects, our graph store is backed by Java primitive type arrays, which are faster in lookups; (3) we store our adjacency list in sorted vertex ID order. Similar to our note above, although baseline comparisons as our Neo4j comparisons are common in the database research community, we think there is little to learn from these experiments. Neo4j is not optimized for the complex subgraph queries we study in this paper.

\begin{table}[H]
	\captionsetup{justification=centering}
	\centering
	\begin{tabular}{ m{0.5cm} m{1cm} m{1.65cm} m{1.65cm} m{1.8cm} }\toprule
		& & Q1 &  Q2 &  Q4 \\
		\midrule
		Am & \texttt{GF} 
		
		\texttt{Neo4j} & 0.7 
		
		332.1 (\textbf{474x}) & 5.7
		
		TL & 4.8
		
		745.2 (\textbf{155x}) \\\midrule
		
		Ep & \texttt{GF}
		
		\texttt{Neo4j} & 0.6
		
		502.4 (\textbf{837x}) & 42.2 
		
		TL & 1.5
		
		TL\\\bottomrule
	\end{tabular}
	\caption{Run-time (secs) of Graphflow (\texttt{GF}) and \texttt{Neo4j}.\\ $TL$ indicates the query did not finish in 30 mins.}
	\label{table:neo4j-vs-gf}
\end{table}
\fi

\end{document}